\documentclass[12pt,letterpaper]{article}

\usepackage{amsmath,amssymb}
\usepackage{iftex}
\ifPDFTeX
  \usepackage[T1]{fontenc}
  \usepackage[utf8]{inputenc}
  \usepackage{textcomp} % provide euro and other symbols
\else % if luatex or xetex
  \usepackage{unicode-math}
  \defaultfontfeatures{Scale=MatchLowercase}
  \defaultfontfeatures[\rmfamily]{Ligatures=TeX,Scale=1}
\fi
\usepackage{lmodern}
\ifPDFTeX\else  
\fi
\IfFileExists{upquote.sty}{\usepackage{upquote}}{}
\IfFileExists{microtype.sty}{% use microtype if available
  \usepackage[]{microtype}
  \UseMicrotypeSet[protrusion]{basicmath} % disable protrusion for tt fonts
}{}
\makeatletter
\@ifundefined{KOMAClassName}{% if non-KOMA class
  \IfFileExists{parskip.sty}{%
    \usepackage{parskip}
  }{% else
    \setlength{\parindent}{0pt}
    \setlength{\parskip}{6pt plus 2pt minus 1pt}}
}{% if KOMA class
  \KOMAoptions{parskip=half}}
\makeatother
\usepackage{xcolor}
\makeatletter
\ifx\paragraph\undefined\else
  \let\oldparagraph\paragraph
  \renewcommand{\paragraph}{
    \@ifstar
      \xxxParagraphStar
      \xxxParagraphNoStar
  }
  \newcommand{\xxxParagraphStar}[1]{\oldparagraph*{#1}\mbox{}}
  \newcommand{\xxxParagraphNoStar}[1]{\oldparagraph{#1}\mbox{}}
\fi
\ifx\subparagraph\undefined\else
  \let\oldsubparagraph\subparagraph
  \renewcommand{\subparagraph}{
    \@ifstar
      \xxxSubParagraphStar
      \xxxSubParagraphNoStar
  }
  \newcommand{\xxxSubParagraphStar}[1]{\oldsubparagraph*{#1}\mbox{}}
  \newcommand{\xxxSubParagraphNoStar}[1]{\oldsubparagraph{#1}\mbox{}}
\fi
\makeatother

\usepackage{longtable,booktabs,array}
\usepackage{calc} % for calculating minipage widths
\usepackage{etoolbox}
\makeatletter
\patchcmd\longtable{\par}{\if@noskipsec\mbox{}\fi\par}{}{}
\makeatother
\IfFileExists{footnotehyper.sty}{\usepackage{footnotehyper}}{\usepackage{footnote}}
\makesavenoteenv{longtable}
\usepackage{graphicx}
\makeatletter
\def\maxwidth{\ifdim\Gin@nat@width>\linewidth\linewidth\else\Gin@nat@width\fi}
\def\maxheight{\ifdim\Gin@nat@height>\textheight\textheight\else\Gin@nat@height\fi}
\makeatother

\def\vv{1.75}

\usepackage{algorithm,algorithmicx,algpseudocode}
\usepackage[authoryear,round,compress,sort]{natbib}
\usepackage{amsmath,amssymb,amsfonts, amsbsy,amsthm, epsfig,  graphicx,rotating, tabularx, booktabs, titling}
\usepackage{amsmath, amssymb, amsfonts, amsbsy, epsfig, graphicx,rotating, subfigure}
\usepackage{multirow,xspace,setspace,booktabs,subfigure}
\usepackage{enumerate}
\usepackage{ulem}
\usepackage{float}
\usepackage{tabularx}
\usepackage{capt-of}
\usepackage{lipsum}
\usepackage{tabulary}
\usepackage{thmtools}
\usepackage{diagbox}
\usepackage{xcolor}
\usepackage{enumitem}
\usepackage{makecell}
\RequirePackage[colorlinks,citecolor=blue,urlcolor=red]{hyperref}
 
 \usepackage{lipsum}       % 示例文本
 \usepackage{sectsty}
 \usepackage{titlesec}
 \usepackage{helvet}
 \usepackage{etoolbox}
 \usepackage{stmaryrd}
 \usepackage{appendix}
 
 \sectionfont{\fontsize{12}{12}\selectfont\sffamily\bfseries}
\titleformat{\subsection}{\normalfont\fontsize{12}{12}\sffamily\itshape}{\thesubsection}{1em}{}
 \titleformat{\subsubsection}{\normalfont\fontsize{12}{12}\itshape}{\thesubsubsection}{1em}{}

\usepackage{fancyhdr}
\usepackage{tikz}

\usepackage[letterpaper, left=0.95in, right=0.95in, top=0.8in, bottom=0.8in]{geometry}

\newcommand{\beq}{\begin{eqnarray*}}
\newcommand{\eeq}{\end{eqnarray*}}
\newcommand{\beqn}{\begin{eqnarray}}
\newcommand{\eeqn}{\end{eqnarray}}

\newcommand{\bi}{\begin{itemize}}
\newcommand{\ei}{\end{itemize}}

\newcommand{\be}{\begin{equation}}
\newcommand{\ee}{\end{equation}}

\newcommand{\bfm}[1]{\ensuremath{\mathbf{#1}}}

\def\bi{\bfm i}

          \def\cR{{\cal  R}}
          \def\cS{{\cal  S}}

\newcommand{\R}{\mathbb{R}}

\newcommand{\clambda}{C_{\lambda}}

\newcommand{\ld}{\sqrt{n}\log n}

\newcommand{\hm}{\hat{m}}

\algrenewcommand\algorithmicrequire{\textbf{Input:}}
\algrenewcommand\algorithmicensure{\textbf{Output:}}

\renewcommand{\[}{\left\[}
\renewcommand{\]}{\right\]}

\newcommand{\s}{\boldsymbol s}

\newcommand{\argmin}{\mathop{\rm arg\min}}

\numberwithin{equation}{section}
\theoremstyle{plain}

\newtheorem{theorem}{Theorem}
\newtheorem{assumption}{Assumption}

\newtheorem{definition}{Definition}

\newtheoremstyle{boldremark}
  {3pt}
  {3pt}
  {\normalfont}
  {}
  {\bfseries}
  {.}
  { }
  {}

\theoremstyle{boldremark}
\newtheorem{remark}{Remark}

\def\be{\begin{equation}}
\def\ee{\end{equation}}

\usepackage{hyperref}
\usepackage{cleveref}

\begin{document}

\renewcommand{\baselinestretch}{\vv}\selectfont

\title{Optimal Spatial Anomaly Detection}

\author{\small Baiyu Wang and Chao Zheng}

\date{\small {School of Mathematical Sciences, University of Southampton}}

\vspace{-1.5cm}

\maketitle

\begin{abstract}
There has been a growing interest in anomaly detection problems recently, whilst their focuses are mostly on anomalies taking place on the time index. In this work, we investigate a new anomaly-in-mean problem in multidimensional spatial lattice, that is, to detect the number and locations of anomaly ``spatial regions'' from the baseline. In addition to the classic minimization over the cost function with a $L_0$ penalization, we introduce an innovative penalty on the area of the minimum convex hull that covers the anomaly regions. We show that the proposed method yields a consistent estimation of the number and locations of spatial anomalies. Under the minimax framework, we characterize the optimal detection error for multidimensional spatial anomaly detection problem and reveal the trade-off between detection performance and the geometric flexibility of anomaly region shapes.  Large-scale Monte Carlo simulations are carried out to examine the numeric performance of the method. The method has a wide range of applications in real-world problems. As an example, we apply it to detect the marine heatwaves using the sea surface temperature data from the European Space Agency.
\end{abstract}


% \section{Introduction}   \lbl{sec:intro}
\section{Introduction} \label{sec:preliminary}

Anomaly detection is a long-standing challenge in engineering, physics, environmental and social sciences, concerned with identifying observations whose values are statistically improbable compared to a given baseline distribution. In many practical applications, anomalies arise in data that are indexed by spatial locations, for example, detecting colorectal cancer by searching tumour regions in  histology slides \citep{gu2023using}; segmenting anomalous areas corresponding to deforestation and burn scars in multi-source satellite imagery \citep{fodor2023rapid};  finding burglary hotspots through crime rates in cities \citep{kalantari2020develop}. In these problems, an anomaly often refers to a collection of points or units on a spatial map, forming regions with possibly very complex shapes.

As a motivating example of this work, oceanographers and climate scientists are interested in studying marine heatwaves (MHWs), which are prolonged, extreme, extensive, and persistent warm water events that occur in the upper layers of the ocean \citep{chapman2022large, holbrook2020kepping}. Accurately identifying  anomalous oceanic regions affected by MHWs, by recovering their locations and spatial extents, is essential for effective monitoring, resource management, and climate impact assessment. This task is challenging especially because the MHW regions have complex and differing spatial shapes, which could be highly non-convex, with internal holes, and consisting of multiple disconnected components. To address this challenge,  we propose a new methodology that can implement automatic detection of complex spatial anomaly regions corresponding to  MHW events.

% There are several existing attempts to address the spatial anomaly detection (SAD) problem, including those based on scan statistics \citep{kulldorff1997scan, patil2003geographic, li2011spatial, tango2005flexibly}, which often 
%   fail to accommodate multiple anomalies with complex shapes, and  lack theoretical guarantees such as detection consistency. Graph-cut methods \citep{boykov2006graph}  do not impose topological constraints on the anomaly region, but are limited to detecting a single anomalous region relative to a baseline. More recently, deep learning techniques \citep{ronneberger2015unet, hansen2022anomaly, wan2022unsupervised} have been introduced to solve related problems. However, these methods usually focus on identifying outliers at the individual point or unit level, which contrasts with the objective of SAD in our study. Moreover, deep learning models generally require a large number of training datasets, which are not available in most real-world applications.

There are several attempts to address the spatial anomaly detection (SAD) problem in existing literature, yet each exhibits certain methodological or theoretical limitations. A major class of methods is based on scan statistics for detecting regularly \citep{kulldorff2006elliptic} and arbitrarily shaped \citep{tango2005flexibly, duczmal2004simulated} anomaly regions on spatial lattices, with subsequent extensions to graphs and networks \citep{sharpnack2013near, arias2011detection}. These methods typically target the identification of a single anomaly region. Although the iterative application of the scan statistic can yield heuristic detection of multiple anomalies, the results are governed by empirically tuned parameters and criteria.
Beyond scan statistics, graph-cut segmentation \citep{boykov2006graph} and level-set boundary methods \citep{li2011level, osher2001level} are proposed to detect multiple anomaly regions, without assessing statistical guarantees for detection accuracy. 
More recently, machine learning techniques have been introduced to solve related problems, including isolation forest \citep{cao2025anoamly}, support vector description \citep{tax2004support}, and neural network models \citep{ronneberger2015unet}. These works usually focus on identifying outliers at the individual or unit level, which contrasts with the objective of SAD in our study.  Another line of work is image segmentation, which can also address problem similar to SAD, such as Meta’s Segment Anything models \citep{kirllov2023segment}, although these typically require large-scale training data and also lack formal statistical guarantees.

To develop a rigorous statistical analysis of multiple SAD problem, we take inspiration from recent techniques in changepoint detection literature,  most of which consider a time-indexed sequence of observations \citep{killick2012optimal, wang2020univariate} and seek a set of changepoints  on the time indices that partition the sequence into  homogeneous segments. 

Direct application of existing changepoint detection methods to SAD is nontrivial, as the spatial scenario fundamentally differs from the timeline setting, particularly when the anomaly regions may exhibit complex geometric and topological structures in general dimensions.  
% \zc{which implies that spatial extent and point count can be measurements of distinct concepts. Considering a region formed by a closed boundary with internal holes, the spatial extent is the area enclosed by the outer boundary, treating the holes as filled, whereas the number of points counts only locations within the region.} Note that in the timeline setting,  anomaly regions are just intervals, hence the spatial extent is defined as the length, which is equivalent to the point count.
%whereas in the timeline setting an anomaly region is a contiguous interval.   Consequently, spatial extent and point count of a spatial anomaly region can be measurements of distinct concepts, e.g., consider a closed spatial anomaly region with internal holes whose area and number of points in the region are different, while in the timeline setting the length and point count of an anomaly region are equivalent.
%spatial extent (i.e., length) and point count of an anomaly region are equivalent in the timeline setting, but they can be measurements of distinct concepts in a spatial anomaly region.    For example, consider a closed spatial anomaly region with internal holes whose area and number of points in the region are different.
In the timeline setting, each segment can be fully characterized by two boundary points. In contrast, identifying spatial regions is substantially more intricate. For instance, when a region has highly irregular boundaries, it may occur that every point on the region is effectively a boundary point. Consequently, SAD necessitates identifying all points within the anomaly region, rather than relying on a small number of boundary points. Furthermore, note that unlike the time index, spatial locations lack a natural total ordering. A large class of sequential timeline detection methods that rely on the search for changepoints along a predefined direction or path, including binary segmentation \citep{vostrikova1981detection,venkatraman1992consistency} and its variants \citep{fryzlewicz2014wild, cho2015multiple, kovacs2023seeded}, are not applicable. Similarly, efficient computational algorithms such as PELT \citep{killick2012optimal}, which achieve linear computational cost by sequentially removing candidate changepoints from future iterations when pruning conditions are met, are also not suitable for the SAD problem. Although artificial orderings, such as row- or column-major order, or partial order based on half- or quarter-plane constructions, can be imposed on spatial data, these do not fundamentally resolve the abovementioned problems.

A number of studies have extended changepoint detection to spatial-temporal scenario, but still assume that changes take place in the temporal domain \citep{gromenk2017detection,moore2025adaptive, zhao2024composite}. In more explicitly spatial settings, \cite{chan2022inference} studies the discrepancy-based statistic over small blocks to identify the boundary of regions with structural change, while \cite{kirch2023scan} develops a method based on contrasts of local-window means to localize spatial anomaly regions in image data. However, these approaches focus solely on identifying the locations of a regional change and do not address the problem of determining the number of such changes. \cite{fan2018approximate} uses $\ell_0$ edge-penalized least squares to recover piecewise constant regions on graphs, where the penalty is imposed on boundary complexity. \cite{madrid2011lattice} applies dyadic classification and regression trees (DCART) to partition the spatial lattice into multiple  constant mean regions under constraints on partition complexity. In related work,  \cite{yu2022optimal} generalizes this framework from regular lattices to graph data.

 % The proposed approach accommodates detection of anomaly regions of arbitrary shapes, where each anomaly region can consist of disconnected components. We establish theoretical analysis demonstrating that DPLS-SAD can consistently estimate the number of anomaly regions and detect their locations within an error range attaining the minimax optimal rate up to a logorithm factor.   

 % and propose an innovative double penalized least square approach for spatial anomaly detection (DPLS-SAD), which yields simultaneous estimation on both the number and the locations of spatial anomalies.

In this paper, we formally develop an anomaly-in-mean model on spatial lattice and propose an innovative double penalized least squares approach for spatial anomaly detection (DPLS-SAD). We provide theoretical guarantees for the consistent detection of anomaly regions with complex geometry, e.g., irregular and nonconvex shapes, internal holes, and disconnected components, including recovering their number and locations within an error bound. Moreover, we systematically establish minimax lower bounds via a new Assouad-type argument, which shows that the detection error achieved by DPLS-SAD is rate-optimal up to a logarithmic factor. In addition, we examine the necessity of imposing structural restrictions on anomaly regions in the SAD problem by establishing a minimax result over an unrestricted class of anomaly regions, which reveals a trade-off between the geometric flexibility of anomaly regions and the achievable detection rate. We further generalize DPLS-SAD to higher-dimensional settings and data with spatial dependence, and demonstrate that it can consistently recover anomaly regions under more complex scenarios.

% \wb{In addition, we address the computational challenges associated with solving the inherently non-convex and NP-hard optimization problem. To this end, we develop a dynamic programming-based search strategy that substantially improves computational efficiency, reducing the complexity from exponential to polynomial in the sample size.}

% We provide theoretical guarantees for the consistent detection of anomaly regions with complex geometry, e.g., irregular shapes, internal holes, and disconnected components, including recovering their number and locations within an error bound that attains the minimax optimal rate up to a logarithm factor. In addition, we address the computational challenges associated with solving the inherently non-convex and NP-hard optimization problem. To this end, we develop a dynamic programming-based search strategy that substantially improves computational efficiency, reducing the complexity from exponential to polynomial in the sample size.

The remainder of the paper is organized as follows. Section \ref{sec:modelset} introduces the SAD model setup and the proposed double penalized cost detection approach DPLS-SAD. Section \ref{sec:consistency} establishes theoretical guarantees for DPLS-SAD, followed by a minimax optimal localization rate analysis in Section \ref{sec:minimax}. Section \ref{sec:extend} extends our SAD problem to more general settings, including higher-dimensional and spatially dependent data.  In Section~\ref{sec:sim}, we present the simulation studies, and Section \ref{sec:realdata} applies our method to detect marine heatwave (MHW) events from the sea surface temperature (SST) data provided by the European Space Agency. Computational algorithms, additional simulation results, and technical proofs are deferred to the Supplementary Material.

\section{Detection of spatial anomalies} \label{sec:modelset}

In this section, we first formulate the spatial anomaly detection problem under an anomaly-in-mean setting and introduce the quantities of interest, such as the number and locations of spatial anomaly regions. Motivated by spatial anomaly patterns arising in real-world applications, we consider a smooth regional class of anomalies that retains substantial flexibility in shape while enforcing mild structural regularity. We then propose a new penalized cost function for estimating anomalies that extends the standard $L_0$-penalized criterion by incorporating a regional penalty on the dispersion of anomaly regions.

\subsection{Model setup and spatial anomalies} \label{subsec:model}
Consider a univariate process $\left\{Y(\s):  \s\in\cS \right\}$, where $\cS = \{(s_1,s_2)\} \subset \R^2$ is a set of spatial locations forming a regular 2D lattice over a rectangular spatial domain. 
Suppose $Y(\s)$ can be decomposed into two components:
\begin{equation*}
	Y(\s) = \mu(\s) + \varepsilon(\s), \qquad \s\in \mathcal{S} = \{1,2\dots, n_1\}\times \{1,2\dots, n_2\},
\end{equation*}
where   $n_1$ and $n_2$ are the sizes of the realized process on the horizontal and vertical coordinates, respectively. The total sample size is $n=n_1\times n_2$. Without loss of generality, we may assume $n$ is a square number and $n_1=n_2=\sqrt{n}$. 

Note that both our methodology and the theoretical results admit a direct extension to higher–dimensional settings, with $\cS=\{(s_1,s_2,\dots, s_d)\}\subset \R^d$, which will be discussed in Section \ref{sec:extend}. In the same section, we further allow for anisotropic growth of the lattice, in the sense that the dimension sizes may diverge at different rates, i.e., $n_1\neq n_2 \cdots \neq n_d$. 

The signal component $\{\mu(\s)\}_{{\s}\in\mathcal{S}}$ are deterministic, on which the anomalies take place. The stochastic error component $\left\{\varepsilon(\s)\right\}_{{\s}\in\mathcal{S}}$ are assumed to be independent sub-Gaussian, as specified in Assumption \ref{assmp:data} below, which is a standard assumption made in many existing timeline changepoint/anomaly detection literatures.
\begin{assumption}\label{assmp:data} (Sub-Gaussian errors) The errors $\{\varepsilon(\s)\}_{\s\in\cS}$ are independent centered sub-Gaussian random variables with $\|\varepsilon(\s)\|_{\psi_2}^2\le \sigma^2$ for all $\s\in\cS$.
\end{assumption}
Here $\|\cdot\|_{\psi_2}$ denotes the Orlicz-$\psi_2$ norm, i.e., 
$\|\varepsilon(\s)\|_{\psi_2}=\inf\{t>0, \mathbb{E} e^{\varepsilon(\s)^2/t^2}\le 2\}$.
%for a random variable $Y$, $\|Y\|_{\psi_2}=\inf\{t>0, \mathbb{E} e^{Y^2/t^2}\le 2\}$. 
The assumption of independent errors may be restrictive for many real-world spatial applications. We relax this assumption in Section \ref{sec:extend} and extend the analysis to accommodate spatially correlated data.

Assume that $\mathcal{S}$ can be partitioned into $m^*+1$ non-overlapping regions: a baseline region ${R}^*_0$ and 
%from which the signal deviates in 
$m^*$ anomaly regions, i.e., ${R}^*_1, {R}^*_2,..., R^*_{m^*}$,  such that $\mu({\s})$ is invariant within each anomaly region, while being different from the baseline:
\begin{equation*}
	\mu({\s}) = \mu^*_{j}, \,\,\,\forall \s \in R^*_j \quad\mbox{and}\quad \mu^*_{j}\neq  \mu^*_{0},  \qquad j=1,\dots, m^*. 
\end{equation*} 
In this way, $\mu({\s})$ is a region-wise constant mean signal. Unlike many existing settings, we do not require anomaly regions to have distinct mean signal levels. Instead, we allow different anomaly regions to exhibit similar or even identical mean signals, thereby accommodating challenging scenarios in real-world data. For instance, in the MHW application, El Niño events may generate anomalies with very close sea surface temperatures while occurring in different locations and years. The number of anomaly regions $m^*$,  their partitions $\{R^*_1,\dots, R^*_{m^*}\}$, and the mean signals $\{\mu^*_1, \dots, \mu^*_{m^*}\}$ are unknown, which are the quantities of interest in SAD problem.

The anomalies $\{R^*_{1}, \dots, R^*_{m^*}\}$ are defined as spatial regions, which are essentially collections of lattice points, see Figure \ref{fig:illustration} below. 
This is similar in spirit to the concept of collective anomalies \citep{fisch2022linear} in the timeline setting (when $d=1$), where each timeline anomaly is an interval and can be identified by its endpoints. However, as discussed in Section \ref{sec:preliminary}, boundary points are not particularly useful for spatial anomalies, and in many SAD settings it is actually necessary to specify all lattice points contained in the region.

\begin{figure}[H]
	\centering
	\includegraphics[scale = 0.1]{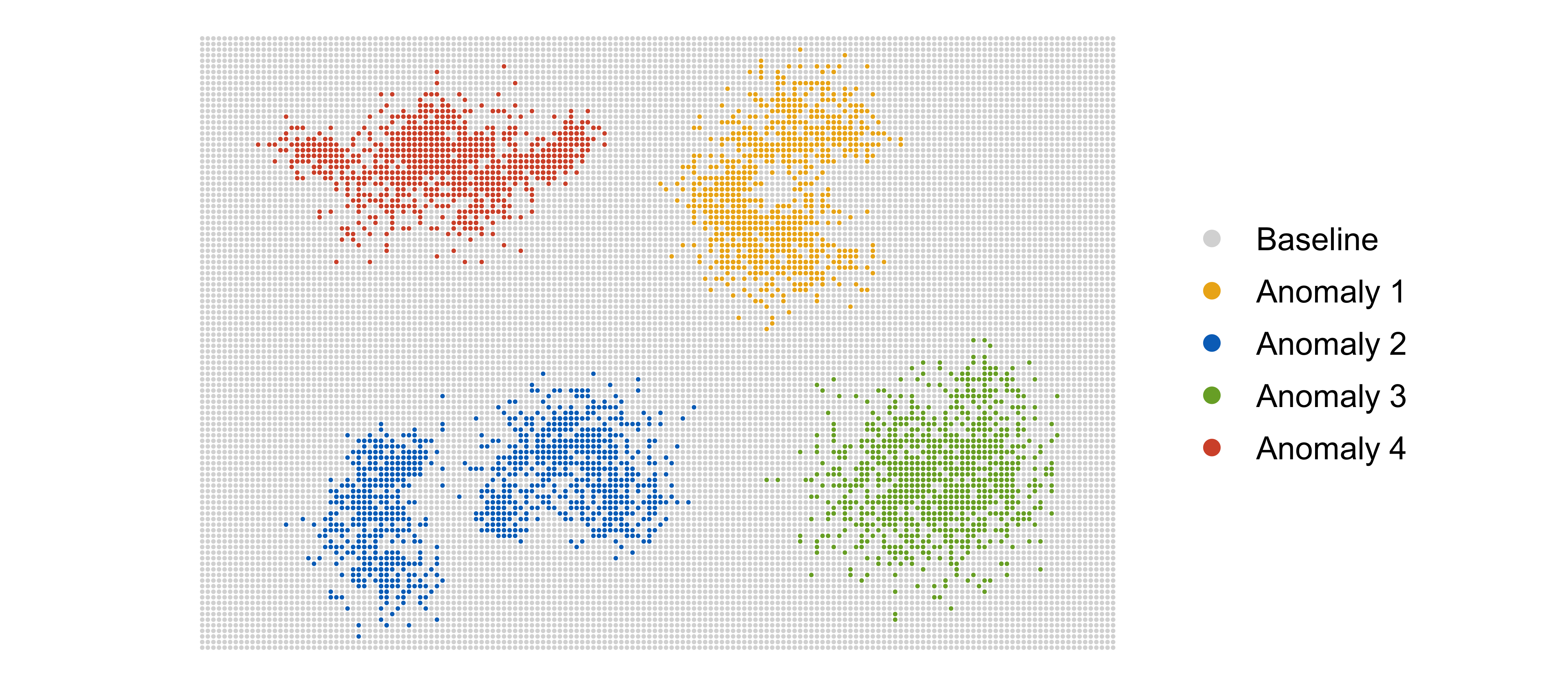}
	\vspace{-0.3cm}
	\caption{Illustration of 2D spatial anomaly regions (each formed by a collection of points, highlighted in same color), and the baseline region is plotted in grey points.}
	\label{fig:illustration}
\end{figure}

Next, we introduce some notations that will be used throughout this paper. For a spatial region $R$, we write $|R|$ as its cardinality, and denote  $\bar{Y}_R = \frac{1}{|R|} \sum_{\s \in R} Y(\s)$ as the regional sample average. We denote $\bar \mu_R= \frac{1}{|R|}\sum_{\s\in R}\mu(\s)$ as the regional average of the mean signals. For two regions $R$ and $R'$, we denote $R\,\backslash \, R'=R\,\backslash\,(R\cap R')$ as the regional  subtraction.  When there is no ambiguity, we sometimes refer to the sets of regions $\{R_1, \dots, R_m\}$ and mean signals $\{\mu_1. \dots, \mu_m\}$ as $R_{1:m}$ and $\mu_{1:m}$, respectively. We denote the underlying true baseline and  anomaly regions, and their mean signals as $\{R^\ast_{0}, R^\ast_{1:m^*}\}$ and $\{\mu_0^\ast, \mu^\ast_{1:m^\ast}\}$, and denote the estimated versions as $\{\widehat R_{0}, \widehat R_{1:\hm}\}$ and $\{\hat \mu_0, \hat \mu_{1:\hm}\}$, where $m^\ast$ is the number of true anomalies and $\hm$ is its estimate. 

\subsection{Smooth regional class}\label{subsec:regionalclass}

Although spatial anomalies in real-world applications may have complex shapes, they rarely appear as highly fragmented collections of isolated locations. For instance, anomaly regions in MHW data may be irregular, non-convex, contain holes, or consist of several disconnected components, yet still exhibit local  geometry. Motivated by these characteristics, we consider the following class of ``smooth'' spatial regions.

\begin{definition}\label{defi:smooth} (Smooth regional class) Define the class of smooth regions as $\cR = \cR_{K}$, where $ 0<K<\infty$ is a finite integer that does not depend on $n$ and $m^*$, such that at any fixed horizontal coordinate $1\le s_1 \le n_1$,  each region in $\cR$ can be divided into at most $K$ consecutive segments. 
\end{definition}

\begin{figure}[H]
	\centering
	\begin{tabular}{cc}
		\includegraphics[scale = 0.22]{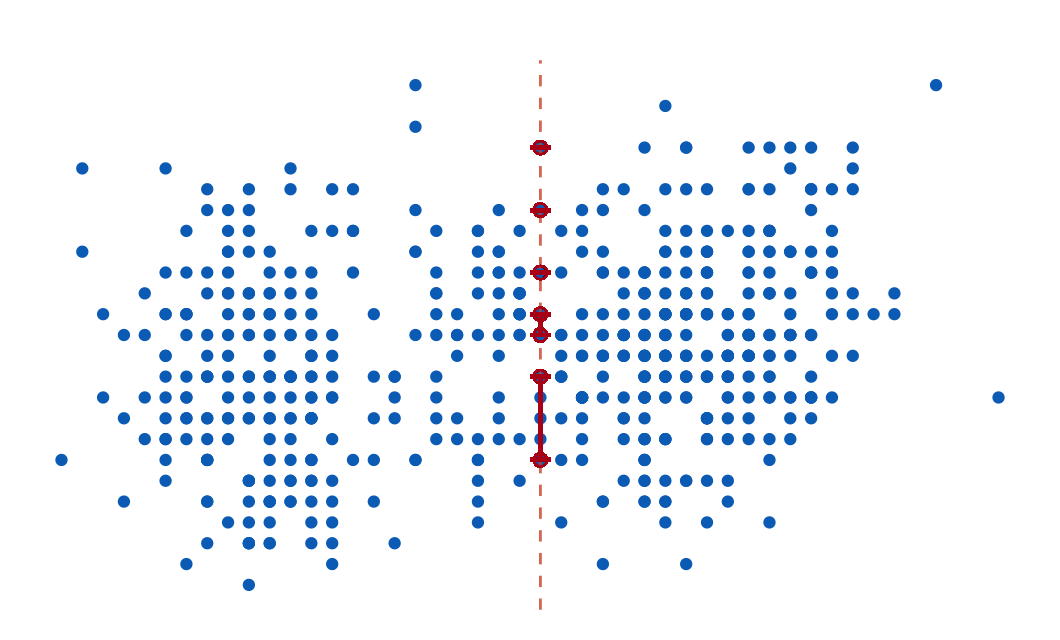} &\includegraphics[scale = 0.22]{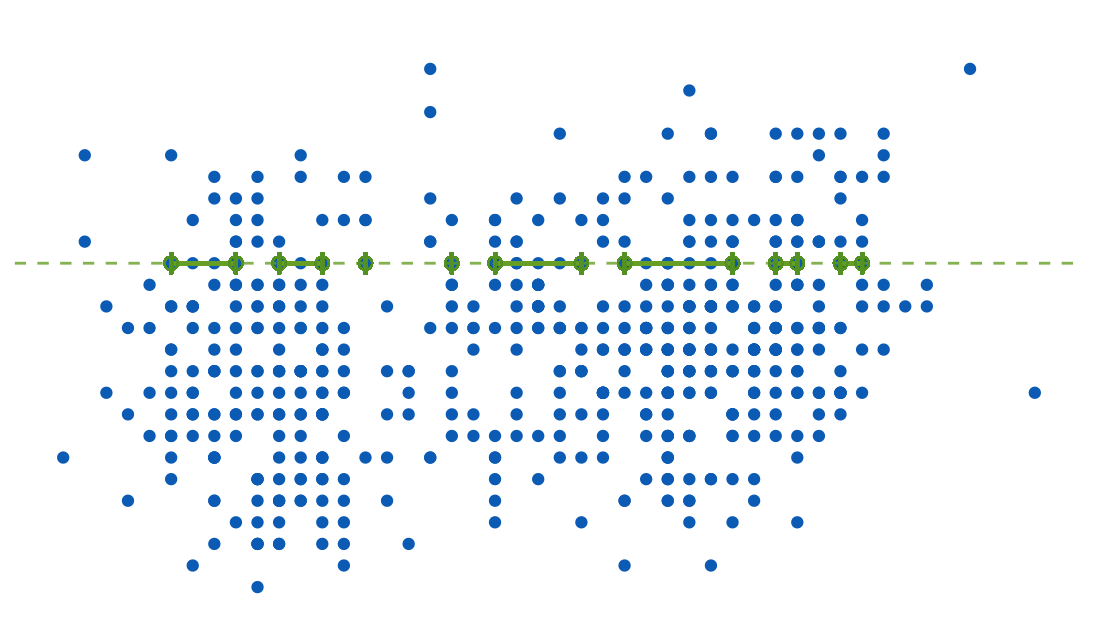}\\
		\hspace{0cm} (a)  & \hspace{0cm} (b) \\
	\end{tabular}
	\caption{ Illustration of smooth region in $\mathcal R$. 
    (a) At a vertical lattice line with a fixed horizontal coordinate  (points on the red line), the anomaly region can be divided into 5 segments, each is a collection of consecutive points. Any isolated point is counted as a single segment.
    (b) At a horizontal lattice line with a fixed vertical coordinate (points on the green line), the anomaly region can be divided into 8 segments.}
	\label{fig:smooth}
\end{figure}

The class $\mathcal{R}$ limits the number of consecutive segments on each vertical lattice line; see Figure~\ref{fig:smooth}\,(a) for an illustration. Equivalently, $\mathcal R$ can also be defined by imposing the same constraint on each horizontal lattice line, as shown in Figure~\ref{fig:smooth}\,(b).

The class $\mathcal{R}$ excludes only regions with excessively irregular geometry, such as highly non-smooth boundaries or an excessive number of holes or disconnected components. These restrictions are mild and preserve considerable flexibility, allowing anomaly regions to contain as many as $O(\sqrt{n})$ holes and disconnected components.

 \begin{assumption}\label{assmp:smooth}
 (Regional smoothness) Each true anomaly region $R_j^*$ belongs to the class of smooth regions, i.e., $R_j^*\in\mathcal{R}$ for  $j\in\{1,...,m^*\}$.
 \end{assumption}

 Assumption~\ref{assmp:smooth} implies that all the true anomaly regions should belong to the smooth regional class. In Section~\ref{sec:minimax}, we  study the SAD problem without imposing this assumption, and show that anomalies are impossible to be consistently detected by any method under the unrestricted setting.

% \wb{Assumption~\ref{assmp:smooth}, together with the definition of $\mathcal R$, specifies the regional target class for the SAD problem. 
% Although a spatial region is represented as a collection of lattice points, the anomaly regions studied here are restricted to collections with controlled geometric complexity. 
% The estimator introduced below is constructed over this regional class, and the subsequent localization and minimax results are established for true anomaly regions satisfying Assumption~\ref{assmp:smooth}.}

\subsection{Regional loss and double penalized cost function} \label{subsec:criterion}

Consider a loss function $L(R;\, \mu)$, which measures the fit to data of a region $R$ with a common mean signal $\mu$.  Often, appropriate losses are specified by parametrically modeling the data in the region, and then setting the  loss to be some seminal measures, e.g., the negative of the log-likelihood or $M$-estimation, under such a model. Throughout this paper, we employ the least squares loss, which has been extensively used in the change-in-mean and anomaly-in-mean detection literature (e.g., see \cite{killick2012optimal, wang2020univariate}). In this way, 
$$
L(R;\, \mu) := \dfrac{1}{\sigma^2} \sum_{\s \in R} \big(Y(\s)-\mu\big)^2.
$$
Minimizing over the mean signal $\mu$ implies  
$$ 
\Bar Y_{R}=\hat \mu_R := \argmin_{\mu} L(R;\,\mu).
$$
Hence, we write the minimized loss in region $R$ as $L(R):=L(R;\, \Bar Y_{R})$.
Next, for a set of anomaly regions $R_{1:m}$, the baseline region is defined as ${R}_0 = \mathcal{S}\setminus \bigcup_{j=1}^m {R}_j$.  
As is standard in the anomaly detection literature, we assume knowledge of the baseline mean value $\mu_0^\ast$ and the variance proxy $\sigma^2$ throughout the rest of the paper. Otherwise, they can be robustly estimated, for example by the sample median and the median absolute deviation, respectively \citep{fisch2022linear}. As a result, we always write $L(R_0)=\sum_{\s \in R} (Y(\s)-\mu_0^\ast)^2/\sigma^2$. 

In a vast number of timeline changepoint/anomaly detection works, 
the number and locations of anomalies are estimated by minimizing the $L_0$ penalized cost
\begin{equation}\label{eq:old_pen_cost}
L(R_{1:m}) + \beta m,
\end{equation}
which prevents overestimation of the number of anomalies by applying a penalty of $\beta$ to each estimated anomaly region; and  $L(R_{1:m}) := L(R_0) + \sum_{j=1}^m L(R_j)$ denotes the least squares loss of the partition $R_{1:m}$. However,  in the context of spatial anomaly detection, minimizing the penalized cost (\ref{eq:old_pen_cost}) does not yield a reliable estimate of $m^*$ and $R^*_{1:m^*}$.   
For example, when two spatially distant anomaly regions exhibit similar mean signal levels, the minimizer of (\ref{eq:old_pen_cost}) fails to distinguish between them. Therefore, we propose an additional regional penalty, which punishes the dispersion of each anomaly region. To define such a penalty, we introduce the following concept of the minimum convex hull. 

\begin{definition}(Minimum convex hull)
The minimum convex hull of a region $R$, denoted by $\text{Co}(R)$, is defined as the convex polygon with the fewest number of points that encloses $R$.
\end{definition}

\begin{figure}[H]
	\centering
    % \vspace*{0.2cm}
	\begin{tabular}{ccc}
		\includegraphics[scale = 0.18]{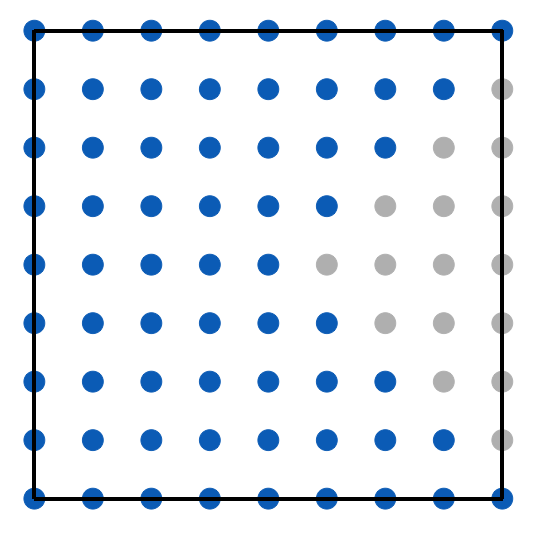} &\includegraphics[scale = 0.18]{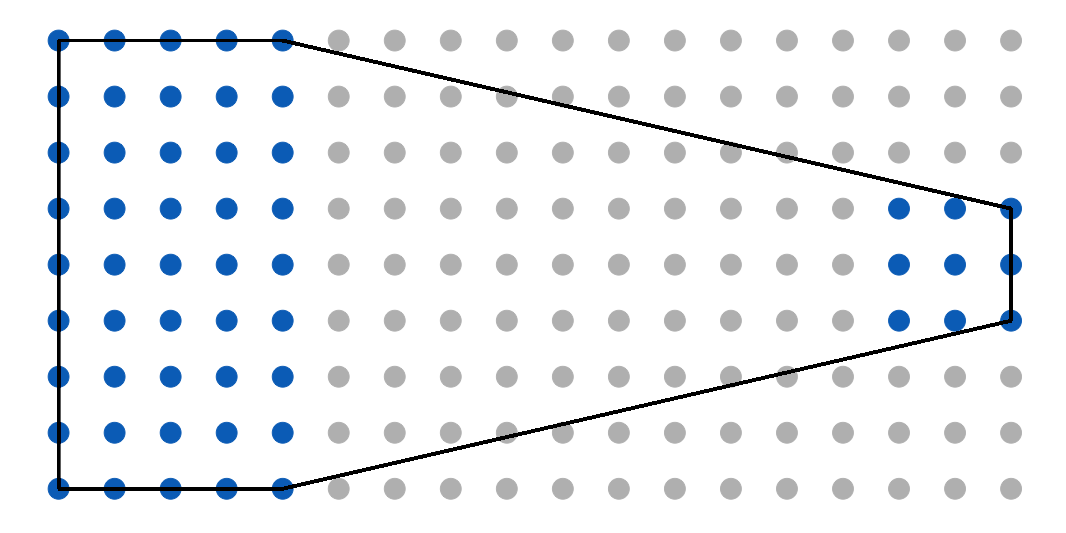} & \multirow{1}{*}[18.5ex]{\includegraphics[scale=0.3]{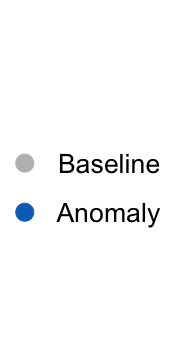}}	\vspace{-0.1cm}\\
	\small (a) & \small (b) &
	\end{tabular}
	\caption{Minimum convex hull (points contained within the solid line) of anomaly regions. (a) The anomaly region is concave, and its minimum convex hull is a square. (b) When add 9 distant points on the right side to the original anomaly region (whose minimum convex hull is a rectangular), the resulted new minimum convex hull encompasses many baseline points.}
	\label{fig:dist_convexhull3D}
\end{figure}

Figure \ref{fig:dist_convexhull3D}  illustrates examples of the minimum convex hull. It effectively captures the spatial dispersion of a region, and its cardinality depends not solely on the number of points within the region. Figure \ref{fig:dist_convexhull3D}\,(b) shows that adding a few distant points can substantially increase the cardinality of the minimum convex hull.

% Figure \ref{fig:dist_convexhull3D} illustrates  examples of the minimum convex hull.
% It efficiently captures the scatteredness of a region while its cardinality is not solely dependent on the number of points within the region. Figure \ref{fig:dist_convexhull3D}\,(b) shows that the cardinality of the minimum convex hull could increase significantly if we add a few distant points to the original region.

In this way, we estimate $m^\ast$ and $R^\ast_{1:m^\ast}$ by minimizing the following cost function, namely the double penalized least squares for spatial anomaly detection (DPLS-SAD):
\begin{equation*}
	C(m;\,{R}_{1:m}): =  L(R_{1:m}) + \beta m + \lambda \sum_{j=1}^m \big|\text{Co}({R}_j)\big|,
\end{equation*}
where both $\beta$ and $\lambda$ are penalty parameters. The $L_0$ penalty term $\beta m$ has been well-studied in the literature, whereas the newly introduced regional penalty, $\lambda \sum_{j=1}^m |\text{Co}({R}_j)|$, serves to regularize the solution by reducing the tendency to merge spatially distant anomaly regions. 
% \red{The regional penalty, $\lambda \sum_{j=1}^m |\mathrm{Co}(R_j)|$, regularizes the solution by penalizing regions whose points are overly scattered, thereby discouraging the merging of spatially distant anomalies. This penalty does not require the estimated regions to be convex. It only controls the overall spatial extent of each region through its convex hull, and therefore still allows flexible regional shapes, such as non-convex regions, regions with holes, and disconnected components.}
% A small $\lambda$ may lead to the incorrect merging of anomalies with close mean signal values, while a large $\lambda$ may cause true anomaly region to be over-segmented. 
The least squares loss and the two penalty terms jointly balance model fit, parsimony in anomaly detection, and the spatial compactness of anomaly regions.

Note that the minimal convex hull in the regional penalty can be replaced by alternative notions of geometric enclosure, such as the $\alpha$-shape, which generalizes the convex hull by permitting more flexible representations of a region.

% \zc{
% Several simple geometric quantities are not suitable as regional penalties. For example, area depends only on the cardinality of points and fails to capture their spatial scatteredness. Perimeter, total variation, and graph-cut size mainly describe local boundary structure and therefore tend to discourage regions with holes or irregular boundaries, both of which are allowed in the smooth regional class. However, if another quantity can capture spatial scatteredness without excluding such admissible geometries, the regional penalty can be replaced. One possbile choice is alpha shape, which generalizes the convex hull by allowing a more flexible polygon shape to enclose region. In this paper, we still adopt the minimum convex hull due to its simplicity, interpretability, and computational feasibility.}

% Whilst an appropriate selection of parameters $\beta$ and $\lambda$ is crucial for the detection results. 

% \zc{comments on the cost function}
% (这个里面是不是应该考虑baseline region，应该是m=0 和 m+1)

% \zc{delete the notation $\mathcal{C}(..)$}

\section{Consistency theory}\label{sec:consistency}

To establish theoretical guarantees for DPLS-SAD, which estimates the number of anomalies and recovers their regions on the lattice, we first introduce some preliminary notations and technical assumptions. 

Define the minimal anomaly region cardinality and the minimal anomaly signal  as
$$
\delta:=\min_{j=1,...,m^\ast}|R_j^\ast|  \quad \text{and} \quad \Delta:=\min_{j=1,...,m^*}|\mu_j^\ast - \mu_0^\ast|,
% \max_{j=1,...,m}|\mu_j - \mu_0| = \kappa_{\max}
$$
respectively. The definition of $\Delta$ ensures that the mean signal of each anomaly region is well separated from the baseline. Note that there is no restriction on the mean signal difference across different anomaly regions. 

Define the intrinsic diameter of a region $R$ as the largest pairwise distance between any two lattice points in $R$:
$$
r_{R} = \max_{\s,\s'\in R} \text{dist}(\s,\s'),
$$
where $\text{dist}(\cdot,\cdot)$ denotes the Euclidean distance, which can be replaced by other well-defined distance metrics depending on the applications. We also define the distance between any two regions as the smallest Euclidean distance between any pair of points from each region:
$$
\text{dist}(R, R') = \min_{\s\in R, \s'\in R'} \text{dist}(\s,\s').
$$

Analogous to classical timeline changepoint and anomaly detection frameworks, we impose the following assumptions to ensure the detectability of spatial anomaly regions. Specifically, we require sufficient signal distinguishability between anomaly and baseline regions, as well as adequate spatial separation between different anomalies.

\begin{assumption} (Signal strength)
	\label{assmp:signal}
	(i) For any $\eta > 0$, there exists a constant $C_{\eta} > 0$, such that
		$$
			\frac{\Delta^2}{\sigma^2} \ge C_{\eta}\cdot\frac{\log^{1+\eta}n}{\sqrt{n}}.
		$$

	(ii) There exists a constant $0 < C_{R} <1$, such that $ \delta \geq C_R\cdot n$. 
	
	(iii) There exist $0<d_A<d_B<1$, such that the maximum intrinsic diameter of true anomaly regions, and the minimum distance between any two true anomaly regions satisfy:
	$$\max_j r_{R_j^\ast} \le d_A \sqrt{n} \quad\text{and}\quad   \min_{i,j}\textnormal{dist}(R_i^\ast, R_j^\ast) \ge d_B \sqrt{n}.
	$$

    %   (iv)\,There exists a constant $C_{\delta}>0$ such that
    % $
    % \big|\text{Co}(R_j^*)\big| - \big|R_j^*\big| \leq C_{\delta}\cdot\delta,
    % %\frac{\delta}{20m^{\ast2} +1},
    % \quad j = 1,...,m^*.
    % $
	%and any two minimum convex hulls of anomaly regions are non-overlapping.
	
	%	(i) There exists  $\eta > 0$, such that 
	%$$\frac{\delta\Delta^2}{\sigma^2}\geq C_{\delta}\cdot\sqrt{n}\log^{1+\eta} n,$$
	%where $C_\delta>0$ is a constant.
\end{assumption}
Assumption \ref{assmp:signal}\,(i) and (ii) ensure that the mean signal of the baseline and that of any anomaly region are distinct enough, and each anomaly region is  sufficiently large. 
Altogether, they lead to a lower bound on the conventional signal-to-noise ratio (SNR) of detection, given by $\frac{\delta\Delta^2}{\sigma^2} \geq C_{\eta} \cdot \sqrt{n} \log^{1+\eta} n$. The SNR rate here is different from the $\log n$ rate in the timeline problem \citep{wang2020univariate}.  In Section \ref{sec:minimax}, we prove this cannot be relaxed to have any consistent detection. 
Assumption \ref{assmp:signal}\,(iii) requires that anomaly regions are sufficiently separated from each other, with distances exceeding the intrinsic diameter of any anomaly region.

% \red{Assumption~\ref{assmp:signal}\,(iv) controls the spatial concentration of the points within each true anomaly region by restricting the difference between a true anomaly region and its convex hull. If the points in an anomaly region are too scattered, they may not form a detectable spatial region.
% The cardinality of this difference is allowed to be as large as $O(n)$, so the condition still permits substantial flexibility in the region shape. 
% As discussed in Section~\ref{subsec:criterion}, the area of the minimum convex hull can also be replaced by another measure, provided that it captures the spatial concentration of points within a region.}

\begin{remark}
Assumption~\ref{assmp:signal}\,(iii) is crucial when multiple anomaly regions exhibit similar or identical mean values. If a stronger SNR condition is imposed, for example, all anomaly regions have distinct mean signals, this assumption can be relaxed or removed. We refer to an alternative set of signal strength assumptions and a detailed discussion in Supplementary Material~\ref{sec:consistencymean}.
\end{remark}

Our final DPLS-SAD estimator is:
\begin{equation} \label{eq:estimator}
	\{\hat{m};\,\widehat{R}_{1:\hm}\} = \argmin_{m; \,R_{1:m} \in \mathcal{R}} C(m;\,{R}_{1:m}),
\end{equation}
where we restrict the minimization within the smooth regional class $\mathcal{R}$ from Definition \ref{defi:smooth}. To measure the fitness of estimated anomaly regions compared to the true ones, we can use the symmetric regional difference, which measures the number of points that have been missed or falsely included in the detection:
\begin{equation}
\label{eq:def-region-distance}
	D\big({R},{R'}\big):=|R\setminus R'|+|R'\setminus R| = |R\cup{R'}| - |{R}\cap{R'}|.
\end{equation}
This distance is equivalent to the Hausdorff localization error $|s-s'|+|e-e'|$ in the timeline scenario, where $(s,e)$ and $(s', e')$ are pairs of starting and ending points of different intervals.
%$R$ and $R'$, respectively.

Now we are ready to present our main result in the following theorem, which shows that the  number of anomalies and their locations can be consistently recovered by DPLS-SAD.
\begin{theorem} (Consistency)
	\label{thm:consistency}
	Suppose Assumptions \ref{assmp:data}, \ref{assmp:smooth} and \ref{assmp:signal} hold.  If we choose $\beta=C_\beta \sqrt{n}\log n$ and $\lambda = \clambda\log n/\sqrt{n}$, where $C_\beta$ and $C_\lambda$ are some absolute constants not depending on $n$ and $m^\ast$. Let  
	$\{\hm;\,\widehat R_{1:\hm}\}$ be the minimizer from solving (\ref{eq:estimator}). There exist constants $c_{\gamma}, C_{\varepsilon} > 0$ such that
	$$ 
	\hm = m^\ast \quad \textnormal{and} \quad D\Big(\widehat R_j, R^\ast_j\Big)\leq \dfrac{C_{\varepsilon}\sigma^2}{\Delta^2}\ld, \quad j=1,\dots, m^* 
	$$
	holds with probability at least $1-2\exp\big(-c_{\gamma}\ld\big)$.
\end{theorem}

Theorem \ref{thm:consistency} provides a non-asymptotic characterization of DPLS-SAD detection. Such characterization leads to straightforward consistent results for different asymptotic regimes. For example, consider the  standard setup in the changepoint literature, where variance proxy $\sigma^2$ and jumping size $\Delta$ are constant. In this case, we can get the detection error rate at $\sqrt{n} \log n$. 
% \zc{add asymptotic regime}

Our detection result is similar to that in the classic timeline anomaly/changepoint detection problem \citep{fisch2022linear,zheng2022consistency}, where the detection rate is $O(\log n)$, which differs from the DPLS-SAD rate $O(\ld)$.  The extra $\sqrt{n}$ can be regarded as the price of extending from timeline to spatial detection. In Section \ref{sec:minimax}, we further prove that this rate is minimax optimal up to a logarithmic factor and therefore cannot be further improved.

In view of Theorem \ref{thm:consistency},  we allow great flexibility on the geometric shape of anomaly regions. This makes the DPLS-SAD method practically useful in many real-world applications, where the shapes of anomalies are typically complex and irregular. It is worthwhile to note that there is a trade-off between the geometric flexibility of anomaly regions and the detection rate. Imposing excessive shape constraints will change the nature of the spatial detection problem. For example, assuming very strict convexity and connectivity restrictions on the anomaly regions makes the spatial detection problem a trivial extension of the timeline counterpart, thus leading to a similar localization rate of $O(\log n)$. If no shape constraint is imposed and anomaly regions are allowed arbitrary geometric flexibility, then no method can consistently detect the spatial anomaly regions. This impossibility result is established in Section~\ref{sec:minimax} through a minimax lower-bound argument.

In reality, sometimes anomaly regions are not spatially well-separated.  We also consider a slightly different detection problem in Supplementary Material \ref{sec:consistencymean}, where we allow anomaly regions to be spatially close to each other, i.e., relaxing Assumption \ref{assmp:signal}\,(iii).  In such case, similar detection rate  can be achieved under stronger SNR conditions.

\section{Minimax optimal spatial detection rate}\label{sec:minimax}

In this section, we provide novel and rigorous minimax arguments to analyze the spatial anomaly detection problem. We first establish an SNR threshold that is related to the signal strength assumption (Assumption~\ref{assmp:signal}), below which consistent detection is impossible for any method. When the SNR exceeds this threshold, consistent detection becomes achievable: we derive an information-theoretic lower bound on the detection error and show that DPLS-SAD attains this minimax-optimal detection rate up to a logarithmic factor. 
In Subsection \ref{subsec:arbitrary.minimax}, We further investigate the minimax lower bound without imposing the regional smoothness assumption (Assumption~\ref{assmp:smooth}), allowing anomaly regions to have arbitrary shapes. We show that, over this unrestricted class of models, the spatial anomaly detection problem is fundamentally ill-posed from a minimax perspective, as no estimator can achieve consistent detection.

\subsection{Minimax optimality of DPLS-SAD}
\label{subsec:minimax.dpls}
% \wb{
% In Theorem \ref{thm:minimax1} below, we give minimax lower bounds below the SNR $\sqrt n\log n$, covering two regimes:
% $$
% \delta\Delta^2/\sigma^2\leq \sqrt n
% \quad\text{and}\quad
% \sqrt n\leq \delta\Delta^2/\sigma^2\leq \sqrt n\log n .
% $$
% In the first regime, no consistent estimator of the spatial anomaly regions exists, e.g., for any detection method we can always find some data scenario such that $\frac{\text{Detection error rate}}{n}$ is bounded away from 0. In the second regime, the lower bound is weaker only by a logarithmic factor. Thus, while it does not strictly rule out consistent detection, the minimax difficulty in this regime remains close to that of the impossible regime.
% }

% We analyse the minimax localization rate over the smooth regional class defined in Definition \ref{defi:smooth}.
    Theorem \ref{thm:minimax1} below characterizes the regime in which consistent detection is impossible. Specifically, suppose the following low SNR condition 
$$
\dfrac{\delta\Delta^2}{\sigma^2}<\sqrt n\log n,
$$
which violates Assumption~\ref{assmp:signal}\,(i) and (ii) jointly, the minimax lower bound of the detection error is  nearly of order $n$. 

% When the SNR is further restricted to
% $$
% \dfrac{\delta\Delta^2}{\sigma^2}<\sqrt n,
% $$
% the minimax lower bound strengthens to order $n$, removing the logarithmic gap and implying that no estimator can achieve consistent localization.}
% In Theorem~\ref{thm:minimax1} below, we first give a minimax lower bound for the regime
% $$
% \dfrac{\delta\Delta^2}{\sigma^2}<\sqrt n\log n,
% $$
% which is an order of $n/\log^2n$. This result gives an impossible regime for consistent localization up to a logarithmic factor. 

% An order $n$ minimax lower bound means that, for any detection method, we can always find some data scenarios such that $\text{Detection error rate}/n$ is bounded away from 0, and hence no consistent estimator of the spatial anomaly regions exists. The theorem also shows that, when the SNR is further restricted to
% $$
% \dfrac{\delta\Delta^2}{\sigma^2}<\sqrt n,
% $$
% the minimax lower bound becomes order $n$, which gives the impossible
% regime for consistent detection.

% \red{In Theorem \ref{thm:minimax1} below, we give the impossible regime of consistent detection. In detail, suppose that the following low SNR condition holds 
% $$
% \dfrac{\delta\Delta^2}{\sigma^2} < \sqrt{n}.
% $$
% No consistent estimator of the spatial anomaly regions exists, e.g., for any detection method we can always find some data scenario such that $\frac{\text{Detection error rate}}{n}$ is bounded away from 0. When the SNR is allowed to increase from $\sqrt n$ up to $\sqrt n\log n$, the minimax lower bound is reduced only by logarithmic factors and still implies a large localization error.}

\begin{theorem}(Impossible regime)
	\label{thm:minimax1}
	Let $\mathcal{Q}$ be a class of distributions satisfying the model setup in Section \ref{subsec:model}, and suppose Assumptions \ref{assmp:data} and \ref{assmp:smooth} hold. As long as $\delta\Delta^2/\sigma^2 < \sqrt{n}\log n$ , for sufficiently large $n$, there exists a positive constant $c$ such that
    $$
	\inf_{\widehat{R}}\sup_{Q\in\mathcal{Q}}\mathbb{E}_Q\bigg\{D\Big(\widehat{{R}},R(Q)\Big)\bigg\}\geq\,
    \dfrac{c n}{\log^2n}.
	$$
    Moreover, if $\delta\Delta^2/\sigma^2 < \sqrt{n}$, the minimax lower bound reaches to $cn/4$.
	% $$
	% \inf_{\widehat{R}}\sup_{Q\in\mathcal{Q}}\mathbb{E}_Q\bigg\{D\Big(\widehat{{R}},R(Q)\Big)\bigg\} \geq 
	% \begin{cases}
 %    \dfrac{cn}{4}, \qquad \quad &\text{if} \quad \delta\Delta^2/\sigma^2 \leq \sqrt{n},\\[1.2ex]
 %    \dfrac{c n}{\log^2 n}, &\text{if}\quad  \sqrt{n}\leq \delta\Delta^2/\sigma^2 \leq \sqrt{n}\log n.
 %    \end{cases}
	% $$
	Here $\widehat R$ denotes the estimator of the set of anomaly regions; $R(Q)$ denotes the true anomaly regions under the distribution $Q$; and the infimum is taken over all estimators. 
\end{theorem}

In the impossible regime where the SNR is not sufficient, the minimax detection rate cannot be reduced to below $O(n)$  up to logarithmic factors. Consequently, for any detection method, there exists a data-generating distribution under which $\frac{\text{Detection error rate}}{n}$ is bounded away from 0, implying that consistent recovery of anomaly regions does not exist. In the next theorem, we will show that when the SNR increases to the regime 
$$
\dfrac{\delta\Delta^2}{\sigma^2} \ge  \sqrt{n}\log n,
$$
which falls within Assumption~\ref{assmp:signal}\,(i) and (ii), the minimax detection rate can be improved to $\sqrt{n}$, which leads to consistent detection, i.e., $\frac{\text{Detection error rate}}{n}\rightarrow 0$ as $n\rightarrow \infty$.
% We can let $c\rightarrow 0$ as $n \rightarrow \infty$

% $\frac{\text{Detection error rate}}{n}$ is bounded away from 0, i.e., no estimator is consistent.

\begin{theorem} (Minimax optimal rate)
	\label{thm:minimax2}
	Let $\mathcal{Q}$ be a class of distributions satisfying the model setup in Section \ref{subsec:model}, and suppose Assumptions \ref{assmp:data} and \ref{assmp:smooth} hold. As long as $\delta\Delta^2/\sigma^2 \geq\ld$, for sufficiently large $n$, there exists a positive constant $c$ such that
	$$
	\inf_{\widehat{R}}\sup_{Q\in\mathcal{Q}}\mathbb{E}_Q\bigg\{D\Big(\widehat{R},R(Q)\Big)\bigg\} \geq 
	\frac{c\sqrt{n}}{8}\frac{\sigma^2}{\Delta^2}\cdot \min\Bigg\{1,\bigg\lceil\frac{\Delta^2}{\sigma^2}\bigg\rceil\exp\bigg(-\frac{\Delta^2}{\sigma^2}\bigg)\Bigg\}. 
	$$
\end{theorem}

In the standard asymptotic regime where $\Delta^2/\sigma^2$ is a constant, the above result implies that the
minimax lower bound is $O(\sqrt n).$
Recall Theorem \ref{thm:consistency}, in which we show DPLS-SAD achieves the detection error 
$$
D\Big(\widehat R_j, R^\ast_j\Big)\leq \dfrac{C_{\varepsilon}\sigma^2}{\Delta^2}\ld.
$$ 
which attains this rate up to a logarithm factor, showing that the proposed method is minimax optimal.

From Theorems \ref{thm:minimax1} and \ref{thm:minimax2}, we can see the minimax detection rate crucially depends on if SNR is less or greater than  $\sqrt{n}\log n$. This is quite similar to the phase transition phenomenon observed in the timeline univariate
change-in-mean problem \citep{wang2020univariate, wang2018high, verzelen2023optimal}, where the minimax localization rate is $O(n)$ in the low SNR regime (${\delta\Delta^2}/{\sigma^2}< \log n$) and improves to $O(1)$ in the high SNR regime (${\delta\Delta^2}/{\sigma^2} \ge\log n$).

% Recall Theorem \ref{thm:consistency}, in which we show DPLS-SAD achieves the detection error 
% $$
% D\Big(\widehat R_j, R^\ast_j\Big)\leq \dfrac{C_{\varepsilon}\sigma^2}{\Delta^2}\ld.
% $$ 
% This matches the minimax optimal rate up to a logarithm factor. 

\begin{remark}
Note that Assumption~\ref{assmp:signal}\,(iii) is not used in the minimax results. It imposes a separation condition between distinct anomalies  and therefore is only relevant in settings involving multiple anomaly regions. However, the minimax lower bound concerns the detection error rate, which is established by constructing a collection of SAD model distributions, each only containing a single anomaly region. For this construction, the separation condition is vacuous, and Assumption~\ref{assmp:signal}\,(iii) plays no role in the argument.
% \red{Note that Assumption~\ref{assmp:signal}\,(iii) imposes separation condition between different anomalies, and is therefore only needed for detecting multiple anomaly regions. In contrast, to derive the minimax lower bound for the SAD problem, it suffices to consider a subclass of distributions with a single anomaly region, for which this assumption is not required.}
% %can be established on the submodel with one single anomaly region. 
% therefore the signal strength depends only on the SNR conditions in Assumption \ref{assmp:signal}\,(i) and (ii).
\end{remark}

\begin{remark}
To derive the minimax lower bounds in Theorems~\ref{thm:minimax1} and~\ref{thm:minimax2}, we develop a novel Assouad-type technique that is essentially different from the commonly used Le Cam method argument in temporal changepoint problems \citep{wang2020univariate}. This construction may be of independent interest to the readers. Le Cam’s method reduces the analysis to testing between two  hypotheses and fails to capture the combinatorial and geometric complexity inherent in spatial anomaly regions. In contrast, our Assouad-type technique accommodates multiple local two-hypothesis constructions over complex regional classes, yielding sharper lower bounds that reflect the intrinsic difficulty of the anomaly detection in spatial settings.
\end{remark}

% \textcolor{red}{
% To derive the minimax lower bounds in Theorems~\ref{thm:minimax1} and~\ref{thm:minimax2}, we develop novel Fano-type and Assouad-type arguments, respectively, rather than relying on the commonly used Le Cam method in temporal changepoint problems. Le Cam’s approach reduces the analysis to testing between two carefully constructed hypotheses and therefore fails to capture the combinatorial and geometric complexity inherent in spatial anomaly regions. In contrast, the Fano-type argument accommodates multi-hypothesis constructions, while the Assouad-type argument accommodates multiple local two-hypothesis constructions over complex region classes, yielding sharper lower bounds that reflect the intrinsic difficulty of spatial localization.}

\subsection{Minimax rate for arbitrarily shaped anomaly regions}
\label{subsec:arbitrary.minimax}
% \red{We next examine how the minimax lower bound changes when the regional smoothness condition in Assumption~\ref{assmp:smooth} is removed. Compared with the setting in the previous subsection, this gives an enlarged class of admissible anomaly regions, in which the regions may be arbitrarily fragmented. Theorem~\ref{thm:minimax-nores} provides the corresponding minimax lower bound.}

In Theorem~\ref{thm:minimax-nores} below, we establish a comparison result for the case where anomaly regions are allowed to take arbitrary shapes. We show that when removing  the regional smoothness restrictions as in Assumption~\ref{assmp:smooth},   the minimax detection  rate is always of order $n$  regardless of the SNR strength, implying that consistent detection is not achievable.

\begin{theorem}
	\label{thm:minimax-nores}
	Let $\mathcal{Q}$ be a class of distributions satisfying the model setup in Section \ref{subsec:model}, and suppose Assumption \ref{assmp:data} holds. Then, for sufficiently large $n$, there exists a positive constant $c$ such that
	$$
	\inf_{\widehat{R}}\sup_{Q\in\mathcal{Q}}\mathbb{E}_Q\bigg\{D\Big(\widehat{R},R(Q)\Big)\bigg\} \geq 
	\begin{cases}
    \dfrac{n}{64}, \qquad \quad &\text{if} \quad \delta\Delta^2/\sigma^2 < n/5,\\[1.2ex]
    \dfrac{cn}{8}\cdot \exp\bigg(-\dfrac{\Delta^2}{\sigma^2}\bigg), &\text{if}\quad   \delta\Delta^2/\sigma^2 \geq n/5.
    \end{cases}   
	$$
\end{theorem}

Theorem~\ref{thm:minimax-nores} shows that removing Assumption~\ref{assmp:smooth} fundamentally changes the minimax behavior of the impossible regime, yielding a $O(n)$ detection error uniformly over all SNR scenario, for example,
% ..... which substantially increases the difficulty of the SAD problem, until detection ....
consider the standard asymptotic regime where $\Delta^2/\sigma^2$ is a constant. 

This result highlights the intrinsic trade-off between shape flexibility and statistical efficiency, as discussed in Section~\ref{sec:consistency}. Assumption~\ref{assmp:smooth} is already very permissive in allowing a wide class of spatially coherent anomaly regions. Relaxing it further effectively eliminates structural constraints on the SAD model space, leading to an ill-posed detection problem.

\section{Extending to general dimensions and spatial dependent data}\label{sec:extend}
We now extend DPLS-SAD to detect anomaly regions in higher dimensions ($d>2$), whilst each dimension is allowed to diverge at a different rate. Consistency and minimax results similar to those established in Sections \ref{sec:consistency} and \ref{sec:minimax} can be derived in this more general setting. Furthermore, we demonstrate that the independent data assumption can be relaxed, showing that our method remains valid for detecting anomaly regions under spatial dependence.

\subsection{Anomaly detection for general dimensional data}\label{subsec:high-dim}

% Consider the process $\{Y(\s):\s\in\mathcal{S}\}$, where $\mathcal{S} = \{(s_1,...,s_d)\}\subset \mathbb{R}^d$ is a set of locations forming a $d$-dimensional regular lattice. 
Consider the process $\{Y(\s):\s\in\mathcal{S}\}$ on a $d$-dimensional regular lattice $\mathcal{S} = \{(s_1,...,s_d)\}\subset \mathbb{R}^d$.
Again, we assume each $Y(\s)$ can be decomposed into a mean signal and random error:
$$
Y(\s) = \mu(\s) + \varepsilon(\s), \qquad \s = \mathcal{S}\in \{1,2,...,n_1\}\times \dots \times\{1,2,...,n_d\},
$$
where $n_i$ is the size of $i$-th dimension, with $n=\prod_{i=1}^dn_i$. Define 
$
{n_{\max}} = \max_{i}n_i,
$
which represents the maximum size of any dimension.  Similar to Section \ref{sec:modelset}, $\mu(\s)$ denotes the mean signal at $\s$, which is invariant within an anomaly region $R$ whenever $\s\in R$.

Consider $\varepsilon(\s)$ being sub-Gaussian as in Assumption \ref{assmp:data}. In the following, we update the Assumptions  \ref{assmp:smooth} (regional smoothness) and \ref{assmp:signal} (signal strength), respectively, to accommodate the general dimensionality. 

\begin{definition}
	\label{defi:smooth_gen_dim} 
	% ~\\
	% (1)  For $n$ observations, the diameter of overall region is $O(\sqrt{n})$.\\
		Define the class of  $d$-dimensional smooth regions $\cR^d=\cR^d_{K}$, where $ 0<K<\infty$ is a finite constant that does not depend on $n$ and $m^*$, such that
	 for any axis–parallel line obtained by fixing $d-1$ coordinates, the set of $\mathcal{S}$  lying on this line can be partitioned into no more than $K$ consecutive  segments (intervals).
	%(ii) The minimum convex hull of each region in $\cR^d$ contains at most $H$ vertices. 
\end{definition}

\begin{assumption}\label{assmp:smooth_gen_dim} 
 Each true anomaly region $R_j^*$, for $j\in\{1,...,m^*\}$, belongs to the class of $d$-dimensional smooth regions, i.e., $R_j^*\in\mathcal{R}^d$ for all $j$.
\end{assumption}

\begin{assumption} 
	\label{assmp:signal_gen_dim}
		(i) For any $\eta > 0$, there exists a constant $C_{\eta} > 0$, such that
	$$
	 \frac{\Delta^2}{\sigma^2} \geq C_{\eta}\cdot\frac{\log^{1+\eta}n}{n_{\max}} .
	$$
	
	(ii) Same as Assumption \ref{assmp:signal}\,(ii).
	
	(iii) There exist $0<2d_A<d_B<1$ such that the maximum intrinsic diameter of anomaly regions and the minimum distance between two anomaly regions satisfy 
	$$\max_j r_{R_j^\ast} \le d_A\cdot n_{\max}\quad \text{and} \quad  \min_{i,j}\textnormal{dist}(R_i^\ast, R_j^\ast) \ge d_B \cdot n_{\max}.$$       
\end{assumption}

Assumption \ref{assmp:signal_gen_dim}\,(iii) imposes a stronger spatial separation condition than Assumption \ref{assmp:signal}\,(iii), requiring $d_B>2d_A$, which means anomaly regions in general dimension must be considerably farther apart from one another

In the DPLS-SAD methodology for general dimensional spatial data, we use the same double penalized cost 
% $$
% C(m; \, R_{1:m}) := L(R_{1:m}) + \beta m + \lambda \sum_{j=1}^m |\text{Co}(R_j)|,
% $$
and  obtain the detected anomalies by minimizing over $\mathcal{R}^d$, i.e.,
\begin{equation}
	\label{eq:pen-cost-gendim}
	\{\hat{m};\,\widehat{R}_{1:\hat{m}}\} = \argmin_{m;\,R_{1:m}\in\mathcal{R}^d} C(m; \, R_{1:m}).
\end{equation}
%$\mathcal{R}_d=\{R_{1:m} | \, R_i \text{ satisfy Assumptions \ref{assmp:signal_gen_dim} and \ref{assmp:smooth_gen_dim}}\}$ 

The following theorem gives a consistency guarantee for the DPLS-SAD method in general dimensions.

\begin{theorem}
	\label{thm:consistency_gen_dim}
	Suppose Assumptions \ref{assmp:data}, \ref{assmp:smooth_gen_dim} and \ref{assmp:signal_gen_dim} hold. If we choose $\beta=C_{\beta, 1}\, n\log n/n_{\max} $ and $\lambda = C_{\lambda, 1}\, \log n/n_{\max}$, where $C_{\beta,1}$ and $C_{\lambda,1}$ are large enough absolute constants not depending on $n$ and $m^*$. Let $\{\hm;\,\widehat R_{1:\hm}\}$ be the minimizer from solving (\ref{eq:pen-cost-gendim}). There exist constants $c_{\gamma,1}, C_{\varepsilon,1}>0$ such that
	$$ 
	\hm = m^\ast \quad \textnormal{and} \quad D\Big(\widehat R_j, R^\ast_j\Big)\leq \dfrac{C_{\varepsilon,1}\sigma^2}{\Delta^2}\cdot \frac{n}
    {n_{\max}}\log n, \quad j=1,\dots, m^* 
	$$
	holds with probability at least $1-2\exp\big(-c_{\gamma,1}\, \frac{n}
    {n_{\max}}\log n\big)$.
\end{theorem}

\begin{remark}Theorem \ref{thm:consistency_gen_dim} shows the consistency result in general dimensional data, where the convergence rate is related to the maximum dimension size $n_{\max}$. When all the dimensions are of equal sizes, i.e., $n_1 = \dots = n_d=\sqrt[d]{n}$, the detection rate is $n^{\frac{d-1}{d}}\log n$. One can observe the phenomenon of the curse of dimensionality, as the detection rate grows to $n$ ( problem becomes non-detectable) when $d$ increases to infinity.
\end{remark}

Theorem \ref{thm:consistency} can be regarded as a special case of above result by setting $d = 2$ and $n_{\max}=\sqrt{n}$. In addition, setting $d = 1$ recovers the $\log n$ error rate in the timeline changepoint/anomaly detection problem, in agreement with results established in the existing literature.

The following theorem extends the corresponding minimax results from the 2D problem to general dimensions.

% \begin{theorem}
% \label{thm:minimax_highdim}
% Let $\mathcal{Q}$ be a class of distributions satisfying the model setup in Section \ref{subsec:high-dim}, and suppose Assumptions \ref{assmp:data} and \ref{assmp:smooth_gen_dim} hold. Then, for sufficiently large $n$,
% $$
% \inf_{\widehat{R}}\sup_{Q\in\mathcal{Q}}
% \mathbb{E}_Q\!\left\{D\big(\widehat{R},R(Q)\big)\right\}
% \;\ge\;
% \begin{cases}
% \dfrac{n}{64}, 
% & \quad \text{if }\, \displaystyle \frac{\delta\Delta^2}{\sigma^2} \;<\; \frac{n}{n_{\max}}\log n,\\[1.2ex]
% \displaystyle \dfrac{\sigma^2}{12\Delta^2}\cdot\frac{n}{n_{\max}},
% & \quad \text{if }\, \displaystyle \frac{\delta\Delta^2}{\sigma^2} \;\ge\; \frac{n}{n_{\max}}\log n.
% \end{cases}
% $$
% \end{theorem}

\begin{theorem}
\label{thm:minimax_highdim}
Let $\mathcal{Q}$ be a class of distributions satisfying the model setup in Section \ref{subsec:high-dim}, and suppose Assumptions \ref{assmp:data} and \ref{assmp:smooth_gen_dim} hold. Then, for sufficiently large $n$, there exists a positive constant $c$ such that
$$
\inf_{\widehat{R}}\sup_{Q\in\mathcal{Q}}
\mathbb{E}_Q\!\left\{D\big(\widehat{R},R(Q)\big)\right\}
\;\ge\;
\begin{cases}
\dfrac{cn}{8n_{\max}} \dfrac{\sigma^2}{\Delta^2}\cdot \min\Bigg\{1,\bigg\lceil\dfrac{\Delta^2}{\sigma^2}\bigg\rceil\exp\bigg(-\dfrac{\Delta^2}{\sigma^2}\bigg)\Bigg\},&\text{if}\quad  \dfrac{\delta\Delta^2}{\sigma^2} \geq \dfrac{n\log n}{n_{\max}},\\[1.2ex]
\dfrac{c n}{\log^2 n}, &\text{if}\quad  \dfrac{\delta\Delta^2}{\sigma^2} < \dfrac{n\log n}{n_{\max}}.
\end{cases}
$$
Moreover, if $\delta\Delta^2/\sigma^2 \leq n/n_{\max}$, the minimax lower bound improves to $cn/4$.
% Here $\widehat R$ denotes the estimator of the set of anomaly regions, e.g., $\widehat R_{1:\hat m}$; $R(Q)$ denotes the true anomaly regions under the distribution $Q$; and the infimum is taken over all estimators $\widehat R_{1:\hat m}$.
\end{theorem}

% \begin{theorem}
% 	\label{thm:minimax1_highdim}
% 	Let $\mathcal{Q}$ be a class of distributions satisfying the model setup in Section \ref{subsec:high-dim} and Assumptions \ref{assmp:data} and \ref{assmp:signal_gen_dim} hold. As long as $\delta\Delta^2/\sigma^2 < \frac{n}{n_{\max}}\log n$ , there exists a sufficiently large $n$, such that
% 	$$
% 	\inf_{\widehat{R}}\sup_{Q\in\mathcal{Q}}\mathbb{E}_Q\bigg\{D\Big(\widehat{{R}},R(Q)\Big)\bigg\} \geq 
% 	\dfrac{n}{64},
% 	$$
% 	where $\widehat R$ denotes the estimator of the set of anomaly regions e.g., $\widehat R_{1;\hat m}$, $R(Q)$ denotes the true anomaly regions under the distribution $Q$, and the infimum is taking over all estimators $\widehat R_{1;\hat m}$. 
% \end{theorem}

% \begin{theorem}
% 	\label{thm:minimax2_highdim}
% 	Let $\mathcal{Q}$ be a class of distributions satisfying the model setup in Section \ref{subsec:high-dim} and Assumptions \ref{assmp:data} and \ref{assmp:signal_gen_dim} hold. As long as $\delta\Delta^2/\sigma^2 \geq\frac{n}{n_{\max}}\log n$, it holds
% 	$$
% 	\inf_{\widehat{R}}\sup_{Q\in\mathcal{Q}}\mathbb{E}_Q\bigg\{D\Big(\widehat{R},R(Q)\Big)\bigg\} \geq 
% 	\dfrac{\sigma^2}{2\Delta^2}\frac{n}{n_{\max}}.   
% 	$$
% \end{theorem}

Similar to the results in Section \ref{subsec:minimax.dpls}, Theorem \ref{thm:minimax_highdim} reveals that the detection rate depends on if SNR is greater than the threshold $\frac{n}{n_{\max}}\log n$ or not. Combining with Theorem \ref{thm:consistency_gen_dim}, we show that DPLS-SAD achieves the minimax optimal detection rate in general dimension, up to logarithmic factors.

\subsection{Anomaly detection for spatially dependent data}\label{subsec:dependent}

Previously, we assume the random errors $\{\varepsilon_{\s}\}$ are independent to each other, which may not  hold in many real-world spatial applications.  We now revise our results, showing that DPLS-SAD still delivers consistent detection under certain spatial correlated data settings. 
% Recalling the penalized cost function under independent assumption, which is
% $$
% \mc(m, R_{1:m}) := L(R_{1:m}) + \beta m + \lambda \sum_{j=1}^m |\text{Co}(R_j)|,
% $$
% where $L(\cdot)$ is the sum of square. In this section, we will use the same penalized cost function to get the estimators. For dependent data, sum of square is not precise to measure the regional loss, since the dependence relationship is not accounted. We will give further discussion about the choice of cost function at the end of this section.

Consider the following spatial dependence, which is the counterpart to Assumption \ref{assmp:data}.

\begin{assumption} 
	\label{assmp:dataset_dependent}
	Let $\varepsilon(\s)$ be centered sub-Gaussian errors with $||\varepsilon(\s)||^2_{\psi_2}\leq \sigma^2$ for all $\s\in\mathcal{S}$. Moreover, for any $R\in\mathcal{R}^d$, assuming that
	$$
	\mathbb{E}\bigg\{ \exp\bigg( \tau\sum_{\s\in R}\varepsilon(\s) \bigg) \bigg\} \leq \exp\big( \tau^2\sigma^2|R|^{\phi} \big),
	$$
	for all $\tau>0$, where $\phi\ge 1$ is a dependence parameter satisfying $n^{\phi-1}\le n_{\max}$. 
\end{assumption}

Assumption \ref{assmp:dataset_dependent} characterizes the spatial dependence structure through the parameter $\phi$. A larger value of $\phi$ corresponds to stronger positive dependence among data points. In particular,  $\phi=1$, corresponds to mutually independent data. The following result states that we can still achieve consistent detection using  DPLS-SAD.

\begin{theorem}
	\label{thm:consistency_dep}
	Suppose Assumptions  \ref{assmp:smooth_gen_dim}, \ref{assmp:signal_gen_dim} and \ref{assmp:dataset_dependent}  hold. If we choose $\beta=C_{\beta,2}\, \frac{n^{\phi}}
	{n_{\max}}\log n$ and $\lambda = C_{\lambda,2}\, 
	\frac{n^{\phi-1}}{n_{\max}}\log n$, where $C_{\beta,2}$ and $C_{\lambda, 2}$ are large enough absolute constants not depending on $n$ and $m^*$. Let 
	$\{\hm;\,\widehat R_{1:\hm}\}$ be the minimizer from solving (\ref{eq:pen-cost-gendim}). There exist constants $c_{\gamma,2}, C_{\varepsilon,2}>0$ such that
	$$ 
	\hm = m^\ast \quad \textnormal{and} \quad D\Big(\widehat R_j, R^\ast_j\Big)\leq \dfrac{C_{\varepsilon,2}\sigma^2}{\Delta^2}\cdot\frac{n^{\phi}}{n_{\max}}\log n, \quad j=1,\dots, m^* 
	$$
	holds with probability at least $1 - 2\exp(-c_{\gamma,2}\frac{n}{n_{\max}}\log n)$.
\end{theorem}

\begin{remark}
Theorem \ref{thm:consistency_dep} shows that the detection rate is related to the dependence parameter $\phi$. When $\phi = 1$, we attain the same  $O\big(\frac{n}{n_{\max}}\log n\big)$ rate as in Theorem \ref{thm:consistency_gen_dim}. Stronger spatial dependence (i.e., as $\phi$ increases) makes detection harder and leads to a larger error.
\end{remark}

% Although we extend our framework to accommodate spatial dependence, we continue to employ the least squares cost function. A more appropriate choice for $L(R_{1:m})$ could be the negative log-likelihood, which can explicitly model spatial dependence. However, this approach is typically intractable and computationally prohibitive. As an alternative, one may adopt the composite log-likelihood approach (e.g., \cite{zhao2024composite}). A detailed investigation of this extension is left for future works.

\section{Simulation studies} \label{sec:sim}

In this section, we evaluate the empirical performance of the proposed DPLS-SAD method. To address the computational challenges arising from the NP-hard optimization problem in \eqref{eq:estimator} and (\ref{eq:pen-cost-gendim}), we develop a dynamic programming–based strategy combined with local neighbourhood segmentation, yielding an approximate solution with complexity $O(n^2)$, and thus making the method scalable to large spatial data applications. Complete algorithmic details, together with pseudocode, are provided in the Supplementary Material ~\ref{sec:algorithm}.

% In this section, we assess the empirical performance of DPLS-SAD and the proposed algorithm. As introduced in Section~\ref{sec:consistency}, DPLS-SAD estimates anomaly regions by solving \eqref{eq:estimator}. However, minimising \eqref{eq:estimator} exactly is computationally challenging, since the objective is highly non-convex and the underlying optimization problem is NP-hard. To this end, we design a heuristic algorithm to estimate anomaly regions, taking inspiration from one-dimensional $k$-means \citep{wang2011ckmeans} and exploiting the spatial structure of the problem. 

To the best of our knowledge, most existing algorithms, such as those in image segmentation and clustering, are not well suited to the SAD problem. These methods typically ignore spatial distance and often impose convexity constraints on regions, which we find usually lead to very  unreliable detection performance. In our experiments, we include DCART \citep{madrid2011lattice} as a benchmark method, while noting that it is also not originally designed for the SAD setting. 
% DCART allows anomaly regions to be close to each other, but requires their mean values to be distinguishable.

\begin{figure}[h]
	\centering
	\begin{tabular}{ccc}
		\includegraphics[scale = 0.1]{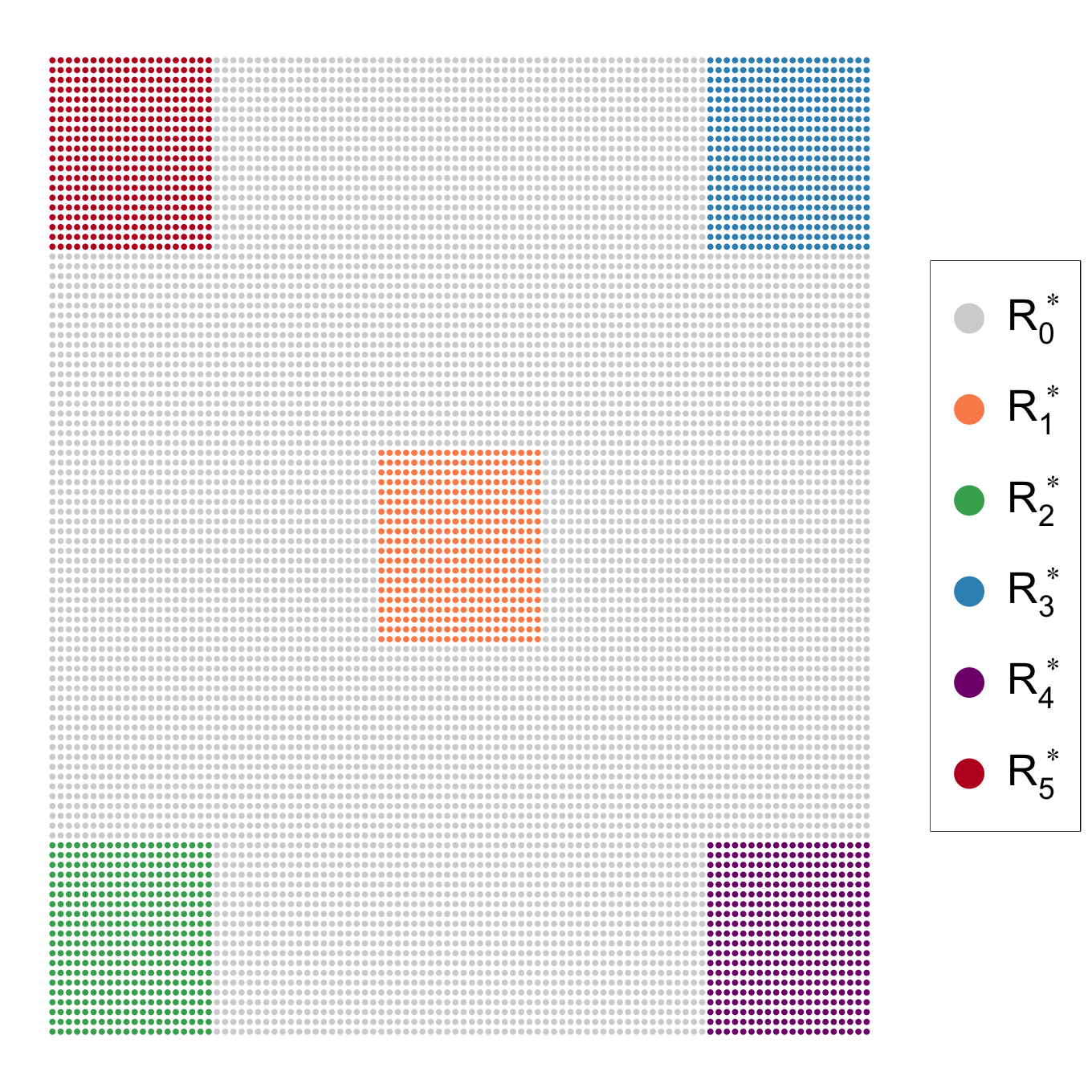} &\includegraphics[scale = 0.1]{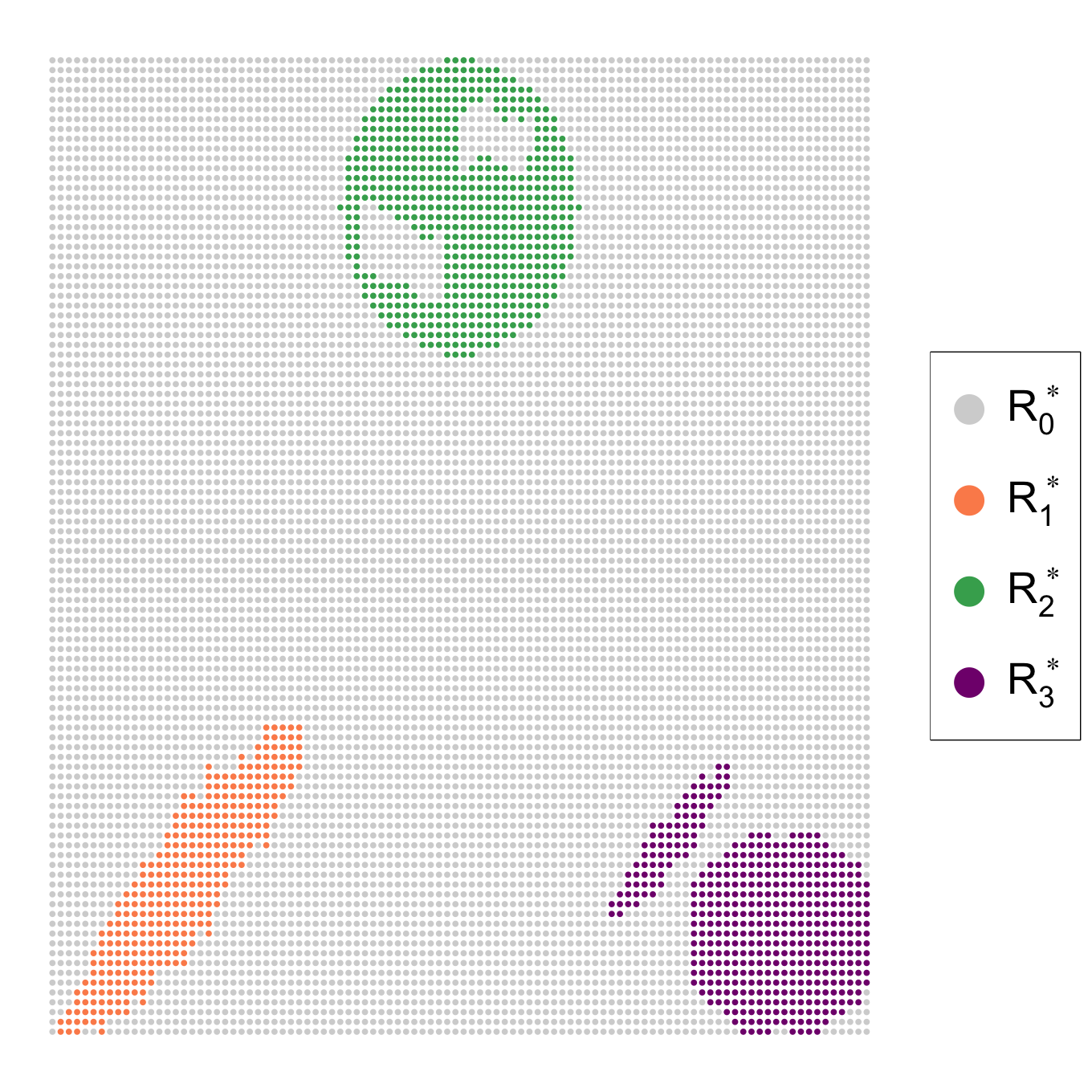} &\includegraphics[scale = 0.1]{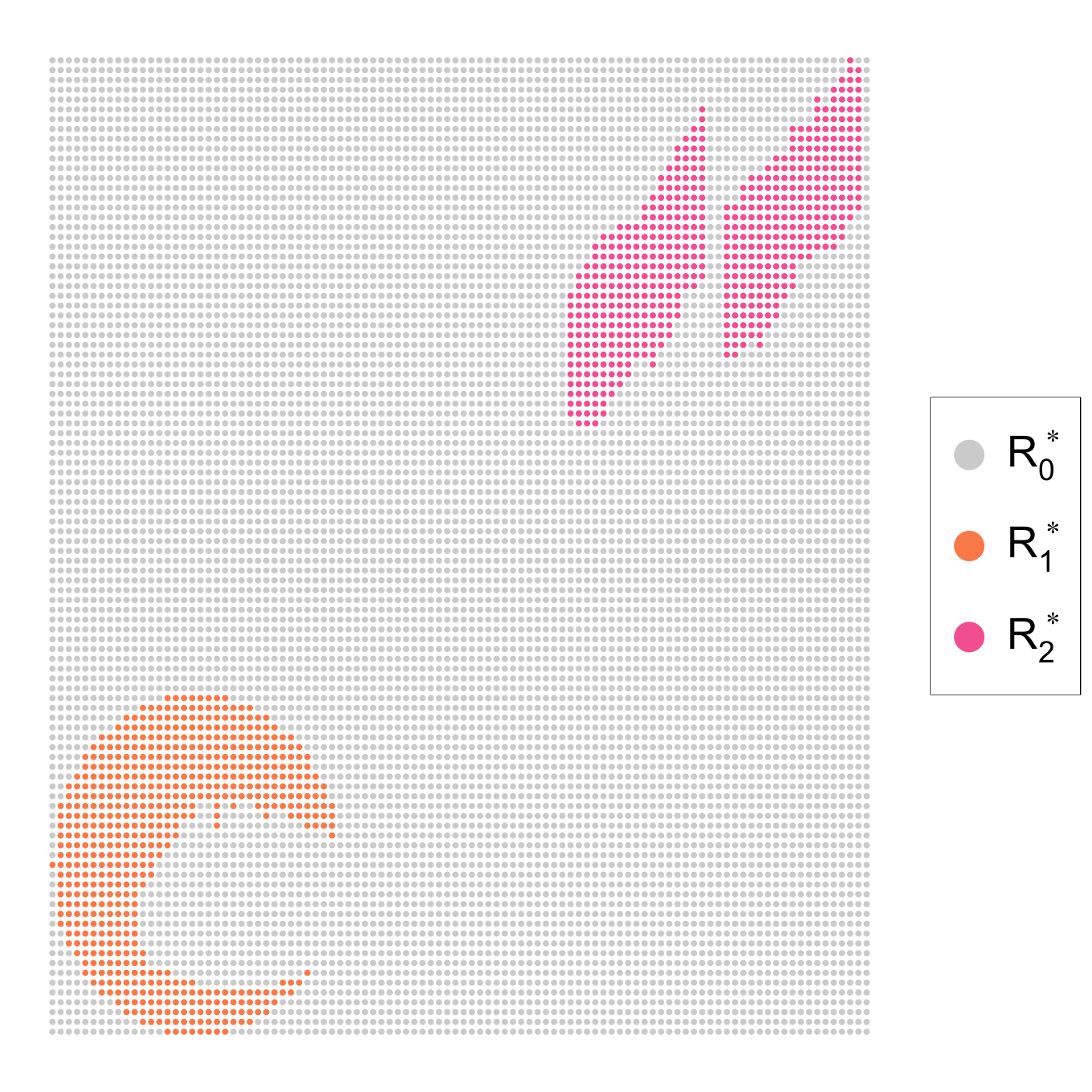}\\
         \vspace{-0.5cm}
         ~\\
		\small  (a1) & \small (a2) & \small  (a3)\\
        ~\\

		\includegraphics[scale = 0.12]{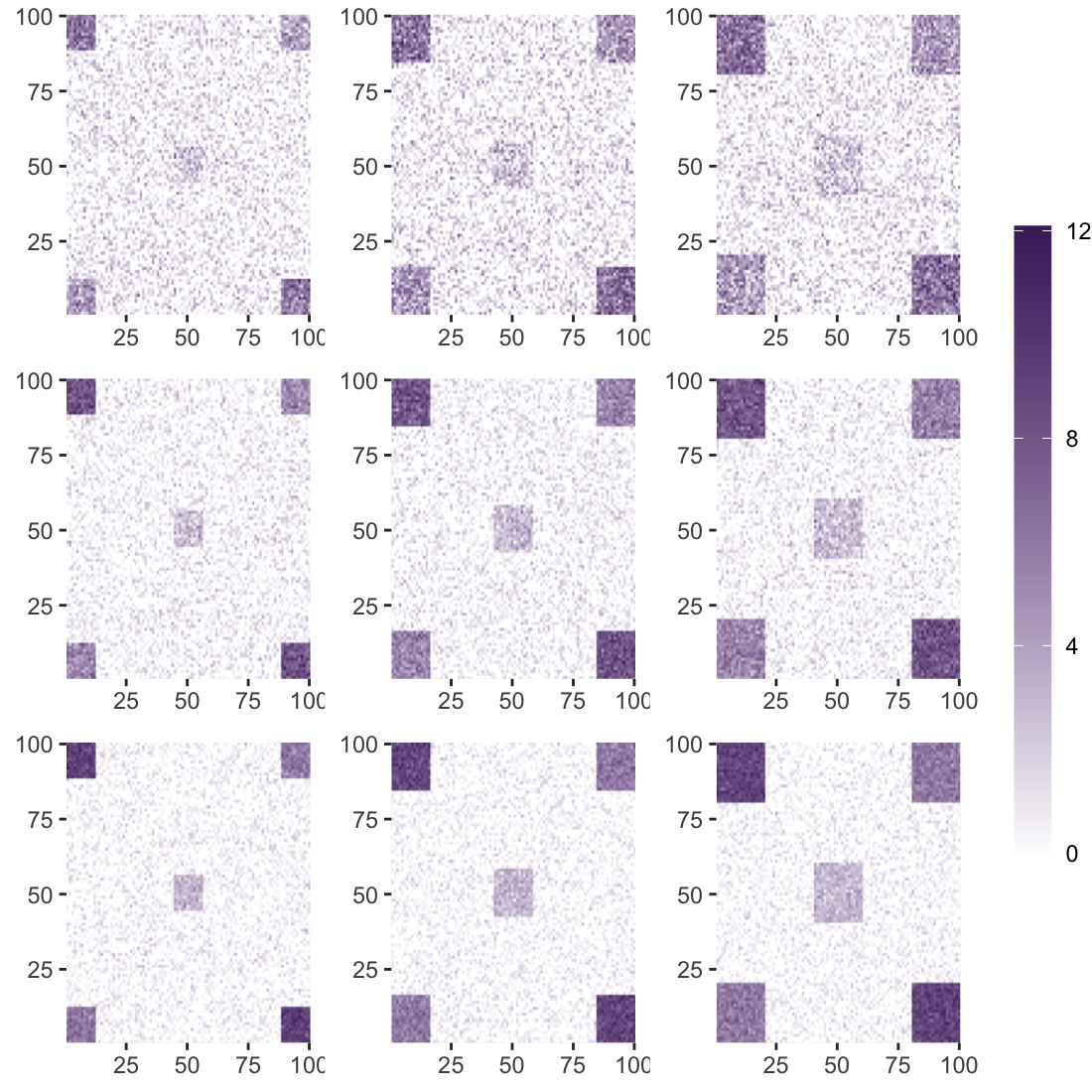} &\includegraphics[scale = 0.12]{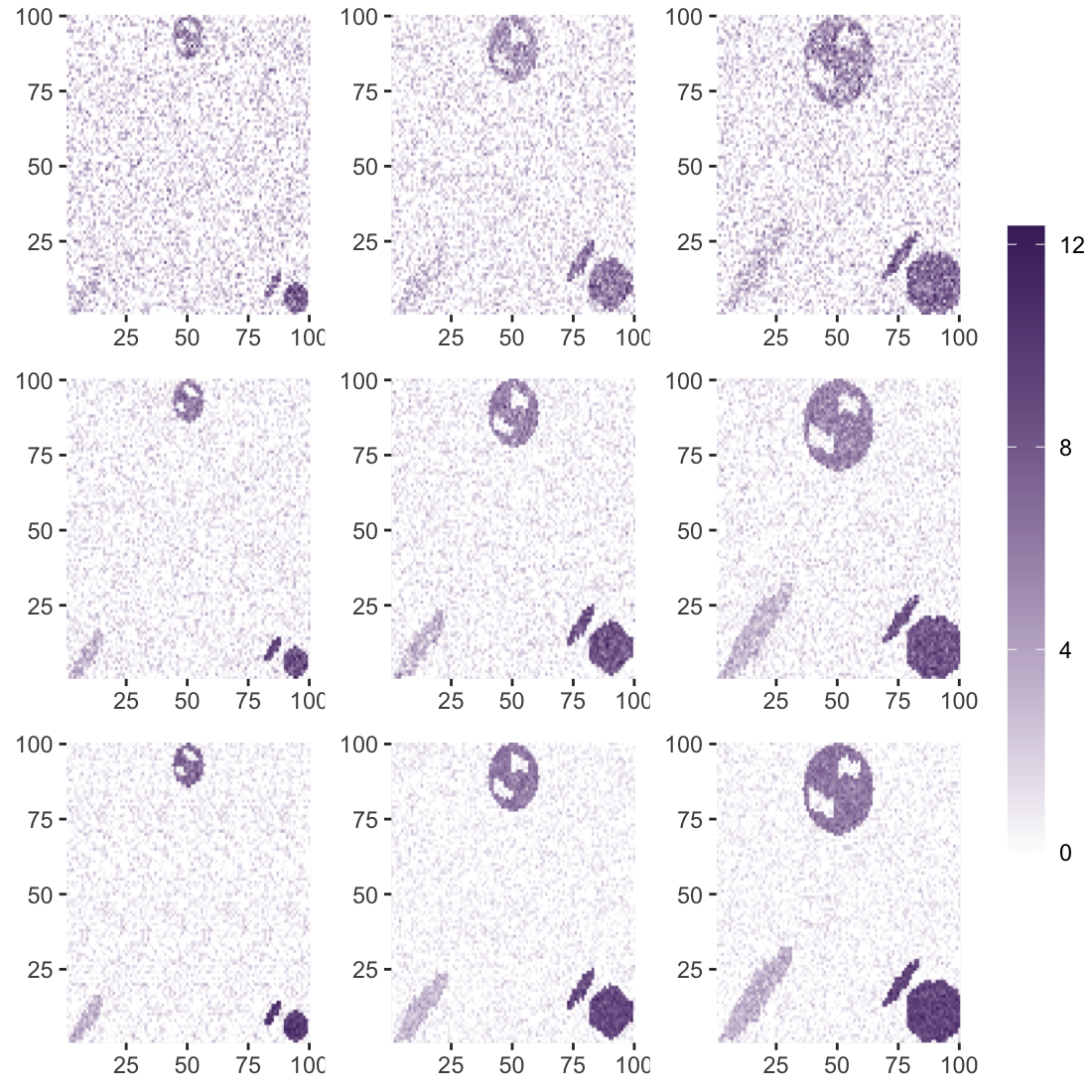} &\includegraphics[scale = 0.12]{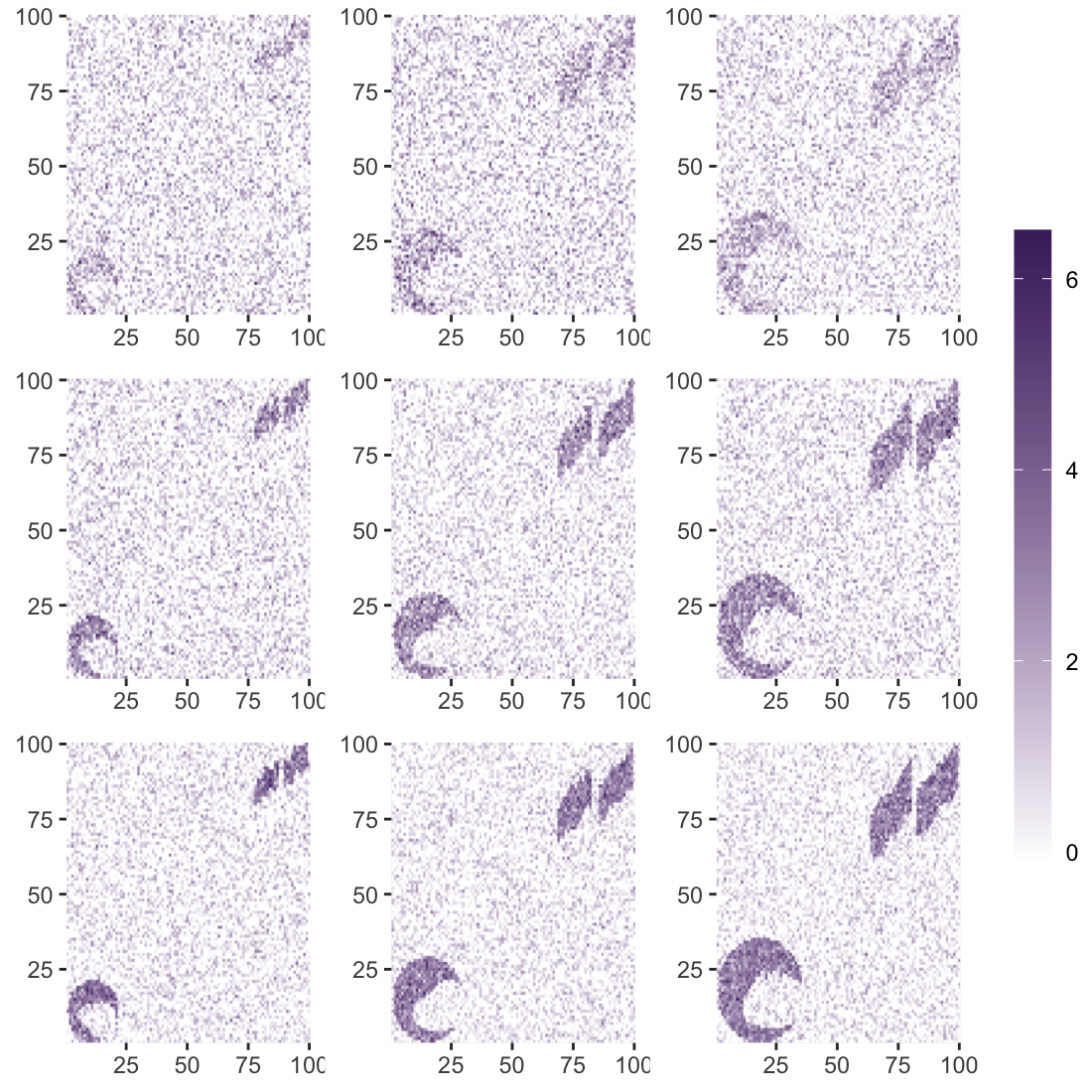}\\
		\small (b1) & \small (b2) & \small (b3)\\
	\end{tabular}
	\caption{Plots of anomaly regions in three different settings (top panel) and realisations of the observed data under different combinations of $\Delta$ and $|R|$ (bottom panel), with sample size $n=10000$ ($100\times 100$).  (a1) Setting 1: five square anomaly regions, where $\mu_1^* = \Delta$, $\mu_2^* =\mu_3^* = 2\Delta$, and $\mu_4^* = \mu_5^*=3\Delta$; (a2) Setting 2: a ellipse anomaly, a circular anomaly with holes, and a disconnected anomaly, where $\mu_1^* = \Delta$, $\mu_2^* = 2\Delta$, and $\mu_3^*=3\Delta$. (a3) Setting 3: a concave anomaly and a disconnected anomaly, where $\mu_1^* = \mu_2^*=\Delta$.
	(b1)-(b3): one time data realisation  under different $\Delta$ and $|R|$ (top to bottom, $\Delta$ increasing; left to right, $|R|$ increasing.)}
	\label{fig:2Dinde-true}
\end{figure}

For the data generating process,  
%We take the sample size to be $n=400$ or $n=2500$ . Additional simulation results for different samples sizes and further discussion on DCART are deferred to \ref{} in Supplementary Material.
we set the random errors $\varepsilon(\s)$ to be identically distributed ${N}(0,1)$ random variables.  As shown in Figure \ref{fig:2Dinde-true}, three different settings of anomaly regions are considered:  1) five equal-sized square anomalies; 2) an ellipse anomaly, a circular anomaly with holes, and a disconnected anomaly; 3) a concave anomaly and a disconnected anomaly. We jitter the boundaries of anomaly regions in Settings 2 and 3, to make them less artificial. The baseline mean signal $\mu_0^*$ is fixed to be 0. Multiple combinations of signal-to-noise ratio, through changing the minimum anomaly mean signal $\Delta$ and the total area of anomalies $|R| = \sum_{j=1}^{m^*}|R_j^*|$, together with different sample sizes $n$ are studied in our simulations.

We adopt two indicators to evaluate the performance of anomaly detection:
$$
\text{NoC} = \dfrac{1}{B}\sum_{b=1}^B\mathbb{I}\{\hat{m}^b = m^*\} \quad \text{and} \quad \text{Err} = \dfrac{1}{B}\sum_{b=1}^{B} \text{Err}\Big(R_{1:m^*}^*,\widehat R_{1:\hat{m}^b}^b\Big),
$$
namely the frequency that we  detect the correct number of anomalies and the averaged detection error, within $B$ times of Monte Carlo simulations, where $\Big\{\hat{m}^b;\,\widehat R_{1:\hat{m}^b}^b\Big\}$ denotes the detected anomalies in $b$-th simulation. The term
$\text{Err}\Big(R_{1:m^*}^*,\widehat R_{1:\hat{m}}\Big)$ consists of the sum of two error components 
adjusted by the total area of anomalies: 
\begin{equation*}
\text{Err}\Big(R_{1:m^*}^* ,\widehat R_{1:\hat{m}}\Big) =\frac{\sum_{i = 1}^{\hat{m}} \min_{j=1,...,m^*}\big|\widehat R_i\setminus R_j^*\big| + \sum_{j=1}^{m^*} \min_{i=1,...,\hat{m}} \big| R_j^* \setminus \widehat R_i \big| }{ \big|R\big|}.
\end{equation*}
The first component measures the error that points in an estimated anomaly do not overlap with the corresponding correct true anomaly region, 
% $$
% \text{Error}\Big(\widehat R_{1:\hat{m}}\setminus R_{1:m^*}^*\Big) = \sum_{i = 1}^{\hat{m}} \min_{j\in\{1,...,m^*\}}\Big|\widehat R_i\setminus \big(R_j^*\cap \widehat R_i\big)\Big|,
% $$
% Each $\widehat{R}_i$ overlap with some true anomaly regions $R_j^*$, $j \in \{1, \dots, m^*\}$. We consider $\max_{j\in\{1, \dots, m^*\}} |R_j^* \cap \widehat{R}_i|$ as the correctly estimated region of $\widehat{R}_i$, and the remaining area is regarded as estimation error. Therefore, $\text{Error}\big(\widehat{R}_{1:\hat{m}} \setminus R_{1:m^*}^*\big)$ is the sum of the errors of $\{\widehat R_1,...,\widehat R_{\hat{m}}\}$.
and the second component measures the error that points in a true anomaly region that have not been detected  correctly.
% $$
% \text{Error}\Big(R_{1:m^*}^*\setminus \widehat R_{1:\hat{m}}\Big) = \sum_{j=1}^{m^*} \min_{i\in \{1,...,\hat{m}\}} \Big| R_j^* \setminus \big( \widehat R_i \cap R_j^* \big) \Big|.
% $$

% The following simulation studies include two-dimensional independent data, two-dimensional dependent data, and three-dimensional independent data. Additional simulations for two-dimensional independent data, involving different sample sizes, unequal dimension lengths of the overall region, and further analysis, are deferred to the Supplementary Material \ref{}.

\subsection{Simulation for independent data}\label{sec:inde-simu}

We carry out $B=100$ simulations under all three settings and different SNR combinations, with sample sizes $n=1225$ and $n=10000$. We also consider settings with $n=400$ and $n=2500$,  which are deferred to the Supplementary Material \ref{sec:add_sim}. 

\begin{figure}[H]
	\centering
	\begin{tabular}{cccc}
		\includegraphics[scale = 0.252]{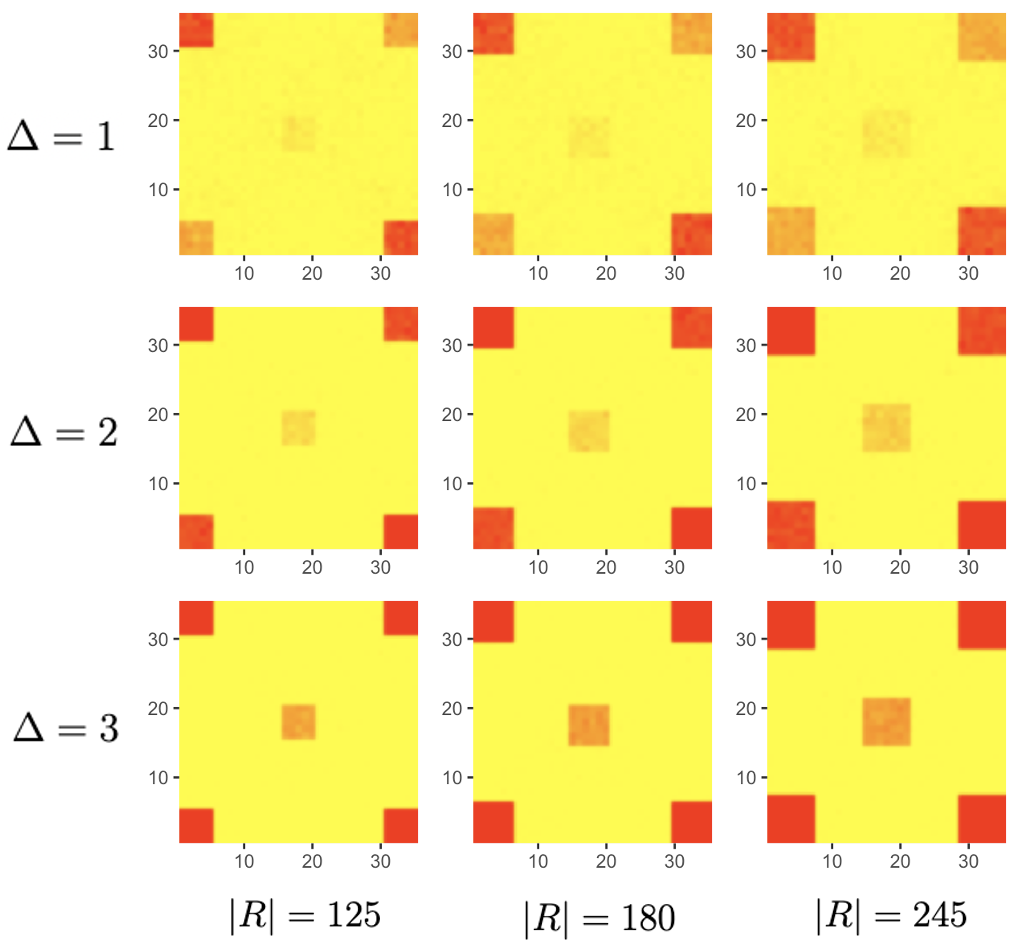} &\includegraphics[scale = 0.252]{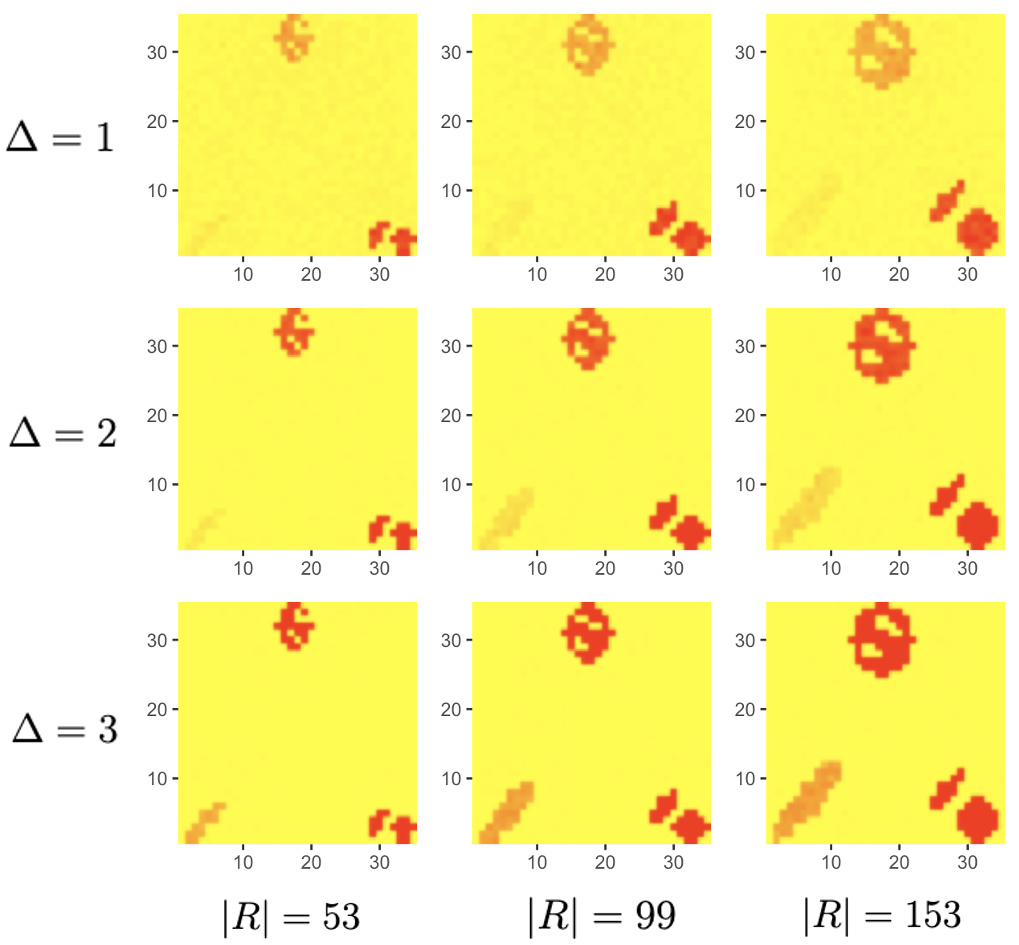} &\includegraphics[scale = 0.252]{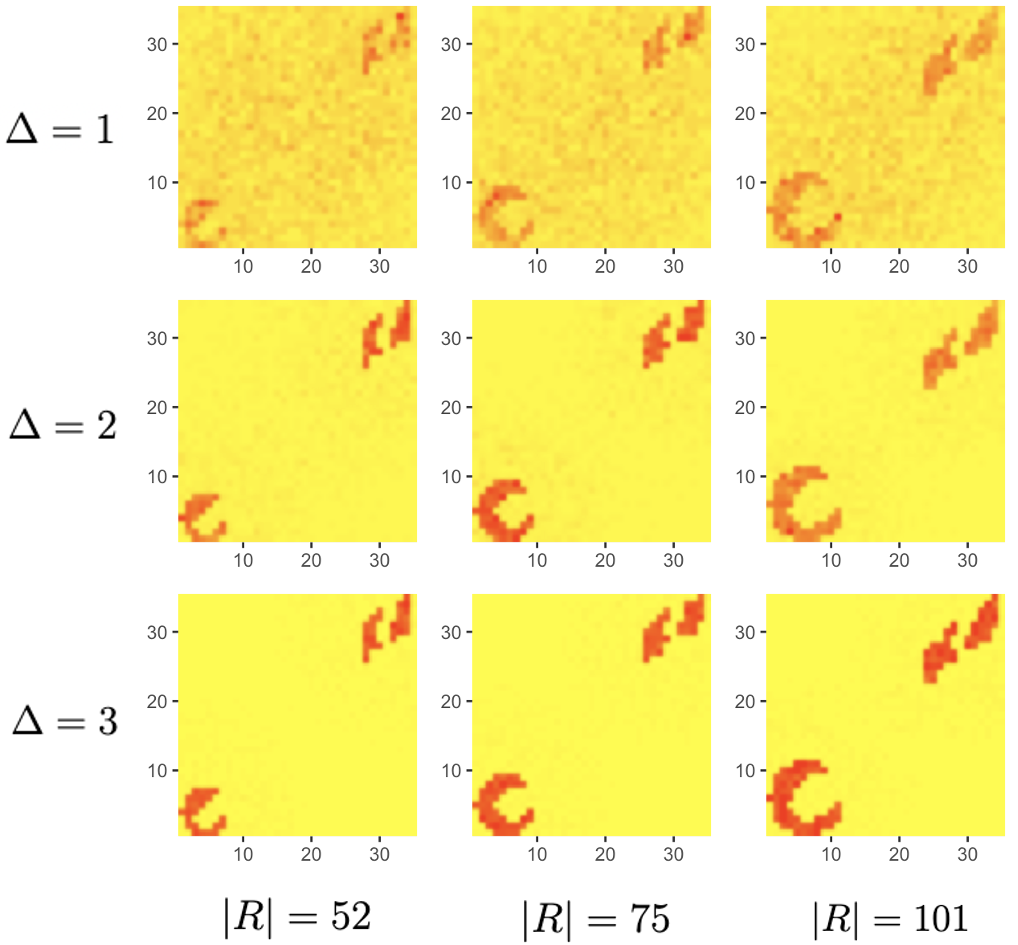}& \multirow{4}{*}[11ex]{\includegraphics[scale=0.3]{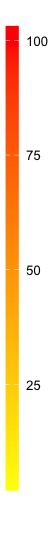}}\\
		\scriptsize \hspace{1cm} (a1)  & \scriptsize \hspace{1cm} (a2) & \scriptsize\hspace{1cm} (a3)  & \\
		~\\
		\includegraphics[scale = 0.24]{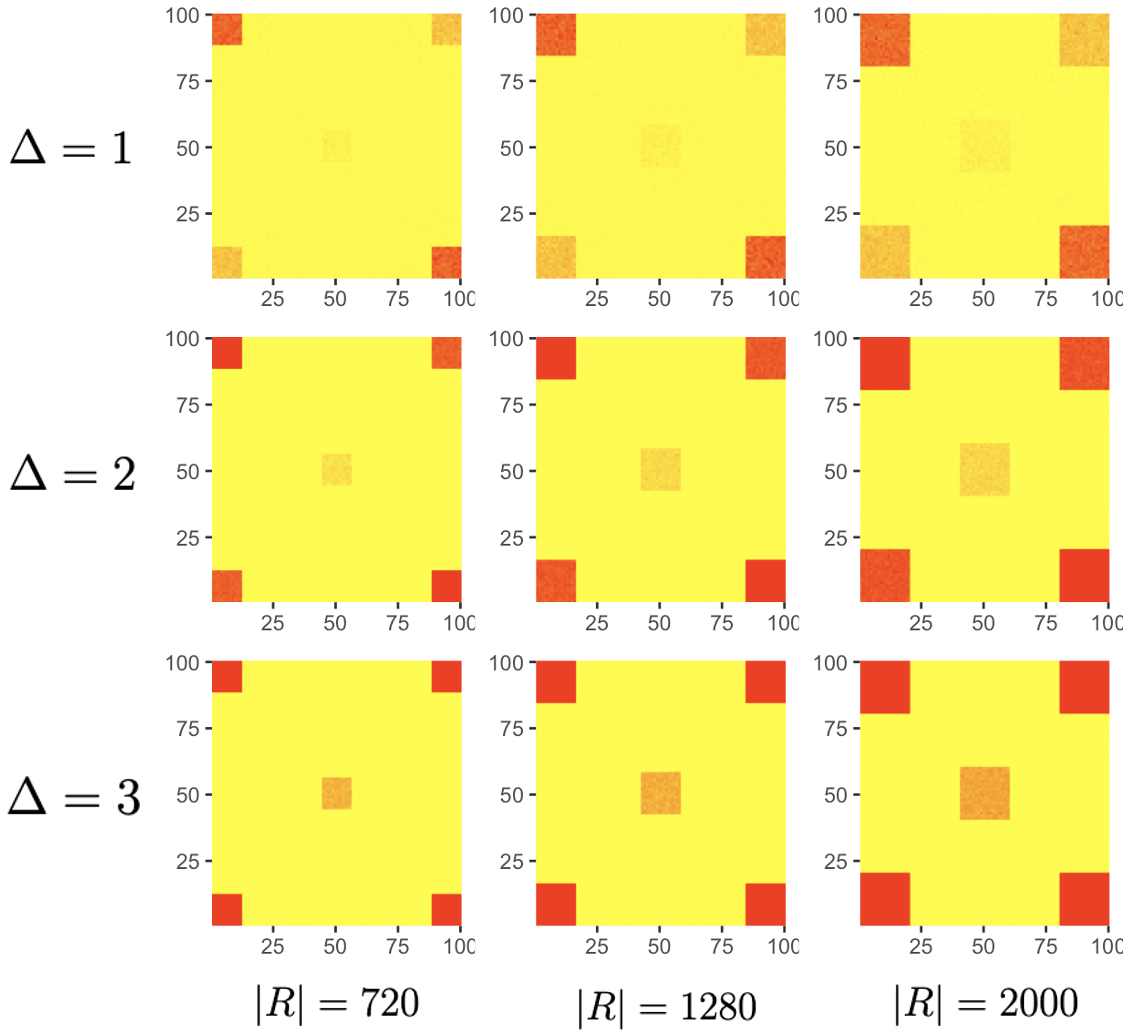} &\includegraphics[scale = 0.24]{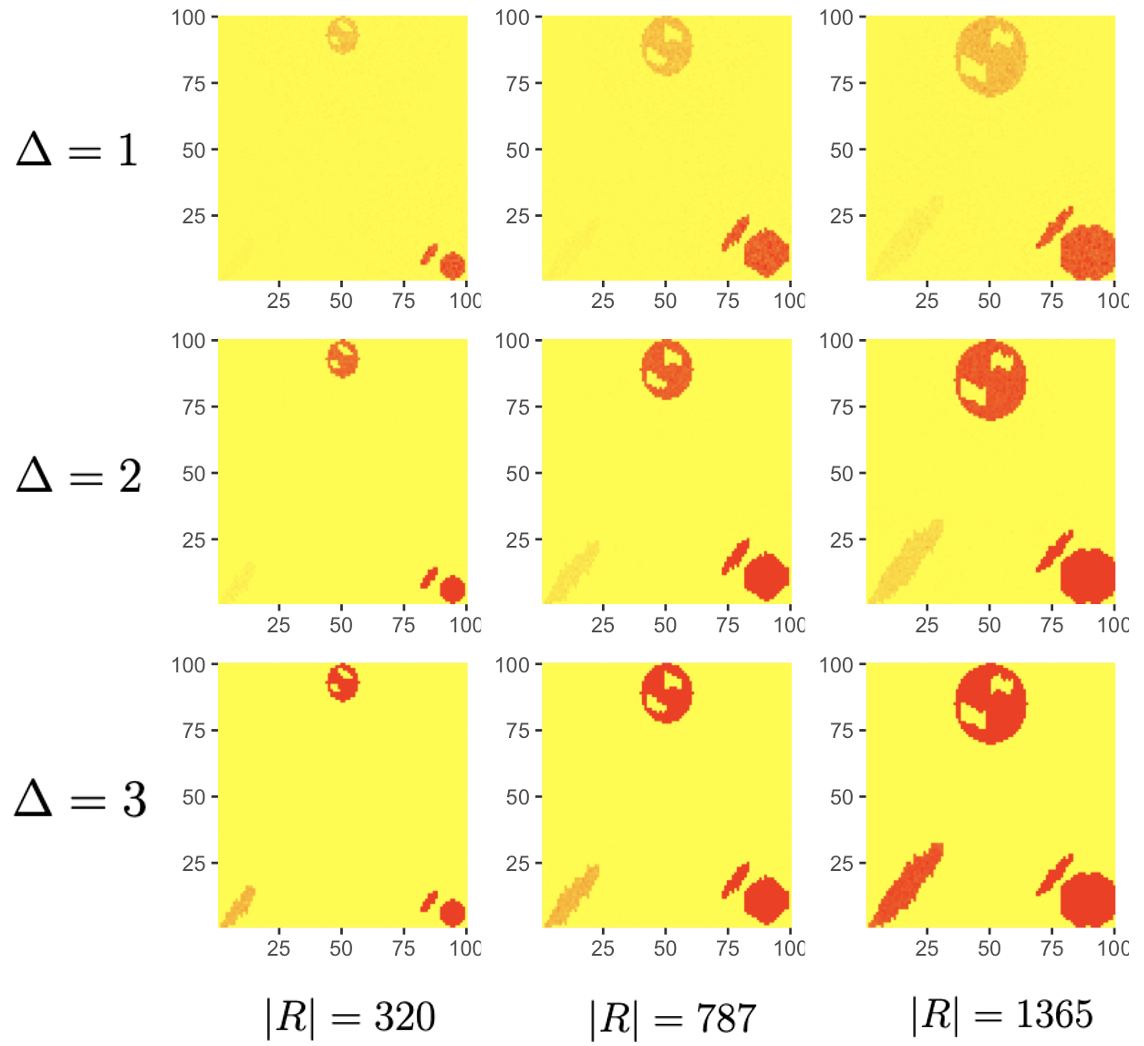} &\includegraphics[scale = 0.24]{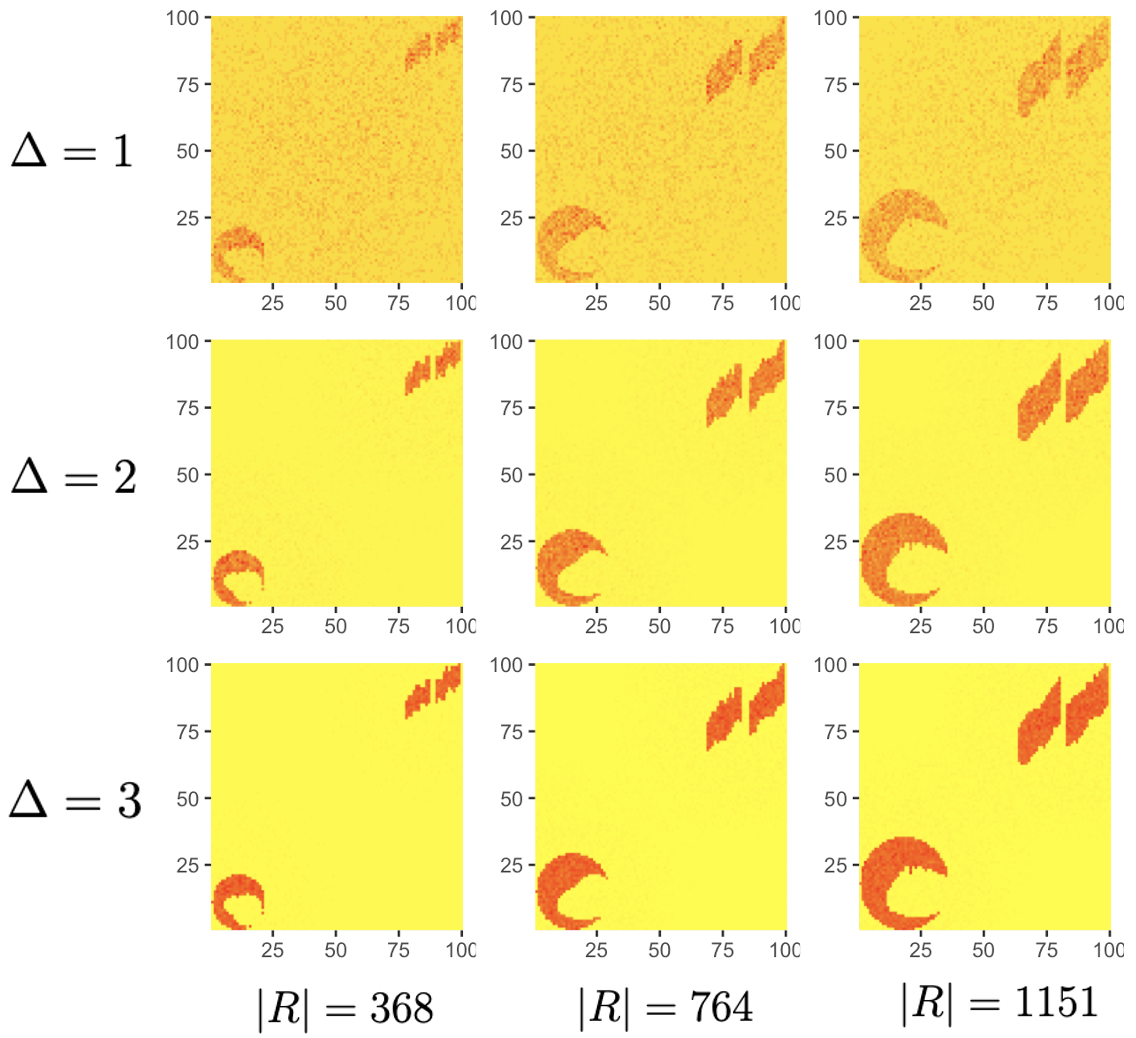}& \\
		\scriptsize \hspace{1cm} (b1)  & \scriptsize \hspace{1cm} (b2) & \scriptsize\hspace{1cm} (b3) & \\
	\end{tabular}
	\caption{ Frequency of points detected as anomalies, varying by 3 settings, with sample sizes $n=1225$ (top panel) and $n=10000$ (bottom panel). Each setting and sample size includes 9 combinations of $\Delta$ and $|R|$ (top to bottom, $\Delta$ increasing; left to right, $|R|$ increasing).}
	\label{result_35_100.plot}
\end{figure}

We observe that the performance of DPLS-SAD is robust to a wide range of penalty parameter values $(\beta, \lambda)$. In theory, we require $\beta < |R_j^{\ast}|\cdot(\mu_j^* - \mu_0^*)^{2}$ to ensure a region is estimated as an anomaly only if doing so results in a sufficient reduction in regional loss. As a result, in most settings, we set $\displaystyle \beta = \Delta\cdot \delta$. From our theorems, $\lambda$ is smaller than $\beta$ roughly by a factor of $n$. Therefore, we fix $\lambda=\beta/n$ in the simulations. In practice, when $(\Delta,\delta)$ are not available, we can select $\beta$ based on sensitivity analysis.
% In the DPLS-SAD algorithm, we always set the threshold as $\xi_m=20\cdot \lfloor\log_{10}(\sqrt{n})\rfloor/m$, where $m$ is the index of the inner loop iteration. 
Our results are summarized in Figure \ref{result_35_100.plot} and Table \ref{tab:1225_10000}, where we also compare with DCART. 

Note that DCART first partitions the lattice into multiple non-overlapping rectangles and then merges partitions with similar mean values to form anomaly regions. This approach tends to perform poorly in settings with complex-shaped anomalies (e.g., our Settings 2 and 3), where a large number of partitions is required to achieve accurate estimation.  Furthermore, DCART struggles to distinguish between anomalies with similar mean signal values, whereas DPLS-SAD is capable of accurately separating them.

From Table \ref{tab:1225_10000}, we can see that DCART only delivers reasonable results in Setting 1 under a few low SNR regimes. It is uniformly outperformed by DPLS-SAD, especially under non-regular anomaly settings, where the DCART often fails to detect any anomalies. The results also reveal that DPLS-SAD becomes more accurate as the mean signal and overall area of anomalies increase, which matches with our theoretical results in Section \ref{sec:consistency}. 

\begin{table}[H]
	\centering
	\renewcommand{\arraystretch}{1.2}
	\setlength{\tabcolsep}{2.9pt}
	\scalebox{0.77}{\begin{tabular}{ccc|ccc|ccc|ccc|ccc|ccc|ccc}
			\toprule[1.5pt]
			\multicolumn{3}{c}{} & \multicolumn{9}{c|}{$n=1225$} & \multicolumn{9}{c}{$n=10000$} \\ \cline{4-21}
			\multicolumn{3}{c}{} & \multicolumn{3}{c|}{Setting 1} & \multicolumn{3}{c|}{Setting 2} & \multicolumn{3}{c|}{Setting 3} &  \multicolumn{3}{c|}{Setting 1} & \multicolumn{3}{c|}{Setting 2} & \multicolumn{3}{c}{Setting 3} \\ \cline{4-21}
			& & \diagbox[width=1.25cm, height=1.2cm]{\hspace{0.3em} $\Delta$}{\multirow{1}{*}{\vspace{-0.7em}$|R|$}} & 125& 180 & \multicolumn{1}{c|}{245}  & 53 &99 & \multicolumn{1}{c|}{153}   & 52 & 75 & 101 & 720 & 1280 & \multicolumn{1}{c|}{2000}  & 320 &787 & \multicolumn{1}{c|}{1365}   & 368 & 764 & 1151 \\
			\bottomrule[1.5pt]
			\multirow{6}{*}{NoC(\%)}& \multirow{3}{*}{\small DPLS-SAD} & 1 & 32 & 34 & 53 & 19 & 23 & 36 & 20 & 24 & 40 & 53 & 71 & 81 & 16 & 38 & 67 & 5 & 7 & 11 \\ 
			& & 2 & 62 & 74 & 88 & 21 & 52 & 74 & 55 & 81 & 83 & 87 & 100 & 100 & 33 & 71 & 99 & 90 & 98 & 100 \\ 
			& & 3 & 97 & 100 & 100 & 76 & 100 & 100 & 100 & 100 & 100 & 100 & 100 & 100 & 97 & 99 & 99 & 100 & 100 & 100 \\
			\cmidrule[1pt]{2-21}
			& \multirow{3}{*}{\small DCART} & 1 & 31 & 31 & 11 & --- &  23 &  34 & --- & --- & --- & 18 & 9 & 5 & 23 & 30 & 32 & --- & --- & --- \\
			& & 2 & 27 & 26 & 32 & 10 & 7 & 15 & --- &--- &  --- & 29 & 28 & 15 & 10 & 29 & 42 & --- & 10 & 6 \\ 
			& & 3 & 30 & 26 & 17 & 1 & 17 & 12 & 8 & 32 & 32 & 35 & 20 & 25 & 41 & 35 & 43 & 29 & 30 & 22 \\
			\bottomrule[1.5pt]
			\multirow{6}{*}{Err(\%)} &\multirow{3}{*}{\small DPLS-SAD} & 1 & \,62\, & \,59\, & \,57\, & 93 & \,74\, & \,69\, & 142 & 123 & 112 & 70 & 68 & 66 & 79 & 72 & 69 & 112 & 102 & 101 \\ 
			& & 2 & 22 & 21 & 20 & 35 & 34 & 31 & 88 & 76 & 76 & 25 & 23 & 21 & 37 & 29 & 26 & 89 & 79 & 77 \\
			& & 3 & 10 & 10 & 9 & 17 & 17 & 16 & 41 & 38 & 36 & 13 & 12 & 10 & 19 & 16 & 3 & 48 & 42 & 39 \\
			\cmidrule[1pt]{2-21}
			& \multirow{3}{*}{\small DCART} & 1 & 42 & 44 & 51 & --- &  56 &  55 & --- & --- & --- & 49 & 57 & 58 & 88 & 71 & 69 & --- & --- & --- \\
			& & 2 & 30 & 32 & 36 & 42 & 40 & 40 & --- & --- &  --- & 36 & 40 & 41 & 47 & 38 & 36 & --- & 83 & 85 \\ 
			& & 3 & 25 & 30 & 35 & 33 & 28 & 27 & 130 & 124 & 110 & 36 & 40 & 38 & 32 & 31 & 25 & 70 & 60 & 61 \\
			\bottomrule[1.5pt]
	\end{tabular}}
	\caption{ Performances of DPLS-SAD and DCART, where   "---" denotes that DCART estimates all the points as baseline in more than 95\% simulations. In Settings 3,  we scale both $\beta$ and $\lambda$ by factors of $0.65$.}
	\label{tab:1225_10000}
\end{table}

Figure \ref{result_35_100.plot} visualizes the frequency that each spatial point is identified as an anomaly point within 100 simulations. The results demonstrate that DPLS-SAD successfully detects anomaly regions even in challenging settings, including cases with complicated anomaly region shapes and distinct regions sharing identical mean values.

\subsection{Simulation for 2D dependent and 3D data}\label{sec:dep-sim}

We extend our experiments to dependent spatial data and 3-dimensional settings. To generate 2D dependent data, we set the covariance between errors ${\varepsilon(\s)}$ and ${\varepsilon(\s')}$ to be $\exp\{-\zeta \cdot \text{dist}(\s, \s')\}$ for any $\s$ and $\s'$, where we consider different dependencies by taking $\zeta\in\{0.01, 0.5, 3\}$. Here we examine data from  Setting 2 with $n = 2500$, under varying values of $\Delta$. Additional results for $n=10000$ can be found in the Supplementary Material \ref{sec:add_sim}. For 3D data, we simulate two anomaly regions with the same mean signal $\Delta$ on a  $12\times 12 \times 12$ lattice, as shown in Figure \ref{fig:true_3D} below.  

\begin{figure}[H]
	\centering
    \vspace*{0.3cm}
	\begin{tabular}{ccc}
	\includegraphics[scale = 0.17]{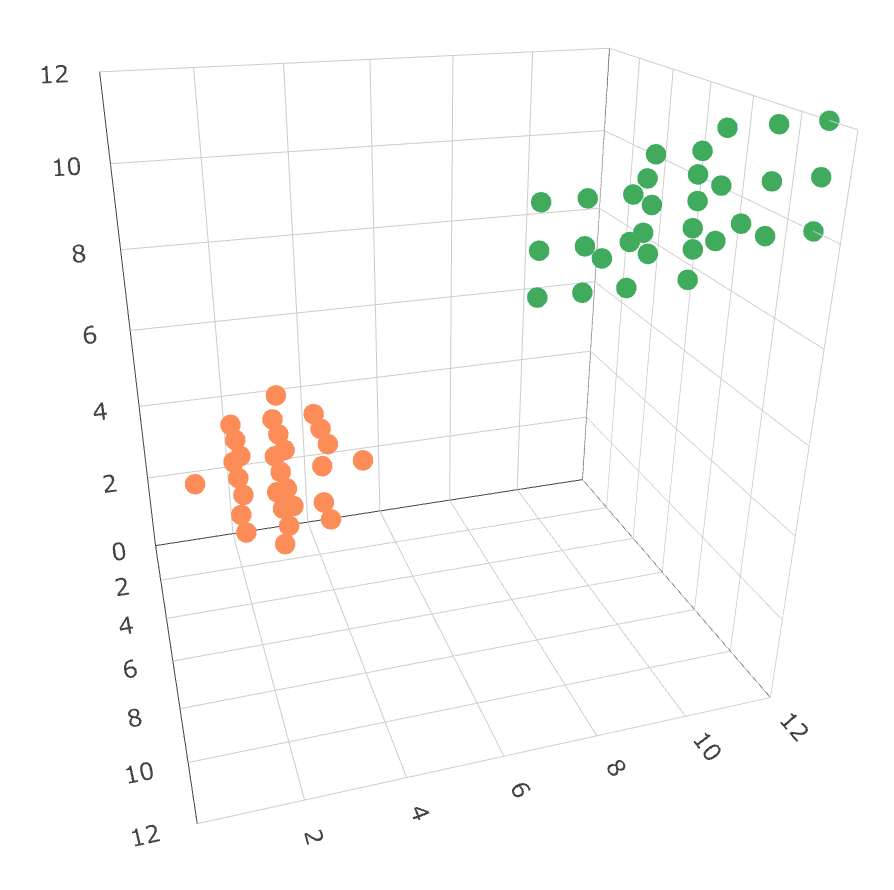}&\,\,& \multirow{1}{*}[28ex]{\includegraphics[scale=0.25]{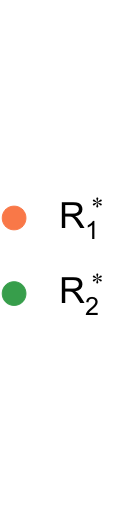}}\\
	\end{tabular}
	\caption{Plot of anomaly regions in three-dimensional data setting, where $R_1$ is a circular anomaly with holes, and $R_2$ is a disconnected anomaly, both have jittered points on the boundary. The baseline data points are not plotted. }
	\label{fig:true_3D}
\end{figure}
% \zc{add legend $R_1$ and $R_2$, delete $s_1$, $s_2$, $s_3$}

\begin{figure}[H]
	\centering
	\begin{tabular}{cccl}
		 \multicolumn{3}{c}{\includegraphics[scale = 0.44]{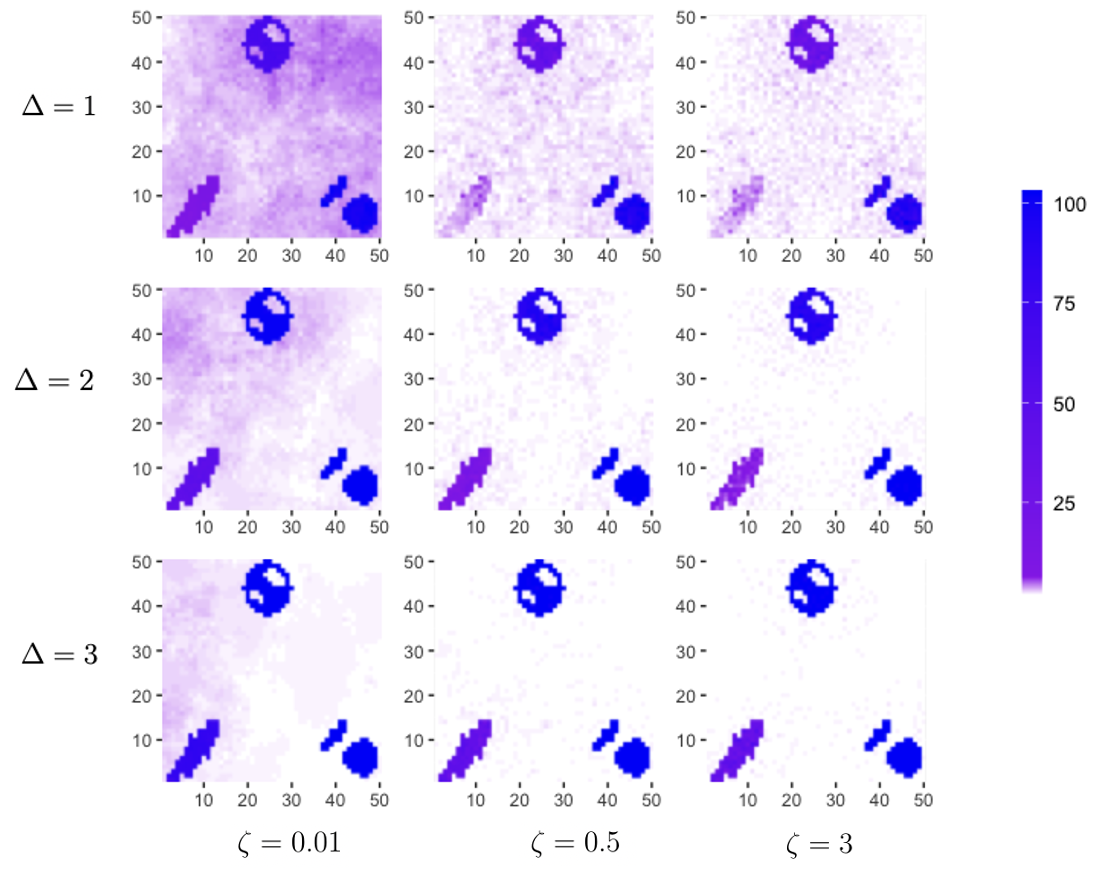}}& \\
		\multicolumn{3}{c}{\small (a)} & \\
		~\\
		 \includegraphics[scale=0.31]{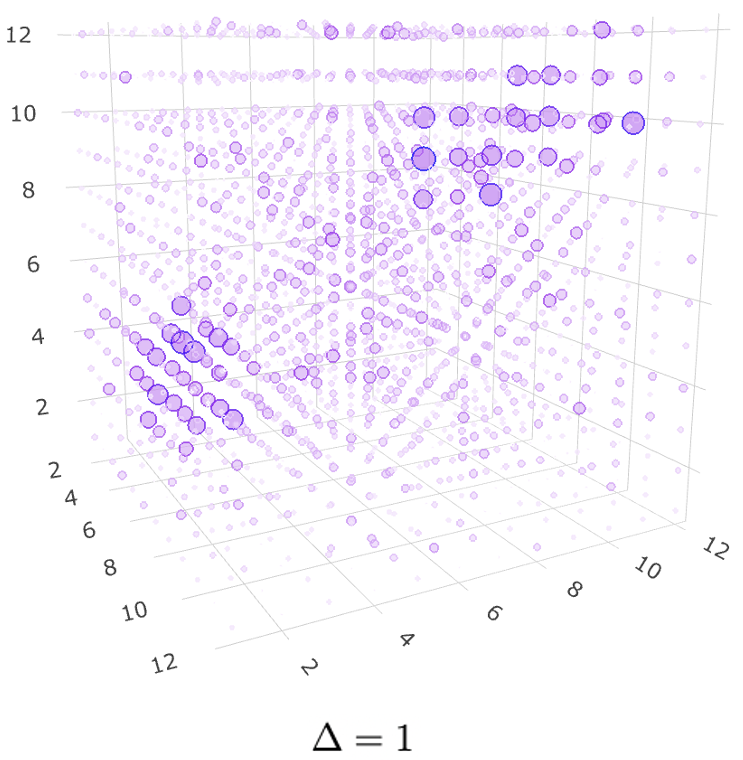} & \includegraphics[scale=0.31]{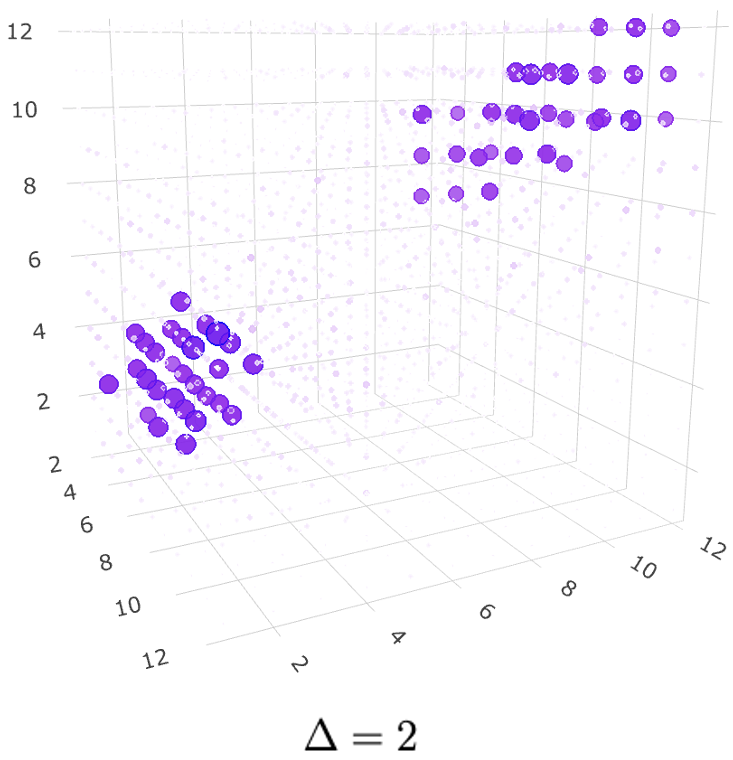} & \includegraphics[scale=0.31]{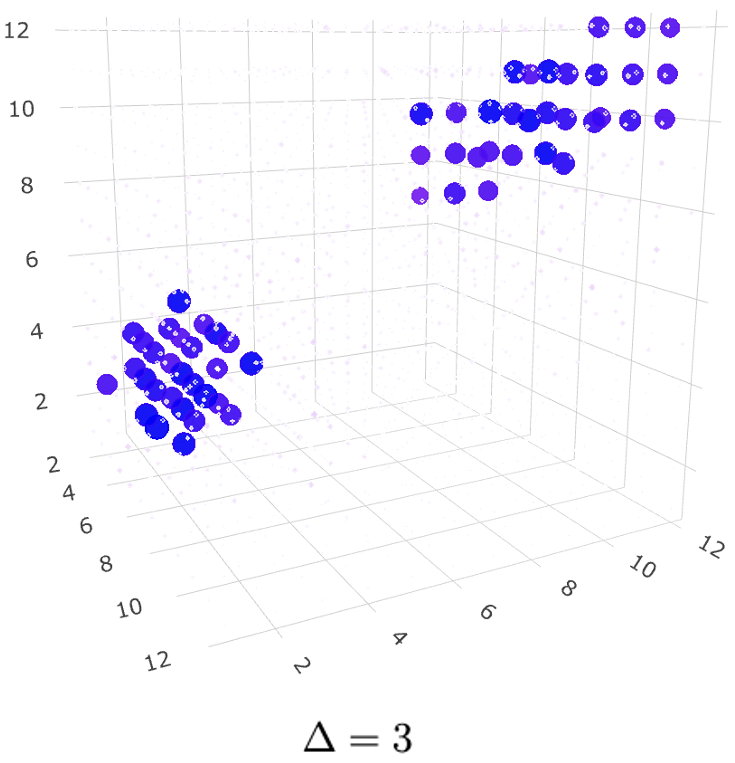}&\multirow{1}{*}[24ex]{\includegraphics[scale = 0.23]{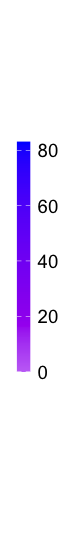}} \\
		 \multicolumn{3}{c}{\small (b)}& \\
	\end{tabular}
	\caption{Frequency of points detected as anomalies for (a) 2D dependent data (n = 2500), and (b) 3D data. For 2D dependent data, we include 9 combinations of $\Delta$ and $\zeta$ (top to bottom, $\Delta$ increasing; left to right, $\zeta$ increasing); Bottom panel show the results for 3D data, varying among $\Delta$ (left to right, $\Delta$ increasing).}
	\label{result_dep_high.plot}
\end{figure}

We adopt the same parameter choices as in Section \ref{sec:inde-simu}, where the $L_0$ penalty and regional penalty parameters are set to $\beta = \Delta \cdot \delta$ and $\lambda = \beta/n$, respectively. Our results are summarized in Figure \ref{result_dep_high.plot} and Table \ref{tab:dep&3D}. Note that other methods cannot handle dependent or 3D spatial data, hence we only present results from DPLS-SAD. It can be seen that the proposed method becomes more accurate as the signal strength $\Delta$ increases. Additionally, Figure \ref{result_dep_high.plot}\,(a) and Table \ref{tab:dep&3D} indicate that a weaker dependence relationship (i.e., larger $\zeta$) leads to better detection outcome, which is consistent with our theory. 
% This trend is less apparent in Figure \ref{result_dep_high.plot} (a1). Although points within true anomalies are darker for smaller $\varphi$, points within true baseline are incorrectly identified as anomalies in many simulations, as Figure \ref{result_dep_high.plot} (a2) shows.

\begin{table}[H]
	\centering
	\renewcommand{\arraystretch}{1.2}
	\vspace*{1.3mm}
	\setlength{\tabcolsep}{2.5pt}
	\scalebox{0.77}{\begin{tabular}{ccccccccccccc}
			\toprule[1.5pt]
			& \multicolumn{7}{c}{2D dependent data} &&\,\,\,\,\,\,\,\,& \multicolumn{3}{c}{3D data}\\ \cline{2-13}
			& \multicolumn{3}{c}{NoC(\%)} &\,\,\,\,& \multicolumn{3}{c}{Err(\%)} &&& \multicolumn{1}{c}{\multirow{2}{1.8cm}[-0.5ex]{\hfil NoC(\%)}} &\,\,& \multirow{2}{*}[-0.5ex]{Err(\%)}\\
			& $\zeta=0.01$ & $\zeta=0.5$ & $\zeta = 3$ & &$\zeta = 0.01$ & $\zeta = 0.5$ & $\zeta = 3$ &&& \multicolumn{1}{c}{}&& \\ 
			\bottomrule[1pt]
			\multicolumn{1}{c}{$\Delta = 1$\,\,\,\,\,\,\,} & 21 & 23 & \multicolumn{1}{c}{27} && 134 & 75 & \multicolumn{1}{c}{73} &&&  15 && 133 \\
			\multicolumn{1}{c}{$\Delta = 2$\,\,\,\,\,\,\,} & 52 & 53 & \multicolumn{1}{c}{66} && 39 & 33 & \multicolumn{1}{c}{33} &&&  52 && 92 \\
			\multicolumn{1}{c}{$\Delta = 3$\,\,\,\,\,\,\,} & 75 & 85 & \multicolumn{1}{c}{100} && 19 & 18 & \multicolumn{1}{c}{17} &&&  92 && 48 \\
			\bottomrule[1.5pt]
	\end{tabular}}
    \caption{Performance of DPLS-SAD for 2D dependent (n = 2500) and 3D data. Results are based on $100$ Monte Carlo simulations, where we fix $|R| = 228$ for 2D dependent data and $|R| = 59$ for 3D data.}
	\label{tab:dep&3D}
\end{table}

\section{Real-world data application}\label{sec:realdata}

We illustrate the proposed method by detecting marine heatwaves (MHWs) over the entire globe. Marine heatwaves have devastating impacts on marine ecosystems, including mass coral bleaching, substantial losses in kelp forests and seagrass, and declines in economically important species such as lobsters, crabs, abalones, and scallops \citep{holbrook2020kepping}. The importance of this research was amplified by anthropogenic warming, which has doubled the occurrence of MHWs since 1982 and increased the total number of days with MHWs by 50\% over the last century \citep{oliver2018longer}. DPLS-SAD can provide an automatic identification of HMWs, whilst in existing literatures these events are usually specified manually by oceanographers.

To carry out our analysis of MHWs, we use the level-4 sea surface temperature (SST) data from the European Space Agency Climate Change Initiative (ESA-CCI) Programme, which provides global and gridded daily mean SST since 1980, derived from combining multiple series of thermal infra-red sensors \citep{embury2024satellite}. We take a coarser version of the SST data with a grid resolution of 1$^\circ$ in longitude and latitude, which equates to a 360 $\times 180$ spatial lattice. Only grid points located in the ocean are retained, resulting in a sample size of 42827.

\begin{figure}[h]
	\centering
	\includegraphics[scale = 0.19]{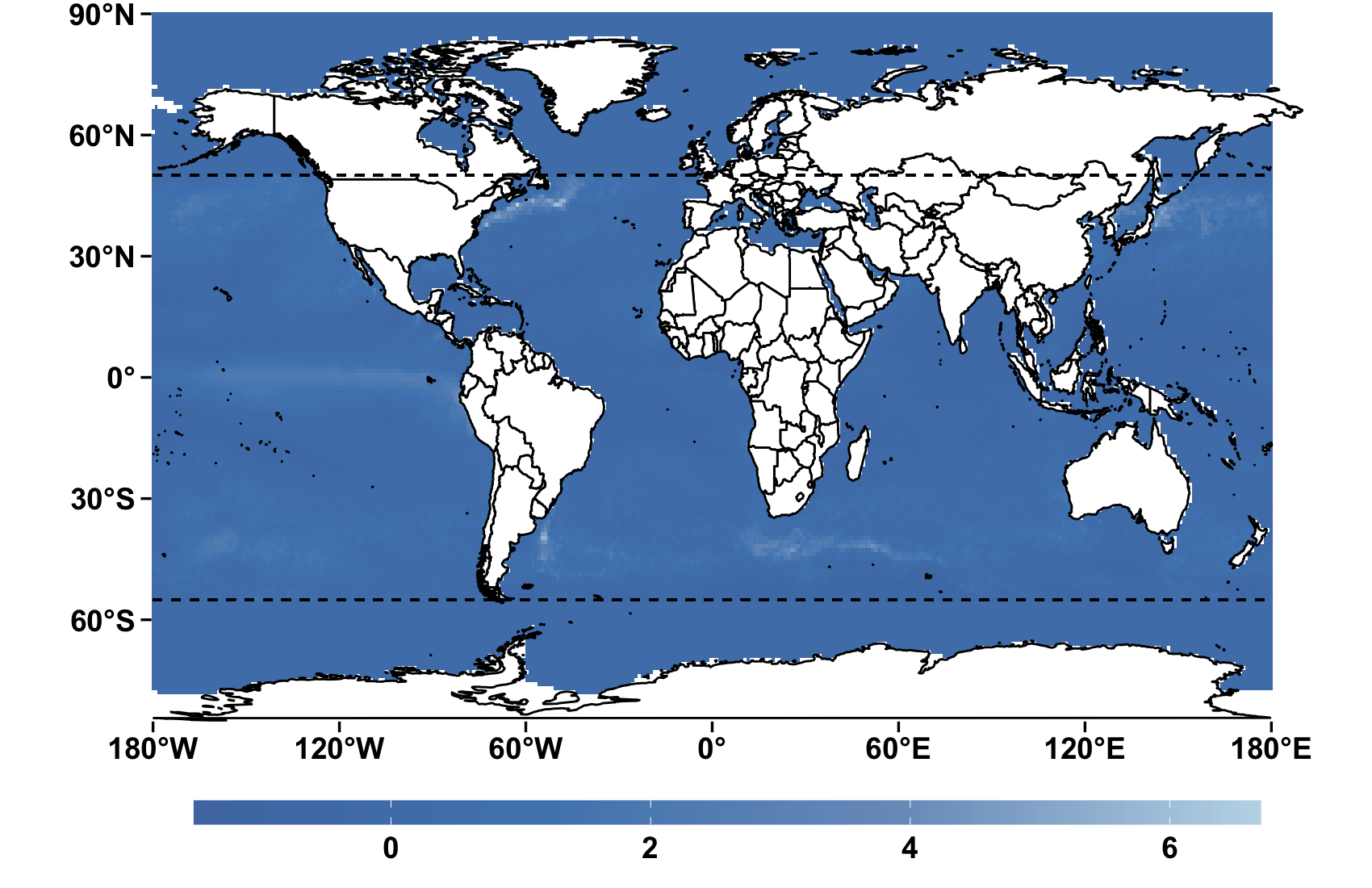}
	\vspace{-0.5cm}
	\caption{Maximum monthly detrended average SST across 2000–2023 (Land temperature are omitted).  Lighter colors indicate higher SST.}
	\label{fig:SSTdata}
\end{figure}

A common linear yearly detrending is applied  on each grid to eliminate seasonal variability and remove the anthropogenic warming trend. As the MHWs are more commonly studied in tropical and temperate regions, we restrict our analysis to latitudes between 55 degree south and 50 degree north. Given that MHWs usually persist for many weeks or months, we then take the maximum monthly average of the detrended SST between the years 2000 and 2023. Figure \ref{fig:SSTdata} below provides a visualization of our final pre-processed data, based on which we aim to simultaneously detect the major MHWs since the 21st century. 

% The PLS-SAD method was applied to pre-processed SST data in order to identify significant marine heatwave regions in the 21st century. The data were pre-processed by fitting a linear regression of SST on year for each day (excluding February 29) and each spatial location from 2000 to 2023, and replacing the original values with the resulting residuals. 
% This removes long-term temperature trends such as global warming and allows focus on spatial anomalies.

For the selection of the tuning parameters, we perform a sensitivity analysis on a grid of $L_0$ penalty parameter $\beta$ based on the scale of $\sqrt{n}\log n$, which indicates that choosing $\beta$  between 450 and 550 yields both stable and reasonable MHW detections. Therefore, we set $\beta = 495$ and $\lambda = \beta/n$. To estimate the mean signal in the baseline region, DPLS-SAD was performed twice, and we update the baseline mean estimate in the second run using the median of the SST values of the detected baseline region from the first run.

\begin{figure}[h]
	\centering
	\includegraphics[scale = 0.405]{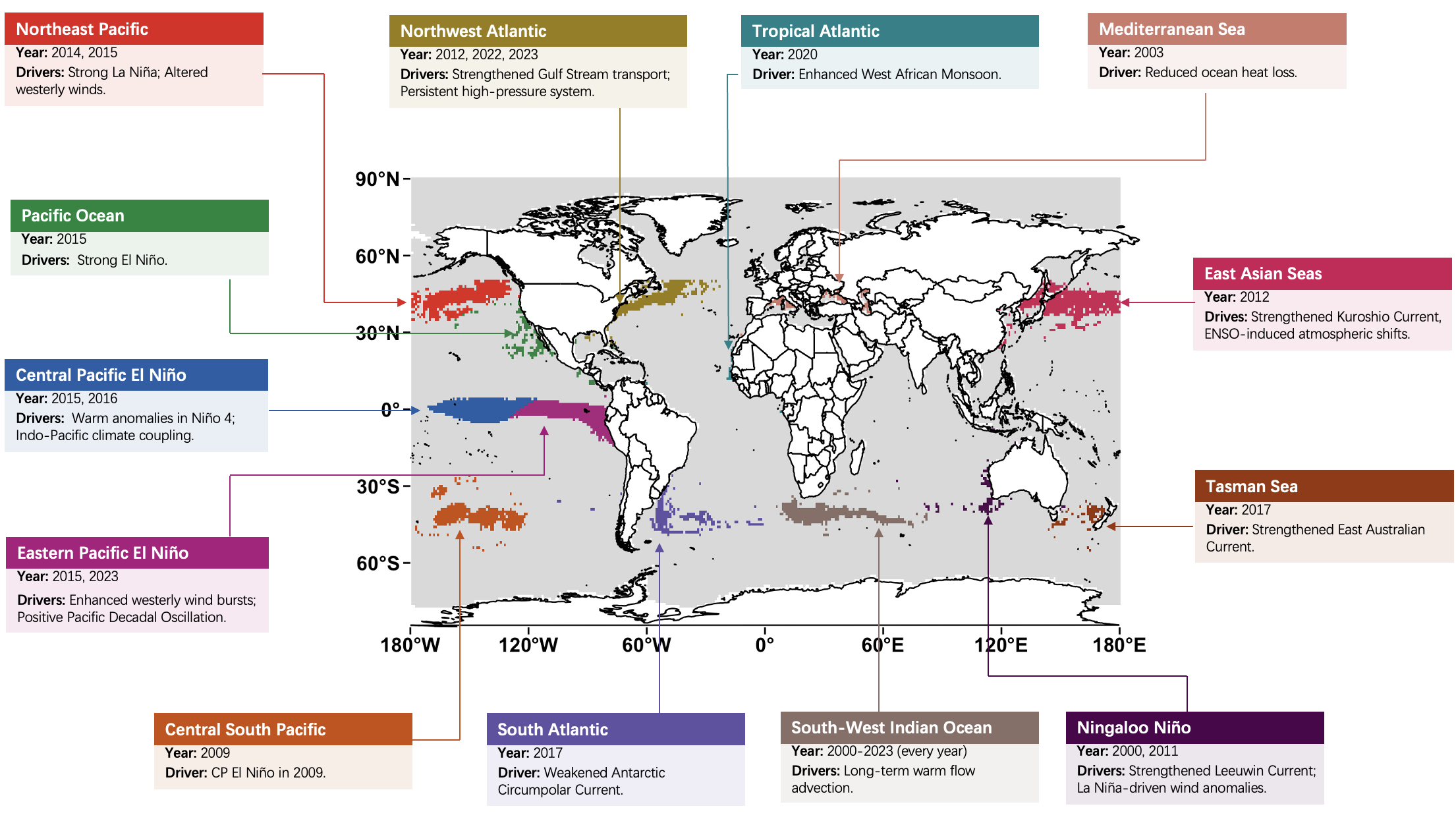}
	\caption{\small Anomaly regions are shown in different colors, with labels marking the active years and drivers of the corresponding MHW events. Detected anomalies are matched with historical major MHWs (with primary peak year and drivers) since 2000, including Northwest Atlantic \citep{mills2013fishers,von2024state}, South Atlantic \citep{manta2018record}, East Asian Sea \citep{miyama2021marine}, Central and Eastern El Ni\~{n}o \citep{lheureux2017observing,lian2023strong}, Pacific Ocean \citep{fewings2019regional}, Mediterranean Sea \citep{olita2003effects}, Tropical Atlantic \citep{pfleiderer2022extreme}, Central South Pacific \citep{lee2010record}, Ningaloo Ni\~{n}o \citep{marshall2015initiation, holbrook2020kepping}, Northeast Pacific Ocean \citep{cavole2016biological}, Tasman Sea \citep{kaitar2022drivers}.}
	\label{fig:SSTresult}
\end{figure}

 In Figure \ref{fig:SSTresult}, we demonstrate the result of applying DPLS-SAD to the pre-processed data, compared to the historical MHW records. Our method provides a highly accurate and reliable detection of major MHWs hotspots since 2000 \citep{oliver2021marine}, recovering their complex shapes. The only exception is the active warm zone in the Southwest Indian Ocean, which is not linked to specific MHW events. However, the area stays warm frequently due to long-term ocean processes like the Agulhas Current and warm eddies, which bring heat from the Indian Ocean into the South Atlantic \citep{beal2011on}. It is interesting to note that we automatically segment the significant MHWs in the Pacific Ocean caused by El Ni\~{n}o into two nearby anomaly regions. The central Pacific region, linked to CP-type El Ni\~{n}o events in 2015 and 2016, is affected by atmospheric changes and shows stronger connections with the southern Indian Ocean.  While the eastern Pacific region is linked to  EP-type El Ni\~{n}o, which was active in 2015 and 2023, and mainly driven by large-scale changes in the thermocline and surface winds that are closely related to the tropical Indian Ocean \citep{kao2009contrasting}.

\bigskip
 {\small
 \renewcommand{\baselinestretch}{1.3}\selectfont

}

\begin{appendices}
\newpage
\setcounter{page}{1}
%\fancyhead[LE,RO]{\bfseries\thepage}
%\fancyhead[LO]{\bfseries\nouppercase{\rightmark}}
%\fancyhead[RE]{\bfseries\nouppercase{\leftmark}}

\setcounter{figure}{0}
\makeatletter
\renewcommand{\thefigure}{S\@arabic\c@figure}
\makeatother

\setcounter{table}{0}
 \makeatletter
\renewcommand{\thetable}{S\@arabic\c@table}
\makeatother

\makeatletter
\@addtoreset{theorem}{section}   % 定理在每个 section 重置
\@addtoreset{assumption}{section}
\makeatother
\renewcommand{\thetheorem}{\thesection.\arabic{theorem}}
\renewcommand{\theassumption}{\thesection.\arabic{assumption}}
\setcounter{theorem}{0}
\setcounter{assumption}{0}
% \lhead{\bfseries SUPPLEMENTARY MATERIAL}
%\renewcommand\theequation{A.\arabic{equation}}
%\setcounter{equation}{0}
%\setcounter{theorem}{0}
%\setcounter{assumption}{0}en
%\setcounter{corollary}{0}
%\setcounter{lemma}{0}
\noindent

\renewcommand{\theHfigure}{S.\arabic{figure}}
\renewcommand{\theHtable}{S.\arabic{table}}
\renewcommand{\theHtheorem}{S.\thesection.\arabic{theorem}}
\renewcommand{\theHassumption}{S.\thesection.\arabic{assumption}}

\setcounter{equation}{0}
\renewcommand{\theHequation}{S.\arabic{equation}}

%===================================================================================================

\section*{\centering {\large Supplementary Material for ``Optimal Spatial Anomaly Detection''}}\label{sec:supp}
%===============================================================================%

\bigskip
\begin{center}
{ Baiyu Wang \,and  \,Chao Zheng}
\end{center}

\bigskip

In the supplementary material, in Section~\ref{sec:consistencymean} we present an alternative SAD problem in which anomaly regions are allowed to be spatially close to each other.
Algorithmic implementation details of the proposed DPLS-SAD are provided in Section~\ref{sec:algorithm}, and additional simulation results complementing Section~\ref{sec:sim} are reported in Section~\ref{sec:add_sim}.

\section{Non-separable spatial anomaly detection }\label{sec:consistencymean} 

In Section \ref{sec:consistency}, we show that spatial anomaly regions can be consistently detected when they are sufficiently separated, as specified in Assumption~\ref{assmp:signal}\,(iii). In practice, however, this assumption may be violated, for example, when anomaly regions share part of their boundaries or when one region is nested within another. In this section, we consider spatial anomaly detection without imposing Assumption~\ref{assmp:signal}\,(iii), while preserving consistent detection guarantees and the same localization rate. To this end, we need to make a stronger assumption on the regional signal difference.
%In Section \ref{sec:consistency}, we discuss the signal strength as outlined in Assumption \ref{assmp:signal}, which comprises two parts: one is related to the difference in mean values between anomaly regions and baseline region, and the other concerns the spatial separation of anomaly regions. In this section, we do not impose additional assumptions on the positions of anomaly regions but instead use a different assumption regarding mean values. 
Recall in Section \ref{sec:consistency}, we define $\Delta$ as the minimum mean difference between any anomaly region and the baseline. In this section, we change its definition to
$$
\Delta := \min_{\mathop { i\neq j\atop i,j=0,...,m^* }} |\mu_i^* - \mu_j^*| ,
$$
where we also consider pairwise mean signal difference between any anomaly regions, i.e., $\Delta_{i,j}:=|\mu_i^* - \mu_j^*|$. The updated detectability  assumptions are as follows:
\begin{assumption}
	\label{assmp:siganlmean}
	%(Signal Strength with Different Mean)
	(i) There exists  $\eta > 0$ such that 
	$$\frac{\Delta^2}{\sigma^2}\geq C_{\eta}\cdot\frac{\log^{1+\eta} n}{\sqrt{n}},$$
	where $C_\eta>0$ is a constant.\\
    (ii) Same as Assumption \ref{assmp:signal}\,(ii).\\
	(iii) For any $i,j \in \{1,...,m^*\}$ and $i\neq j$, there exist constants $C_{low}$ and $C_{up}$ satisfied
	$$
	C_{low} \cdot \Delta \leq \Delta_{i,j} \leq C_{up} \cdot \Delta.
	$$
\end{assumption}
Similar to Section \ref{sec:consistency}, Assumption \ref{assmp:siganlmean}\,(i) and (ii) impose a lower bound on the SNR. Assumption \ref{assmp:siganlmean}\,(iii) is an additional assumption that constrains the scale of mean signal differences.

%Comparing Assumption \ref{assmp:signal} with Assumption \ref{assmp:siganlmean}, we find that the only difference lies in the definition of $\Delta$. In Assumption \ref{assmp:signal}, $\Delta$ considers only the difference of mean value between anomaly regions and baseline region, However, in Assumption \ref{assmp:siganlmean}, it additionally encompasses the mean values between each of the anomaly regions. Therefore, Assumption \ref{assmp:siganlmean} shifts the 'well separated' assumption from spatial position to mean values.

Relaxing Assumption \ref{assmp:signal}\,(iii) removes constraints on the distance between anomaly regions, therefore the regional penalty is no longer necessary. We can obtain an estimator by minimizing the classic $L_0$ penalized cost function, i.e.,
\begin{equation}
	\label{eq:estimator_mean}
	\{\hat{m};\,\widehat{R}_{1:\hm}\} = \argmin_{m; \,R_{1:m} \in \mathcal{R}} \big\{L(R_{1:m}) + \beta m\big\},
\end{equation}
which is similar to the timeline setting, where $\mathcal{R}$ is defined as the class of smooth regions in Definition \ref{defi:smooth}.

The following Theorem \ref{thm:meanconsistency} shows that solving (\ref{eq:estimator_mean}) yields consistent estimators of the spatial anomaly regions, with localization error at the rate of $O(\sqrt{n}\log n)$.

\begin{theorem}
	\label{thm:meanconsistency}
    Suppose Assumptions \ref{assmp:data}, \ref{assmp:smooth} and \ref{assmp:siganlmean} hold. If we choose $\beta=C_{\beta,3} \sqrt{n}\log n$, where $C_{\beta,3}$ is a large enough absolute constant not depending on $n$ and $m^*$. Let $\{\hm;\,\widehat R_{1:\hm}\}$ be the minimizer from solving (\ref{eq:estimator_mean}). There exist constants $c_{\gamma}, C_{\varepsilon,3}>0$ such that
	$$ 
	\hm = m^\ast \quad \textnormal{and} \quad D\Big(\widehat R_j, R^\ast_j\Big)\leq \dfrac{C_{\varepsilon,3}\sigma^2}{\Delta^2}\ld, \quad j=1,\dots, m^*
	$$
	holds with probability at least $1-2\exp\big(-c_{\gamma}\ld\big)$.
\end{theorem}

\bigskip

\section{Algorithm for fast detection}\label{sec:algorithm}

We first consider detecting spatial anomaly regions in 2D data, which requires numerically solving the optimization problem (\ref{eq:estimator}). Classical changepoint algorithms, such as dynamic programming \citep{jackson2005algorithm, killick2012optimal-ap} or pruning-based methods \citep{maidstone2017optimal}, are designed for sequential data, which are not applicable. 
The minimization here is highly non-convex and NP-hard, making the problem computationally challenging, with a cost of $O(n^n)$. Therefore, we propose an efficient search strategy, inspired by one-dimensional $k$-means clustering algorithm \citep{wang2011ckmeans-ex}, that computes an approximate instead of exact solution to (\ref{eq:estimator}), with a reduced computational cost of $O(n^2)$. 

% $1-D$ $k$-means clustering algorithm divide data $Y_1>Y_2>...>Y_n$ into $k$ groups by minimising the sum of square within groups, and the cost function is denoted as $\mathcal{C}(k,n)$. \red{cite paper} use the following recurrence equation to find the optimal partition with $O(n^2 k)$ computational cost:
% \begin{equation}
% 	\label{recur_equation}
% 	\mathcal{C}(l,p) = \min_{l\leq q\leq p}\Big\{\mathcal{C}(l-1,q-1) + \sum_{j = q}^p (Y_j - \Bar Y_{q:p})^2\Big\}
% \end{equation}
% where $\Bar Y_{q:p}$ is the mean value of $\{Y_p,...,Y_q\}$. The algorithm improves computational efficiency by updating the cost function of new partitions using information from the previous partitions.

% PELT is an efficient algorithm for changepoint detection in time series, designed to find the minimizer of a penalized cost function in linear time. The algorithm employs a pruning technique to improve the efficiency of global search by reducing computations at points that meet specific conditions.

 Consider a new sequence $\{Y_i\}_{i=1}^n$, with $|Y_1| \geq |Y_2| \geq ... \geq |Y_n|$, as a rearrangement of $\big\{ Y(\s) - \mu_0^* \big\}_{\s\in\mathcal{S}}$. In this way, all the baseline points are likely to be at the end of the new sequence. Denote $\s_i$ as the corresponding lattice location of $Y_i$, and define $\mathcal{S}_{\mathcal{I}}=\{\s_i: i \in \mathcal{I}\}$ for some index set $\mathcal{I}$. Consider the following minimization problem:
\begin{equation}\label{eq:approx.pc.1}
\min_{1\leq N \leq n}\Bigg[ L\big(R^N_0\big) + \min_m\bigg\{\min_{R^N_{1:m}}\Big(\sum_{j=1}^m L\big(R_j^{N}\big) + \lambda \sum_{j=1}^{m}\big|\text{Co}\big(R_j^{N}\big)\big|\Big)+ \beta m\bigg\} \Bigg],
\end{equation}
where $R^N_{1:m}$ are $m$ non-overlapping regions that form a segmentation of $\mathcal{S}_{1:N}$, i.e., $\bigcup_{j=1}^m R_j^N = \mathcal{S}_{1:N}$, and $R_0^N=\mathcal{S}_{N+1:n}$.
%with $L(R^N_j)=L(R^N_j, \bar{Y}_{R^N_j})$. 
The resulting minimizer gives an approximate solution to the original problem (\ref{eq:estimator}). 

In problem (\ref{eq:approx.pc.1}), for fixed $m$ and $N$, solving the minimization over any segmentation $R^N_{1:m}$ on $\mathcal{S}_{1:N}$ is still not straightforward. Instead of obtaining $R^N_{1:m}$ through minimization, we introduce the following Local Neighborhood Segmentation (LNS) algorithm that provides a reasonable estimate efficiently by exploiting the spatial information of anomaly regions.

\begin{algorithm}[H]
\caption{Local Neighborhood Segmentation (LNS)}
\begin{algorithmic}[1]
	\Require
	$(Y_{1:N}$, $m$, $\xi)$
	,  $\mathcal{N} = \{1,...,N\}$
	\For {$k=1,\dots,m$}
		\State  Pick $i = \min \mathcal{N}$;
	    \State  Calculate $\widetilde R_k^N = \mathcal{S_{\mathcal{N}}} \cap \mathcal{B}\big(\s_i,\sqrt{\frac{n}{m\pi}}\big)$
	    \State Update $\mathcal{N} = \mathcal{N}\setminus \{j : \s_j\in \widetilde R_k^N\}$
	    \If{$|\widetilde R_k^N| \geq \xi$}
	    	\State  $k = k+1$
	    \EndIf
	\EndFor% \State $k = k-1$
	\Ensure  $\widetilde R_{1:m}^N$ 
\end{algorithmic}
\label{alg:CRS}
\end{algorithm}
\vspace{-1em}
% \zc{think about $\cup_{j \in \mathcal{N}}\s_j$ and $\{j : \s_j\in R_k^N\}$}
In the Algorithm \ref{alg:CRS}, $\mathcal{B}(\s, r)$ denotes a spatial ball on $\mathcal{S}$ centered at $\s$ with radius $r$, and $\xi$ is a pre-defined threshold on the size of anomaly regions. We always set the threshold as $\xi=20\cdot \lfloor\log_{10}(\sqrt{n})\rfloor/m$. Recall in Assumption \ref{assmp:signal}\,(ii) and (iii), we require that anomaly regions are large enough and well-separated (distant from each other). Inspired by this assumption, in each iteration we divide the remaining points into a circular region and  restrict their sizes to be no larger than $n/m$. We then intersect each of them with candidate lattice points to approximate an anomaly. If additional information about the anomaly regions is available, the shape of $\mathcal{B}(\s,r)$ can be adapted, for example, by using an ellipse or rectangle instead of a ball.
%When we have prior knowledge about the spatial anomalies, the shape of $\mathcal{B}(\s, r)$ can be modified, e.g. anomalies are long and thin, $\mathcal{B}(\s, r)$ can use ellipse by applying different weights along each coordinate axis when computing distances.

From Algorithm \ref{alg:CRS}, we obtain a reasonable $\widetilde R_{1:m}^N$ without solving the minimization problem over all possible segmentations.  The rest of points are all considered as baseline, denoted by $\widetilde R^N_0$. Next, we only need to search over combinations of $(m, N)$, to find the best $\widetilde R_{1:m}^N$ that minimizes the following cost:
\begin{equation*}
C(m,N)=\min_{1\leq N \leq n}\Bigg[ L\big(\widetilde R^N_0\big) + \min_m\bigg\{\sum_{j=1}^m L\big(\widetilde R_j^{N}\big) + \lambda \sum_{j=1}^{m}\big|\text{Co}\big(\widetilde R_j^{N}\big)\big|+ \beta m\bigg\} \Bigg].
\end{equation*}

In this way, we propose our algorithm for DPLS-SAD below. 
\begin{algorithm}[H] 
	\caption{Approximated Double penalized Least Squares for Spatial Anomaly Detection (DPLS-SAD)}
	\begin{algorithmic}[1]
		\Require 
		$(Y_{1:n},\beta, \lambda, M)$,
		\For{$N = 1,...,n$}
		\For{$m = 1,...,M$}
		\State  $\widetilde R_{1:m}^N=\textbf{LNS}(Y_{1:N}, m, \xi_m)$;
		\State Calculate 
        \State $\quad \widetilde R^N_0=\mathcal S\setminus \cup_{j=1}^m \widetilde R^N_j$
        \State $\quad {C}(m, N) = L(\widetilde R^N_0) + \sum_{j=1}^mL(\widetilde R_j^N) +\lambda\sum_{j=1}^m\big|\text{Co}(\widetilde  R_j^N)\big|+\beta m$
		%		\Else
		%		\State $\mathcal{C}^*(a,p,q) = \infty$
		%		\EndIf
		\EndFor
		% \State ${C}(m, N) = \mathcal{C}_{part}(m, \cup_{j=1}^m R_j^N) + \beta m + L(R^0_p) $
		\EndFor		
		\State $(\tilde m, \tilde N) = \arg\min_{1\leq m\leq n,  1\leq N\leq n} {C}(m, N)$ 
		\Ensure
		$(\tilde m, \widetilde R_{1: \tilde m}^{\tilde N})$
	\end{algorithmic}
    \label{alg:PLS-SAD}
\end{algorithm}
\vspace{-1em}
Algorithm \ref{alg:PLS-SAD} approximates the number and locations of anomaly regions by solving (\ref{eq:estimator}). Its computational cost is $O(M^2 n^2)$, where $M$  is a pre-specified upper bound on the number of anomaly regions. Consequently, when $M$ is finite, the overall complexity is quadratic in $n$.
 Algorithm \ref{alg:PLS-SAD} can be easily extended to higher dimensions, by changing $\mathcal{B}(\s, r)$ to a $d$-dimensional ball and setting the  radius $r=\sqrt[d]{\frac{n \Gamma(d/2+1)}{m(\pi)^{d/2}}}$ in the CRS algorithm, where $\Gamma(\cdot)$ is the gamma function.

\section{Additional simulations}\label{sec:add_sim}

We report additional simulations for both two-dimensional independent and dependent data as a supplement to Section \ref{sec:inde-simu} and Section \ref{sec:dep-sim}. For the independent setting, we examine the same 2D settings as in Section~\ref{sec:sim} at sample sizes $n=400$ and $n=2500$ with different SNR combinations. The results are summarized in Figure~\ref{result_20_50.plot} and Table~\ref{tab:400_2500} below. 

% \subsection{Simulation for independent data with $n = 400$ and $n = 2500$}

% We report additional simulations for 2-dimensional independent data at sample size $n=400$ and $n = 2500$ as a supplement to Section \ref{sec:inde-simu}. We examine the same 2D settings as in Section \ref{sec:sim}, with 9 different SNR combinations and applying the same parameter selection criterion from Section \ref{sec:inde-simu}. The results are summarized in Table \ref{tab:400_2500} and Figure \ref{result_20_50.plot}.

\begin{figure}[H]
	\centering
	\begin{tabular}{cccc}
		\includegraphics[scale = 0.248]{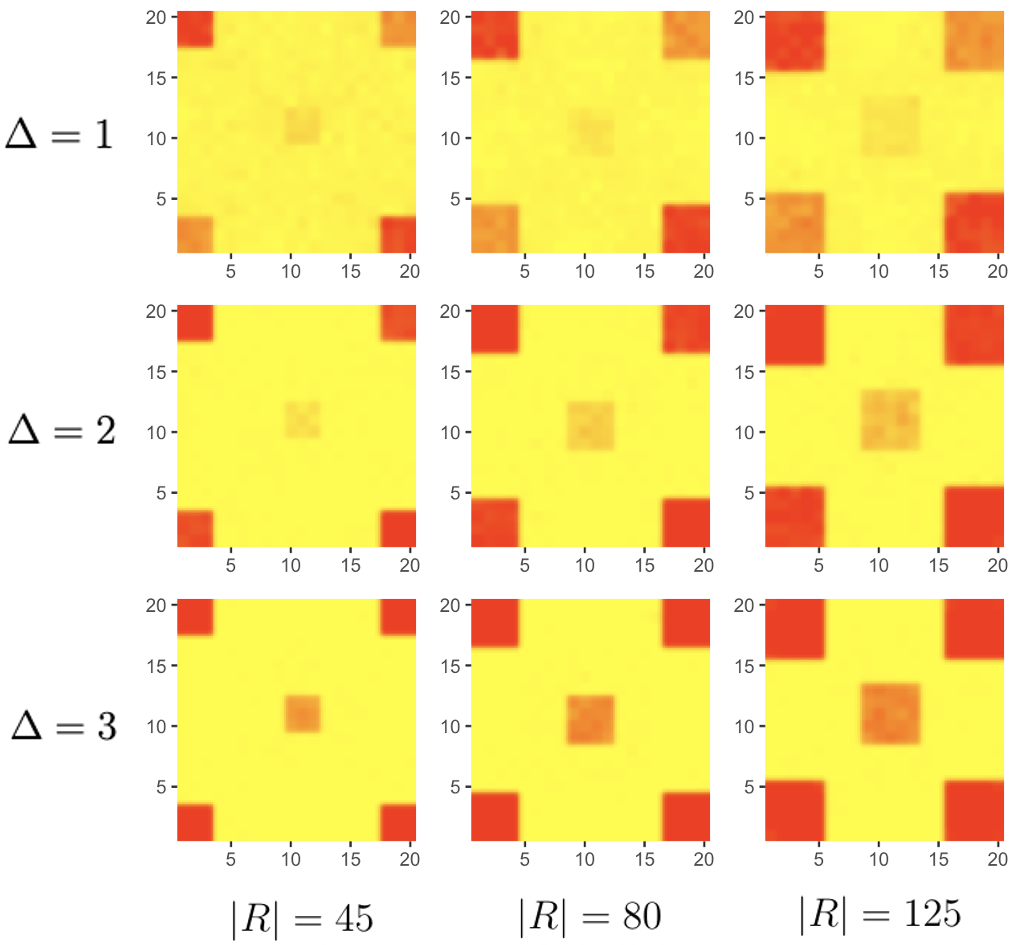} &\includegraphics[scale = 0.248]{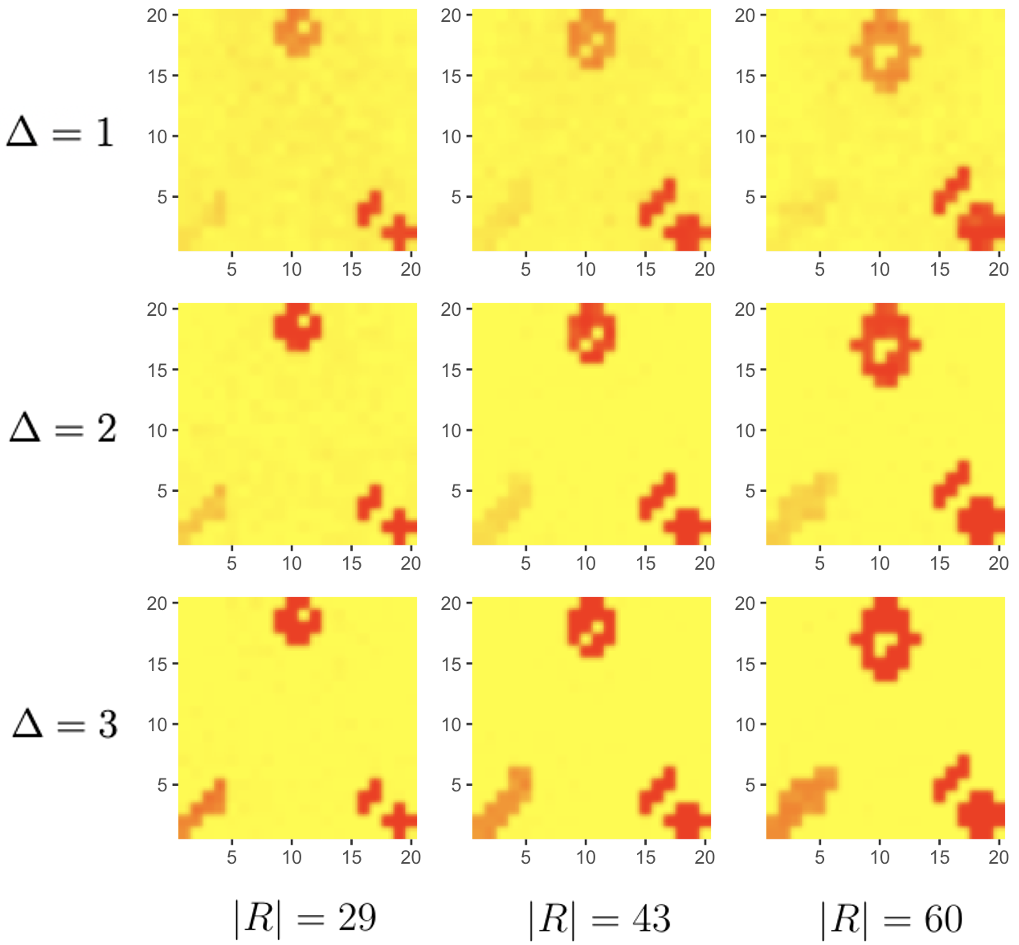} &\includegraphics[scale = 0.248]{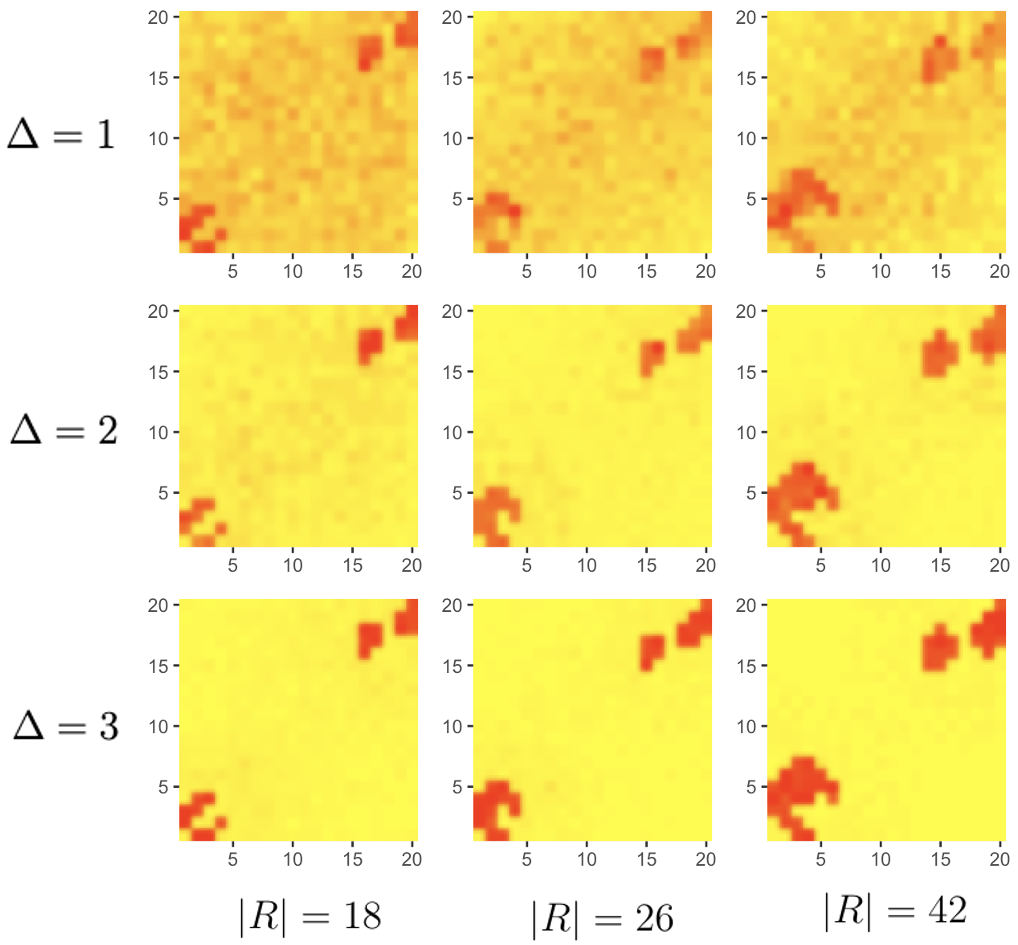}& \multirow{4}{*}[11ex]{\includegraphics[scale=0.3]{simulation_plot/legend_esti_1.png}}\\
		\scriptsize \hspace{1cm} (a1)  & \scriptsize \hspace{1cm} (a2) & \scriptsize\hspace{1cm} (a3)  & \\
		~\\
		\includegraphics[scale = 0.248]{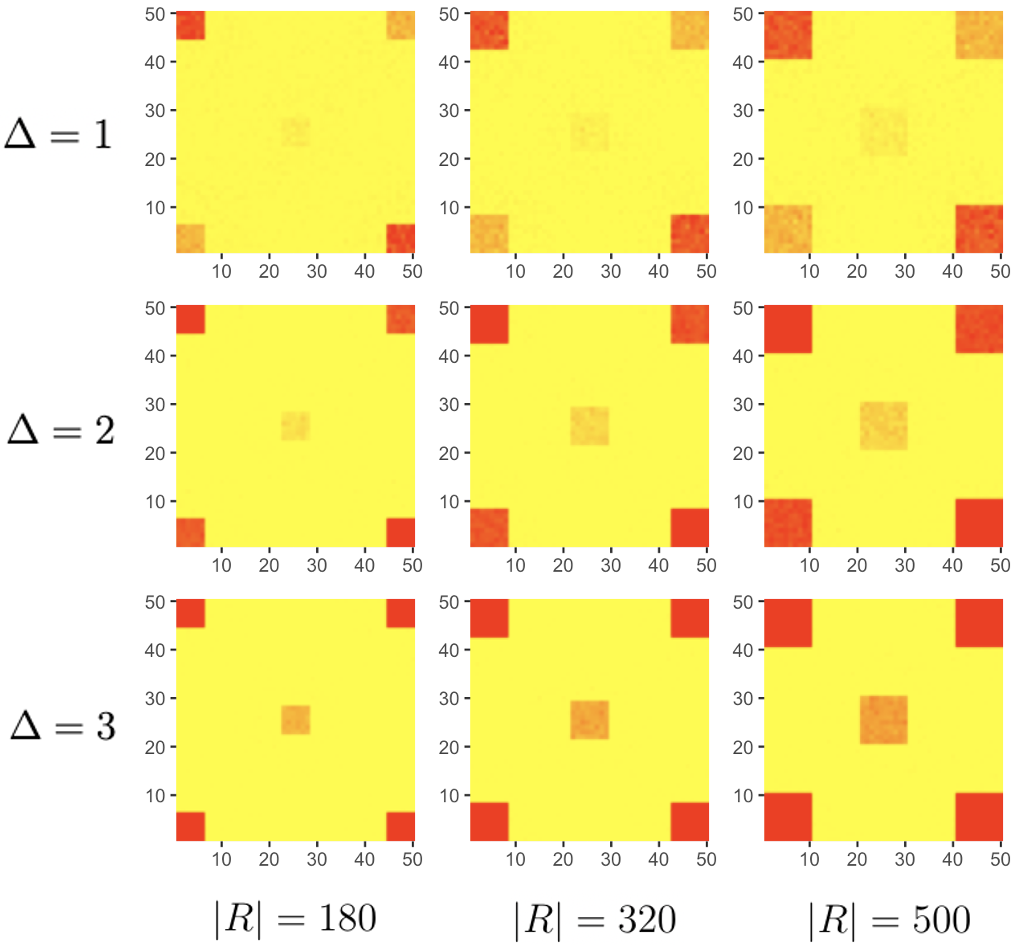} &\includegraphics[scale = 0.248]{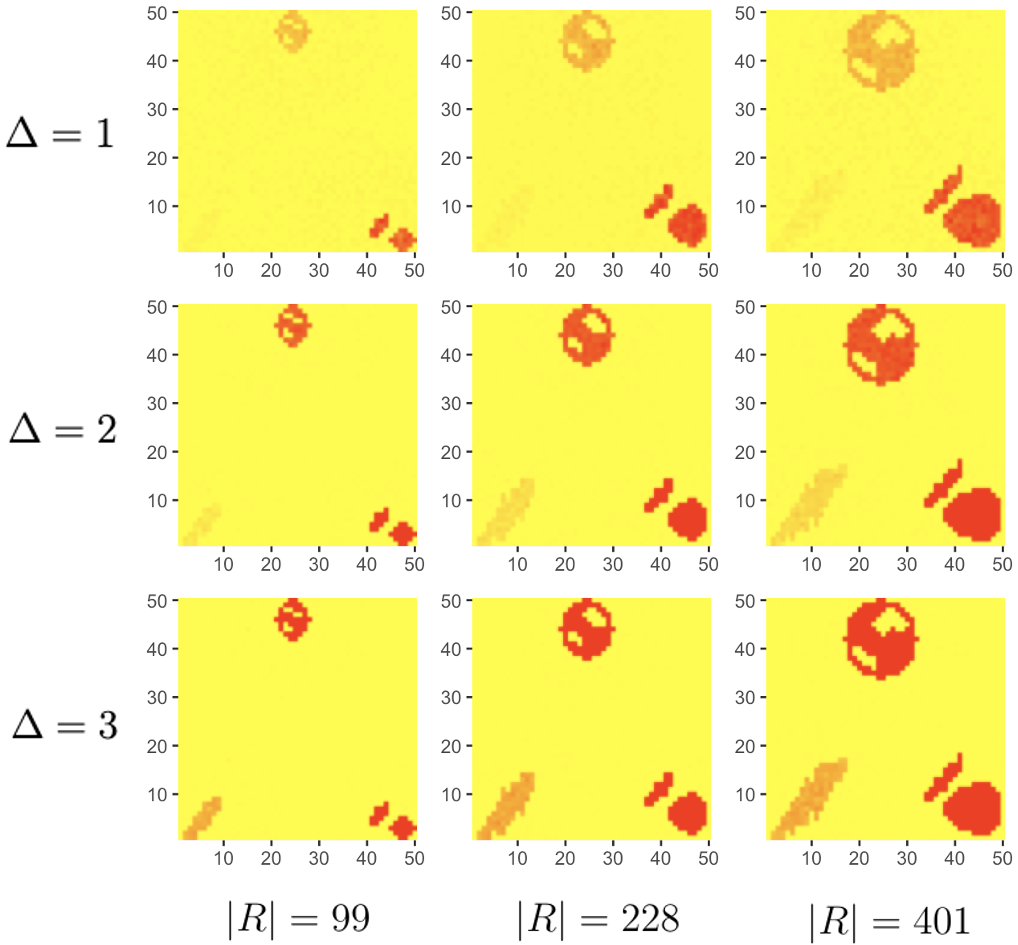} &\includegraphics[scale = 0.248]{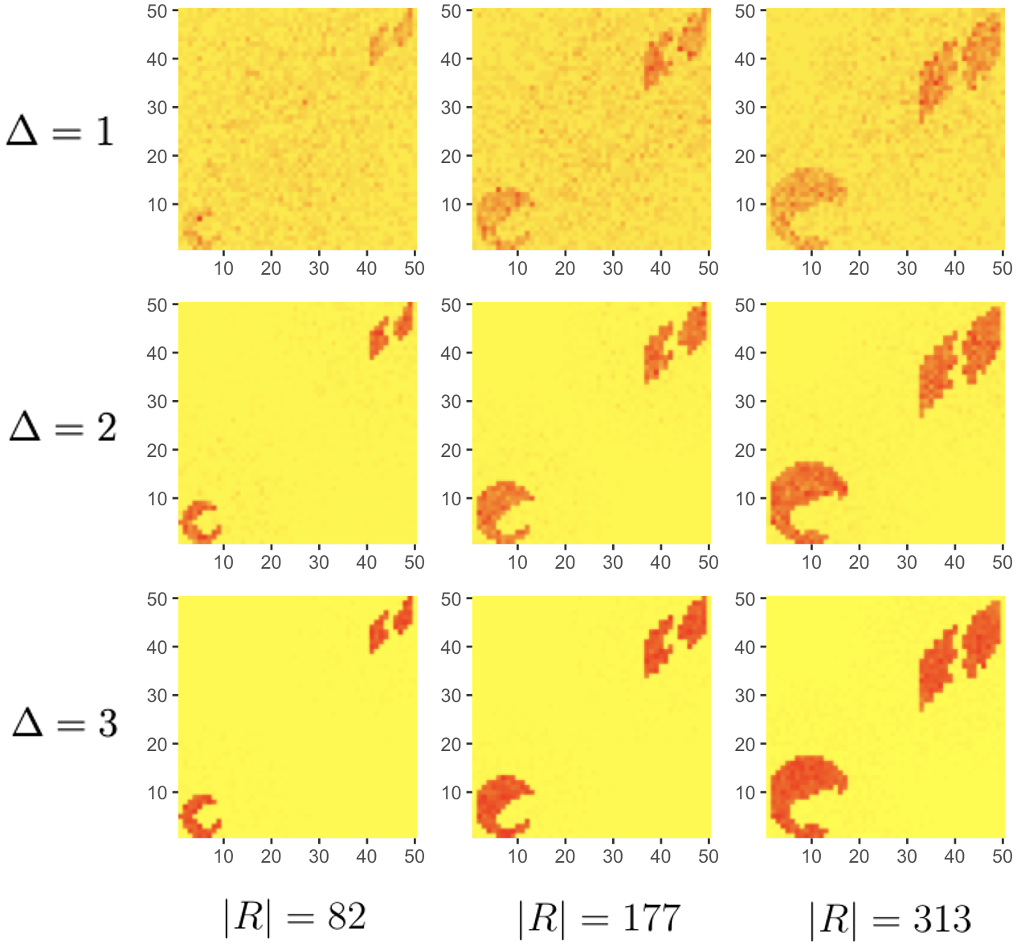}& \\
		\scriptsize \hspace{1cm} (b1)  & \scriptsize \hspace{1cm} (b2) & \scriptsize\hspace{1cm} (b3) & \\
	\end{tabular}
	\caption{ Frequency of points detected as anomalies, varying by 3 settings, with sample sizes $n=400$ (top panel) and $n=2500$ (bottom panel). Each setting and sample size includes 9 combinations of $\Delta$ and $|R|$ (top to bottom, $\Delta$ increasing; left to right, $|R|$ increasing).}
	\label{result_20_50.plot}
\end{figure}

\begin{table}[H]
	\centering
	\renewcommand{\arraystretch}{1.2}
	\setlength{\tabcolsep}{2.9pt}
	\scalebox{0.78}{\begin{tabular}{ccc|ccc|ccc|ccc|ccc|ccc|ccc}
			\toprule[1.5pt]
			\multicolumn{3}{c}{} & \multicolumn{9}{c|}{$n=400$} & \multicolumn{9}{c}{$n=2500$} \\ \cline{4-21}
			\multicolumn{3}{c}{} & \multicolumn{3}{c|}{Setting 1} & \multicolumn{3}{c|}{Setting 2} & \multicolumn{3}{c|}{Setting 3} &  \multicolumn{3}{c|}{Setting 1} & \multicolumn{3}{c|}{Setting 2} & \multicolumn{3}{c}{Setting 3} \\ \cline{4-21}
			& & \diagbox[width=1.25cm, height=1.2cm]{\hspace{0.3em} $\Delta$}{\multirow{1}{*}{\vspace{-0.7em}$|R|$}} & 45& 80 & \multicolumn{1}{c|}{125}  & 29 &43 & \multicolumn{1}{c|}{60}   & 18 & 26 & 42 & 180 & 320 & 500 & 99 & 228 & 401 & 82 & 177 & 313  \\
			\bottomrule[1.5pt]
			\multirow{6}{*}{NoC(\%)}& \multirow{3}{*}{\small DPLS-SAD} & 1 & 11 & 28 & 41 & 18 & 24 & 26 & 5 & 16 & 32 & 37 & 44 & 63 & 19 & 25 & 48 & 9 & 14 & 24 \\ 
			& & 2 & 26 & 54 & 80 & 31 & 34 & 52 & 25 & 54 & 78 & 59 & 87 & 97 & 37 & 68 & 86 & 68 & 85 & 92 \\ 
			& & 3 & 80 & 94 & 99 & 88 & 89 & 96 & 55 & 88 & 99 & 98 & 98 & 100 & 96 & 100 & 100 & 99 & 100 & 100 \\
			\cmidrule[1pt]{2-21}
			& \multirow{3}{*}{\small DCART} & 1 & 31 & 20 & 15 & --- &  --- &  --- & --- & --- & --- & 29 & 17 & 4 & 11 & 39 & 40 & --- & --- & --- \\
			& & 2 & 29 & 34 & 13 & 15 & 18 & 42 & --- &--- &  --- & 20 & 30 & 14 & 20 & 21 & 42 & --- & 25 & 10 \\ 
			& & 3 & 31 & 33 & 33 & 12 & 8 & 19 & --- & 22 & 33 & 35 & 25 & 10 & 51 & 44 & 56 & 23 & 32 & 31 \\
			\bottomrule[1.5pt]
			\multirow{6}{*}{Err(\%)} &\multirow{3}{*}{\small DPLS-SAD} & 1 & \,75\, & \,54\, & \,49\, & 114 & \,82\, & \,69\, & 247 & 191 & 132 & 67 & 63 & 61 & 82 & 73 & 63 & 126 & 107 & 100 \\ 
			& & 2 & 24 & 19 & 17 & 54 & 35 & 30 & 159 & 97 & 71 & 24 & 22 & 20 & 38 & 33 & 25 & 86 & 78 & 73 \\
			& & 3 & 10 & 8 & 8 & 20 & 17 & 13 & 80 & 41 & 33 & 12 & 11 & 10 & 19 & 17 & 14 & 45 & 39 & 37 \\
			\cmidrule[1pt]{2-21}
			& \multirow{3}{*}{\small DCART} & 1 & 48 & 41 & 44 & --- &  --- &  --- & --- & --- & --- & 41 & 50 & 56 & 103 & 89 & 73 & --- & --- & --- \\
			& & 2 & 37 & 33 & 40 & 111 & 94 & 70 & --- & --- &  --- & 29 & 35 & 40 & 72 & 60 & 49 & --- & 103 & 96 \\ 
			& & 3 & 25 & 25 & 31 & 87 & 72 & 58 & --- & 95 & 98 & 26 & 32 & 40 & 54 & 50 & 43 & 127 & 98 & 74 \\
			\bottomrule[1.5pt]
	\end{tabular}}
	\caption{ Performances of DPLS-SAD and DCART, where   "---" denotes that DCART estimates all the points as baseline in more than 95\% simulations. In Settings 3,  we scale both $\beta$ and $\lambda$ by factors of $0.65$.}
	\label{tab:400_2500}
\end{table}

% \subsection{Simulation for dependent data with $n = 10000$\wb{need to change}}
For the dependent data simulation, we report additional results at sample size $n=10000$. We examine the same setting as in Section~\ref{sec:dep-sim}, and consider different combinations of $\zeta$ and $\Delta$. The results are summarized in Figure~\ref{result_dep&3D-100} and Table~\ref{tab:dep&3D-100} below.

\begin{figure}[H]
    \centering
    \includegraphics[scale = 0.45]{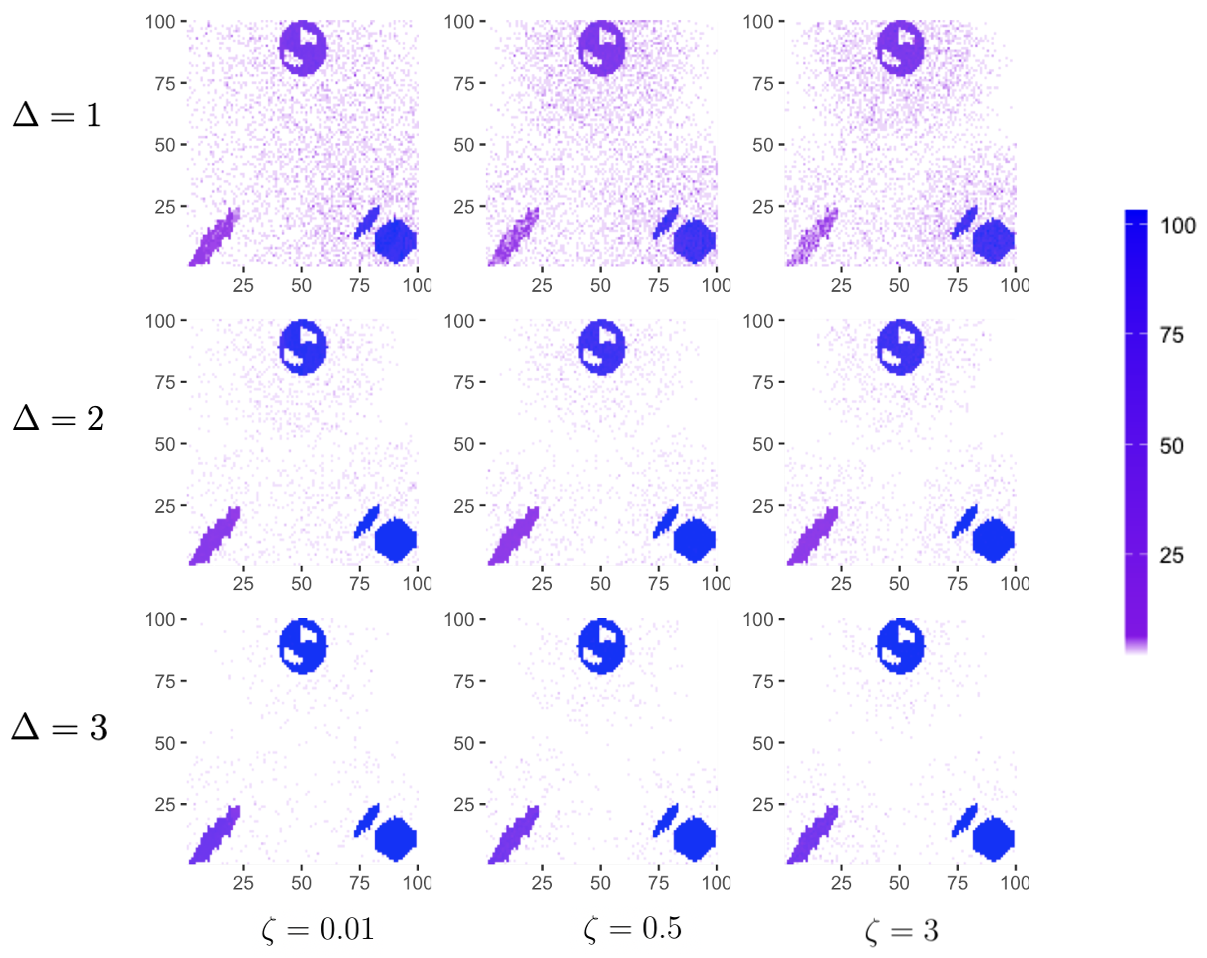}
    \caption{Frequency of points detected as anomalies for 2D dependent data with $n = 10000$. For 2D dependent data, we include 9 combinations of $\Delta$ and $\zeta$ (top to bottom, $\Delta$ increasing; left to right, $\zeta$ increasing).}
\label{result_dep&3D-100}
\end{figure}

\begin{table}[H]
	\centering
	\renewcommand{\arraystretch}{1.3}
	\vspace*{1.3mm}
	\setlength{\tabcolsep}{2.5pt}
	\scalebox{0.9}{\begin{tabular}{ccccccccccccc}
			\toprule[1.5pt]
			& \multicolumn{7}{c}{2D dependent data} \\ \cline{2-8}
			& \multicolumn{3}{c}{NoC(\%)} &\,\,\,\,& \multicolumn{3}{c}{Err(\%)}\\
			& $\zeta=0.05$ & $\zeta=0.5$ & $\zeta = 3$ & &$\zeta = 0.05$ & $\zeta = 0.5$ & $\zeta = 3$\\ 
			\bottomrule[1pt]
			\multicolumn{1}{c}{$\Delta = 1$\,\,\,\,\,\,\,} & 18 & 40 & \multicolumn{1}{c}{42} && 71 & 71 & \multicolumn{1}{c}{70}\\
			\multicolumn{1}{c}{$\Delta = 2$\,\,\,\,\,\,\,} & 30 & 59 & \multicolumn{1}{c}{69} && 25 & 29 & \multicolumn{1}{c}{29}\\
			\multicolumn{1}{c}{$\Delta = 3$\,\,\,\,\,\,\,} & 53 & 92 & \multicolumn{1}{c}{100} && 15 & 15 & \multicolumn{1}{c}{15}\\
			\bottomrule[1.5pt]
	\end{tabular}}
    \caption{Performance of DPLS-SAD for 2D dependent with $n = 10000$. Results are based on $100$ Monte Carlo simulations, where we fix $|R| = 787$ for 2D dependent data.}
	\label{tab:dep&3D-100}
\end{table}

\section{Proof of theorems}
In the rest of the supplementary material, we include proofs of Theorems \ref{thm:consistency}\,-\,\ref{thm:consistency_dep} and Theorem \ref{thm:meanconsistency}, which are available upon request

\end{appendices}

\end{document}